\documentclass[useAMS,usenatbib]{mn2e}
\usepackage{epsfig}
\usepackage{natbib}
\usepackage{subfigure}
\usepackage{multirow}
\usepackage{amssymb}

\newcommand{\msun}{\mbox{M$_{\odot}$}}

\title{The Role of Dissipation in the Scaling Relations of Cosmological Merger Remnants}

\author[Covington et al.]
{M. D. Covington$^{1,2,3}$, 
J.~R. Primack$^{1,2}$\footnotemark[1], 
L.~A. Porter$^{1,2}$, 
D.~J. Croton$^{4}$, 
R.~S. Somerville$^{5,6}$,\newauthor 
and A. Dekel$^{2,7}$ \\
        $^1$Department of Physics, University of California, Santa Cruz, California 95064, USA\\
	$^2$Santa Cruz Institute for Particle Physics, University of California, Santa Cruz, California 95064, USA\\	
	$^3$NSF International Research Fellow, Karst Research Institute ZRC SAZU,
Titov trg 2, SI-6230 Postojna, Slovenia\\
        $^4$Centre for Astrophysics and Supercomputing, Swinburne University of Technology, Melbourne, Australia\\
        $^5$Space Telescope Science Institute, 3700 San Martin Drive, Baltimore,  Maryland 21218, USA\\
        $^6$Department of Physics and Astronomy, Johns Hopkins University, Baltimore, Maryland 21218, USA\\
	$^7$Racah Institute of Physics, The Hebrew University, Jerusalem, 91904, Israel\\        
	}

\begin{document}
\maketitle
\begin{abstract}
There are strong correlations between the three structural properties of elliptical galaxies -- stellar mass, velocity dispersion and size -- in the form of a tight ``fundamental plane" and a ``scaling relation" between each pair. Major mergers of disk galaxies are assumed to be a mechanism for producing ellipticals, but {\bf  semi-analytic galaxy formation models (SAM) have} encountered apparent difficulties in reproducing the observed slope and scatter of the size-mass relation. We study the scaling relations of merger remnants using progenitor properties from two SAMs. We apply a simple merger model that includes gas dissipation and star formation based on theoretical considerations and simulations.  {\bf Combining the SAMs   and the merger model allows calculation of the structural properties of the remnants of major mergers that enter the population of elliptical galaxies at a given redshift.} Without tuning the merger model parameters for each SAM, the results roughly match the slope and scatter in the observed scaling relations and their evolution in the redshift range $z=0-3$. Within this model, the observed scaling relations, including the tilt of the fundamental plane relative to the virial plane, result primarily from the decrease of gas fraction with increasing progenitor mass.  The scatter in the size-mass relation {\bf of the remnants} is {\bf reduced from that of the progenitors} because of a correlation between progenitor size and gas fraction at a given mass.

\end{abstract}
\begin{keywords}
galaxies: interactions -- galaxies: evolution -- galaxies: elliptical and lenticular, cD -- galaxies: formation -- methods: .
\end{keywords}
	\footnotetext[1]{email: joel@scipp.ucsc.edu}

\section{Introduction}
\label{sec:intro2}

The merging of disk galaxies is one of the main hypothesized
mechanisms for the formation of elliptical galaxies. Simulations have
shown that mergers of disks with similar masses can effectively
disrupt the ordered rotation in the disks and convert it into random
velocity support, creating merger remnants that appear similar to
observed elliptical galaxies \citep{TT72, T77, Barnes92, MH94dsc}.
Furthermore, the $\Lambda$CDM cosmology predicts the hierarchical
buildup of galaxies through a sequence of mergers.  These results
suggest that merging is a likely mechanism for the production of
elliptical galaxies.

However, observed ellipticals follow a number of scaling relations,
including relatively tight relations between stellar mass and velocity
dispersion, the Faber-Jackson relation \citep{FJ}, and between size
and stellar mass \citep{Kormendy77}.  Furthermore, observed
ellipticals fall on a tight plane, the fundamental plane (FP), in the
three-dimensional space of stellar mass, size, and velocity dispersion
\citep{Djorgovski87, Dressler87}.  Recent studies of the Sloan Digital
Sky Survey (SDSS) have provided excellent statistics on these scaling
relations in the local universe \citep{Shen03, Bernardi03a,
  Bernardi03b, Padmanabhan04, Gallazzi06, Shankar10b}, and studies
using high redshift surveys have provided evidence for the evolution
of these relations over cosmological time \citep{Barden05, McIntosh05,
  Trujillo06,Trujillo07, Cimatti08,vanderWel08,Buitrago08, Saracco09,
  vanDokkum10, Fan10, Mancini10, Williams10}.  If mergers are a major
mechanism for the production of elliptical galaxies then they must be
able to produce the correct scaling relations as well as the evolution
of these scaling relations over time. Theoretical {\bf high-resolution, hydrodynamical} studies have shown
that simulations of gas-rich galaxy mergers are capable of reproducing
the observed scaling relations of elliptical galaxies if the correct
progenitor properties are used \citep{Dekel06, RobertsonFP,
  HopkinsFP,Bournaud:2011a}.  {\bf These studies have been successful in creating high-redshift compact ellipticals from major mergers of gas-rich compact disk galaxies \citep{Bournaud:2010a, Wuyts:2010b}.} This is a step toward verifying the production of
scaling laws through mergers. However, current computing power only
allows the simulation of relatively small numbers of mergers, and the
space of possible merger initial conditions and progenitor properties
is quite large.  Furthermore, these simulations are not placed within
a cosmological context, making it more difficult to explore in detail
the origin and evolution of scaling relations.

Currently, the primary theoretical tool for studying the evolution of
statistical samples of galaxies over cosmological timescales is
semi-analytic modeling (SAM) \citep{kwg1993, Cole94, SP1999, Cole00,
  Galics03, Croton06, DeLucia06, Bower06, S08, Fontanot09, Benson10,
  BensonBower10, Cook10, Guo10}.  These models combine dark matter
halo merger trees with analytic recipes for populating the halos with
galaxies.  However, semi-analytic models (SAMs) do not currently
incorporate realistic formulas for predicting the properties of the
remnants of galaxy mergers including the effects of dissipation.  The
agreement between current SAMs and observed early-type scaling
relations is not impressive.  In particular, the observed size-mass
relation of early-types is steeper than that of their potential
late-type progenitors \citep{Shen03}, and the scatter in the observed
size-mass relation for early-types is remarkably small \citep{Shen03,
  Nair10}.  The disspationless merger models currently implemented
within SAMs have thus far been unable to reproduce these features
\citep[e.g.,][]{Shankar10a,Guo10}.  {\bf Using a simple power law
  dissipation model \citet{KhochfarSilk06} were able to reproduce the
  redshift-size evolution of elliptical galaxies, which, along with
  the high-resolution hydrodynamical simulations described above,
  suggests that dissipative effects may play an important role in
  determining elliptical galaxy scaling relations.}

\citet{remnants} recently developed a physically-motivated analytic
model for predicting the stellar half-mass radii and central velocity
dispersions of merger remnants \citep{remnants}.  The parameters in
this new merger model were calibrated using a suite of {\bf high-resolution hydrodynamical} galaxy merger
simulations (see Section \ref{sec:methods2}).  Here we implement a
simplified version of this model using post-processing of merger
outputs from the SAMs developed by \citet{Croton06}, based on the
Millennium simulation \citep{Springel05}, and \citet{S08}.  This
results in a population of tens of thousands of merger remnants over a
large range of redshifts ($0<z<3$) complete with predicted values of
size, stellar mass, and velocity dispersion.  Comparison of the
modeled population of ellipticals with the observed scaling relations
provides an important test of the merger hypothesis as well as
physical insight into the origin and evolution of these relations via
merging.  {\bf It is important to note that the model presented here is calibrated only for major mergers of disk-dominated galaxies, and that we make no attempt to model subsequent evolution following the formative major merger.}  Future work will {\bf expand our model to include minor mergers and mergers of bulge-dominated galaxies.  We will also} implement our merger model
self-consistently within the semi-analytic machinery rather than by
post-processing.

Section 2.1 describes our analytic merger model for calculating the
properties of stellar spheroids from galaxy mergers including energy
losses from dissipation.  Section 2.2 explains how we implement our
merger model using outputs from the \citet{Croton06} and \citet{S08}
SAMs.  Section 3 systematically explores the effects of the merger
model.  In order to turn the rather shallow size-mass relation of disk
galaxies into the steeper size-mass relation of observed early-type
galaxies, we find that it is essential to include both dissipation and
the decreasing gas content of more massive progenitor disk galaxies.
In Section 4 we summarize the observational results that we compare
with our model outputs, focusing especially on Sloan Digital Sky
Survey (SDSS) data for nearby galaxy size vs.  mass and other scaling
relations, and on data from several surveys \citep{Trujillo06} for the
evolution of these relations to higher redshifts.  In section 5 we use
the predicted properties of progenitor disk galaxies from the two
semi-analytic models to predict the size-mass and other relations for
spheroids formed in major gas-rich mergers and compare them with the
observations out to redshift $z=3$.  Section 6 summarizes these
results and discusses their implications and some follow-on studies
that are in progress.  Finally, an Appendix describes an improvement
in the treatment of the central dark matter in the analytic merger
model of \citet{remnants} that we used in \citet{Covington08} and in
the present paper.

\section{Methodology}
\label{sec:methods2}
We use a combination of modeling approaches to construct a theoretical
framework for predicting the evolution of early-type scaling relations
over cosmological time.  In previous work, a large suite of
hydrodynamical galaxy merger simulations were developed \citep{thesis,
  Cox05, Cox08}.  These simulations were performed using the
N-body/SPH code GADGET \citep{SpGad,SH03}, and include hydrodynamics,
star formation, and stellar feedback.  The simulation suite contains
mergers with a wide variety of progenitor properties, mass ratios, and
merger orbits.  Variations in progenitor properties include a range of
stellar masses, gas fractions, dark matter halo concentrations, bulge
fractions, baryonic fractions, and gas disk sizes.

In subsequent work, Covington et al.~(2008, hereafter C08) constructed
a physically-motivated analytic galaxy merger model capable of
predicting the half-mass radii, stellar masses, and velocity
dispersions of galaxy merger remnants given the properties of the
{\bf disk} progenitor galaxies and the initial orbits of the mergers.  This model
was calibrated using the galaxy merger suite described above.  Unlike
previous similar models \citep{Cole00, Galics03}, this model includes
the effects of star formation and energy loss due to dissipation.
Here we combine a simplified version of this new galaxy merger model
with merger rates and progenitor properties predicted by two
semi-analytic models (SAMs) in order to explore the creation and
evolution of the scaling relations of early-type galaxies.

\subsection{Description of the Merger Model}
\label{ssec:model}

Previous models for predicting the sizes and velocity dispersions of
galaxy merger remnants employed a combination of energy conservation
and the virial theorem \citep{Cole00, Galics03}.  Assuming homology
between progenitors and remnants then allows a straightforward
calculation of remnant properties.  However, the most common mergers
at higher redshifts are `wet' mergers of gaseous galaxies, and the
approach used previously does not account for energy losses due to gas
processes.  Recent work has suggested that these dissipative effects
play an important role in the formation of elliptical galaxies
\citep{Cox05, RobertsonFP, Dekel06, Ciotti07, Cox08, HopkinsFP}{\bf, though there is also evidence that disks may preserved in major, gas-rich mergers \citep{Springel:2005c,Robertson:2006b,Hopkins:2009d}}.  C08
develops a framework for modifying the energy conservation approach to
include dissipative losses due to gas processes.

In order to account for dissipative losses, we include a radiated
energy term within the energy conservation equation.  Furthermore,
since stars form during the merger, we include the mass of gas that
will form stars when calculating the internal energies of the
progenitor galaxies.  C08 found that the energy lost and the number of
new stars formed are a function of both the gas fraction and the
initial merger orbit.  Closer, more disruptive encounters will result
in greater total energy losses and larger numbers of stars forming.
However, the set of orbital parameters in the simulations used to
calibrate the model was not intended to be statistically
representative of mergers in the real Universe.  Rather, it was
intended to sparsely span the range of reasonable values.  In the
present work, we find that when the merger model of C08 is applied to
SAMs using the distribution of merger orbital parameters observed in
cosmological N-body simulations \citep{Benson05} there is relatively
small scatter in remnant properties as the result of orbit.
Consequently, for this paper we use a simplified version of the model
in C08 where remnant properties are not a function of orbit. This
simplified model is calibrated to the same suite of galaxy merger
simulations \citep{thesis}, effectively averaging over the range of
orbits within the suite. 

Additionally, the model for calculating remnant central dark matter
fractions developed in C08, when applied within SAMs, fails for a
subset of cases with large radii and high gas fractions.  Therefore,
here we use a new model for calculating central dark matter fractions
that has a broader range of applicability (see Appendix A for
details).  Together, the modifications described result in a simpler
model that captures the core dynamics of the more complicated model of
C08.  This has the added benefit of reducing the number of unknown
model parameters.  However, the scaling relations produced by the two
models are not identical, and further work now underway implementing these
models within SAMs will be needed to determine whether the simpler
model presented here is sufficient (see Section 5.3 for a discussion
of differences).

The first step in the new model is to calculate the star formation
efficiency of the merger, $e$.  Following \citet{SPF} we use 
\begin{equation}
e= e_{\rm 1:1}\left( \frac{M_{\rm sat}}{M_{\rm primary}} \right) ^\gamma .
\end{equation}
This formula was calibrated to our merger simulations in \citet{Cox08}
resulting in parameter values of $e_{\rm 1:1}=0.55$ and $\gamma=0.69$.
For each merger $M_{\rm sat}$ is taken to be the total mass (dark
matter plus baryons) of the less massive progenitor and $M_{\rm
  primary}$ is taken to be the total mass of the more massive
progenitor.  For each progenitor
\begin{equation}
M_{\rm ns}= e M_{\rm gas},
\end{equation}
where $M_{\rm ns}$ is the mass of new stars formed in the merger
and $M_{\rm gas}$ is the initial (cold) gas mass of the progenitor.

As in C08, we use an energy conservation equation to calculate the
half-mass radius of the merger remnant.  This equation includes final
and initial internal energy terms and a term for energy loss due to
gas dissipation,
\begin{equation}
E_{\rm final}= E_{\rm init} + E_{\rm rad}.
\end{equation}
The energy terms are calculated as follows:
\begin{equation}
E_{\rm init}= G \left( \frac{(M_{\rm s,1} + M_{\rm
    ns,1})^2}{R_{\rm 1}} + \frac{(M_{\rm s,2} + M_{\rm
    ns,2})^2}{R_{\rm 2}} \right),
\end{equation}
\begin{equation}
E_{\rm final}= G \frac{(M_{\rm s,1} + M_{\rm ns,1} +
  M_{\rm s,2} + M_{\rm ns,2})^2}{R_{\rm final}},
\end{equation}
and
\begin{equation}
E_{\rm rad}= C_{\rm rad} f_{\rm g} E_{\rm init},
\end{equation}
where $M_{\rm s,i}$ is the initial stellar mass of progenitor $i$,
$R_{\rm i}$ is the initial 3D {\bf stellar} half-mass radius of progenitor $i$, and
$R_{\rm final}$ is the 3D {\bf stellar} half-mass radius of the remnant.  $C_{\rm
  rad}=2.75$ is a constant parameter calibrated using the merger
simulations, and
\begin{equation}
f_{\rm g}= \frac{M_{\rm g,1} + M_{\rm g,2}}{M_{\rm s,1} + M_{\rm g,1} + M_{\rm s,2} + M_{\rm g,2}}
\end{equation}
is the merger gas fraction {\bf where $M_{\rm s,i}$ and $M_{\rm g,i}$
  are the stellar and gas masses, respectively, of the progenitors}.
{\bf Unlike in C08, here orbital energy is neglected, because we find
  that, using orbital parameters observed in cosmological N-body
  simulations \citep{Benson05}, the scatter resulting from orbital
  effects is small.}  {\bf In the limit of no dissipation (i.e.$f_{\rm
    g}=0$), Eqs. 3-5 in our paper reduce to Eqs. 1 and 2 of
  Naab et al. (2009), if one assumes the proportionality of effective
  and gravitational radii as we do here.}

\begin{figure*}
\begin{center}
\subfigure[][]{\resizebox{8cm}{!}{\includegraphics{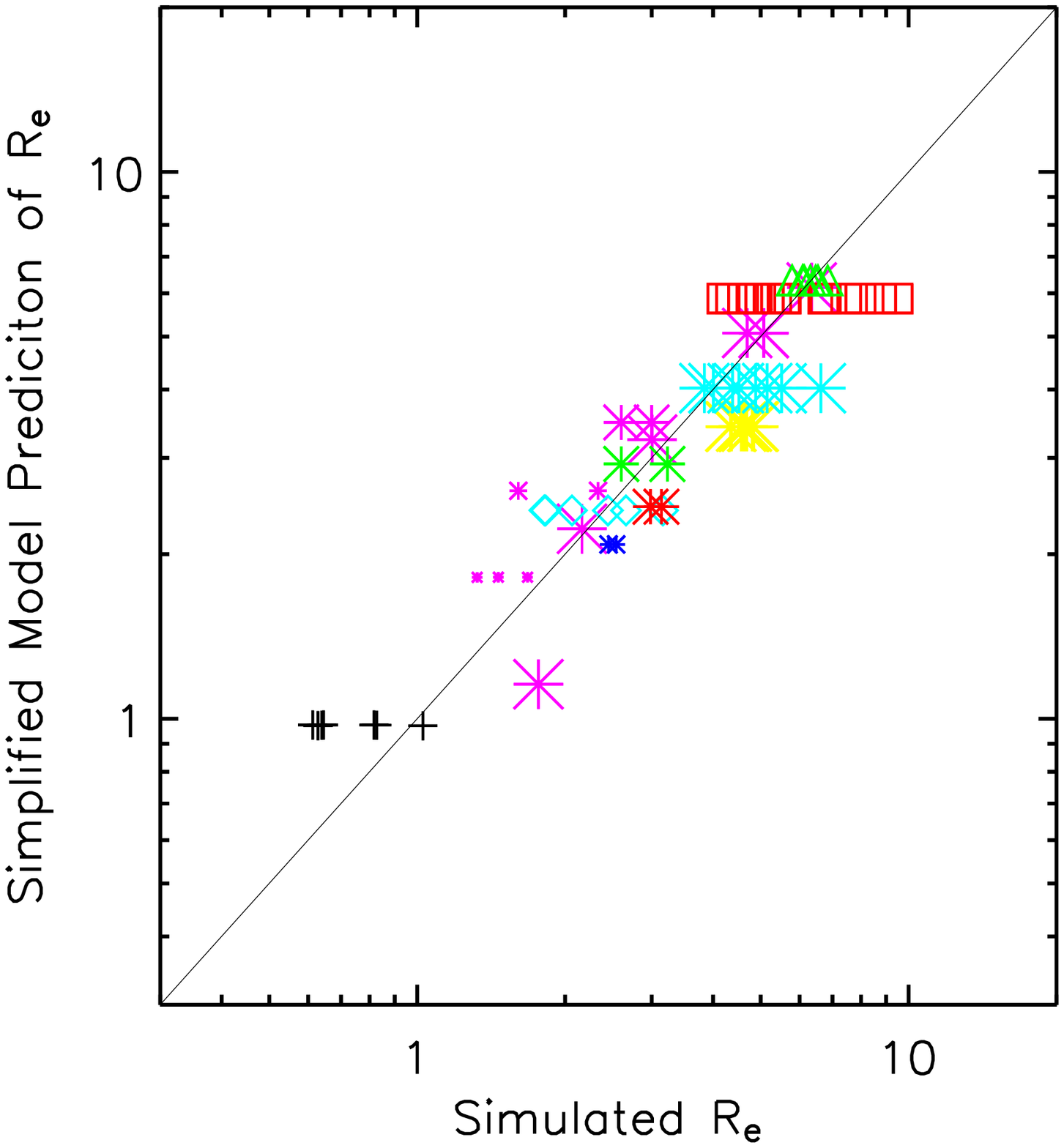}}\hspace{0.0cm}}
\subfigure[][]{\resizebox{8cm}{!}{\includegraphics{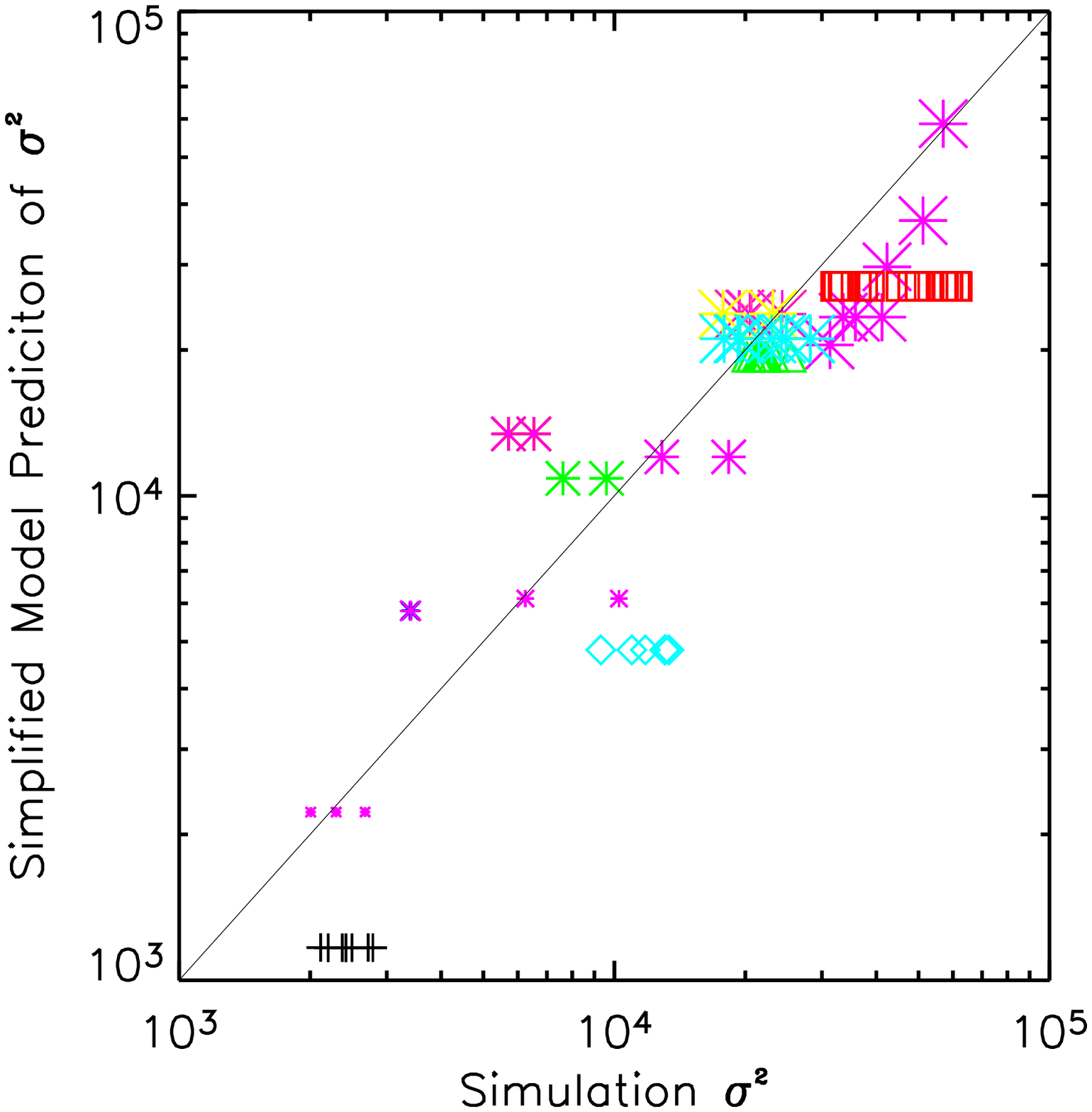}}\hspace{0.0cm}}\\
\subfigure[][]{\resizebox{8cm}{!}{\includegraphics{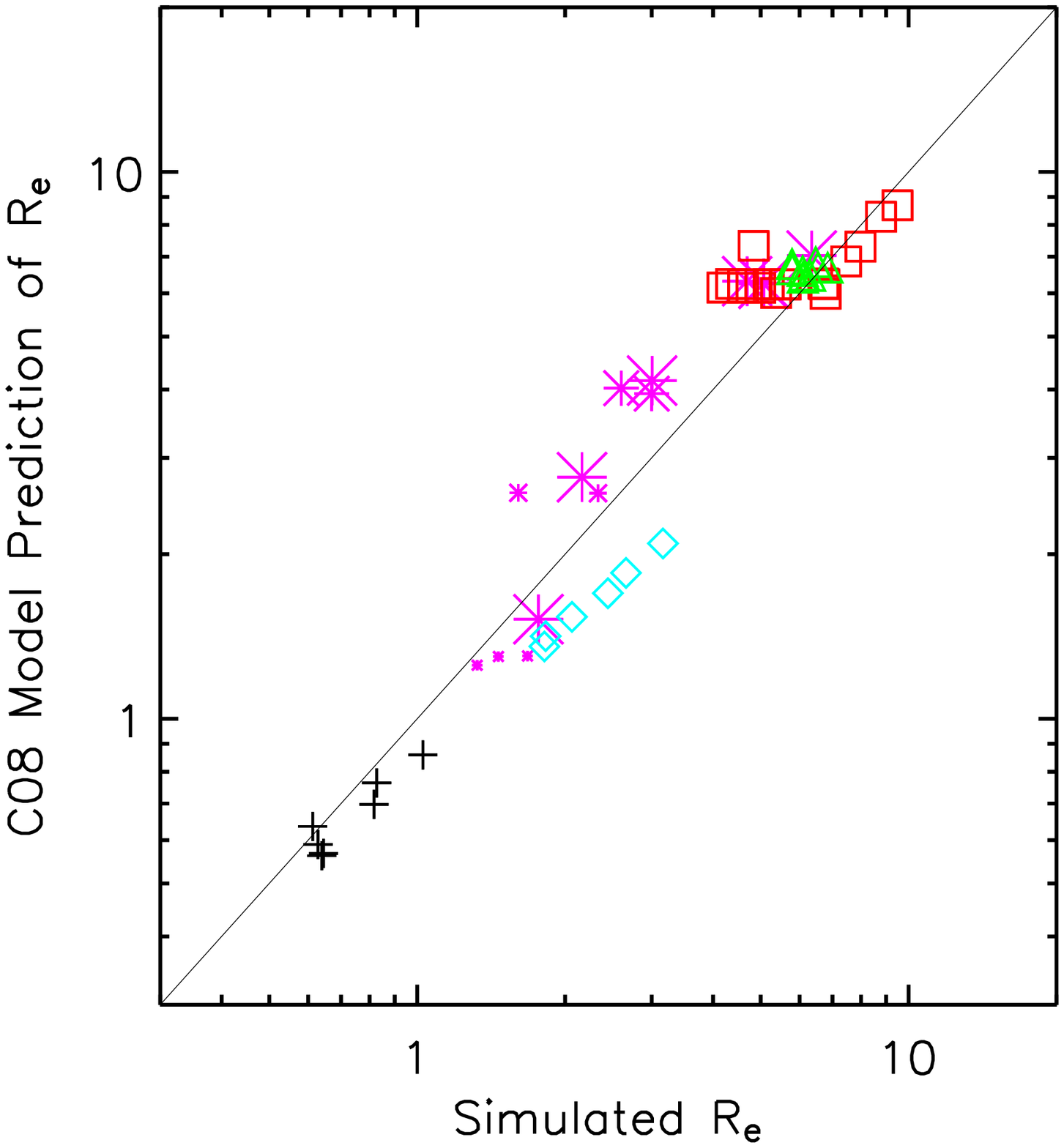}}\hspace{0.0cm}}
\subfigure[][]{\resizebox{8cm}{!}{\includegraphics{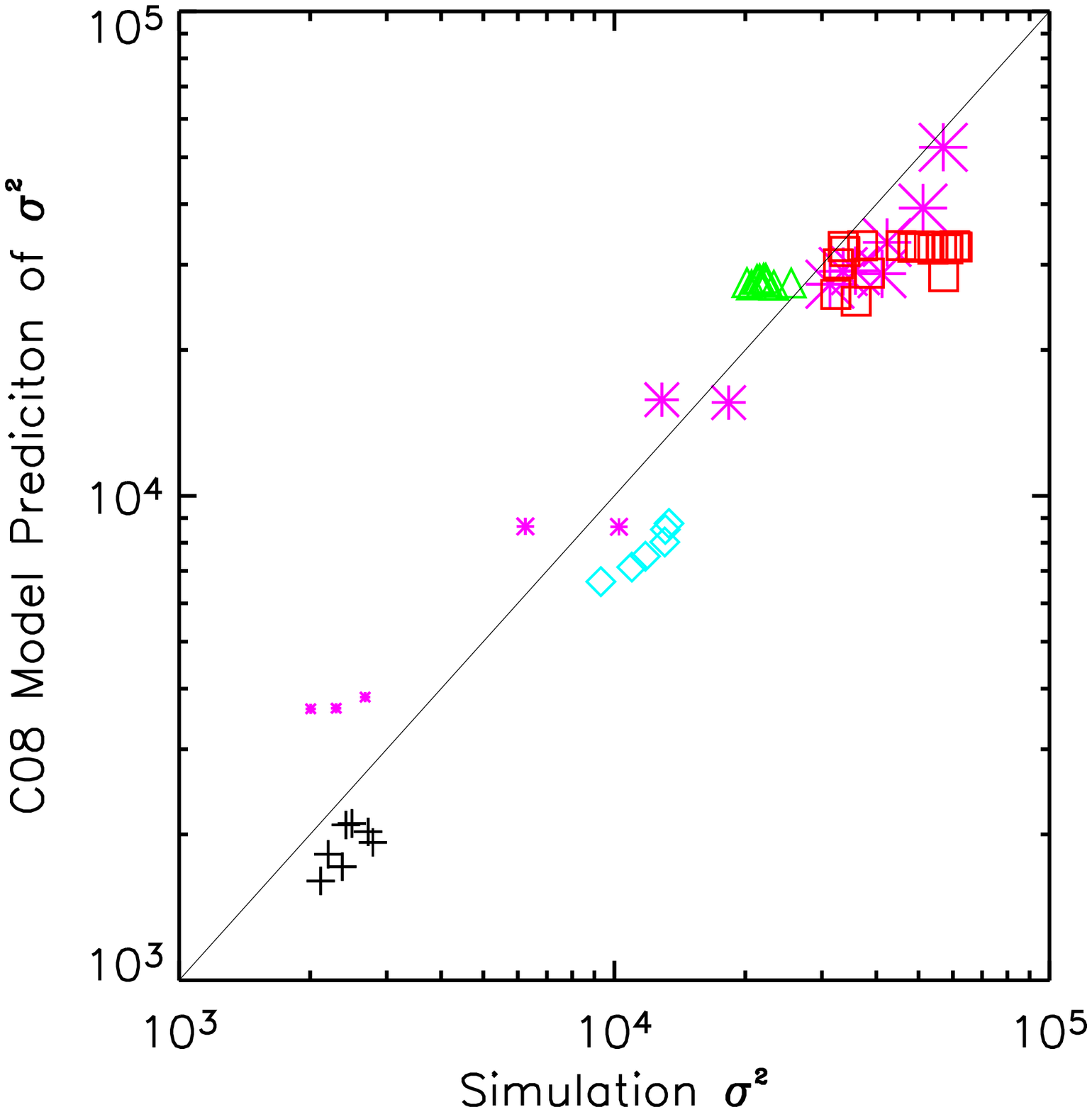}}}\\
\caption{Half-mass radii of merger remnants from our simulation suite
  plotted against radii predicted from our simplified analytic merger
  model, (a), and the full C08 model with central dark matter density
  calculated as in Appendix A (c). Panels (b) and (d) show the
  predictions for velocity dispersions for the simple model and C08
  plus Appendix A, respectively.  Different symbols represent
  different types of progenitor galaxies, {\bf described in detail in C08}.  This new, simpler merger
  model does not capture variations due to initial orbit, {\bf  which is seen in the horizontal spread within symbols of the same size, shape, and color,} but we find
  that the spread resulting from this mechanism is insignificant when
  drawing orbits from the distribution observed in N-body simulations.
  The simple model was calibrated with a few more simulations that
  were not available during the initial calibration in C08.}
\label{fig:simp_mod}
\end{center}
\end{figure*}

Once we have calculated the final radius, we can compute the
velocity dispersion.  First, we calculate the central dark matter
fraction of the merger remnant, $f_{\rm dm,f}$ (see Appendix A for
details).  Then, as in C08, we calculate the remnant velocity
dispersion using the virial relation,
\begin{equation}
\sigma^2= \frac{G C_{\rm sig}M_{\rm s,f}}{R_{\rm f} (1 - f_{\rm dm,f})}.
\end{equation}
$M_{\rm s,f}$ is the stellar mass of the remnant, and $C_{\rm
  sig}=0.15$ is a constant parameter fit to the simulations.

In order to determine the parameters $C_{\rm rad}$ and $C_{\rm vir}$,
we calculated a best-fit of the model to the merger remnants in our
simulation suite.  The fit of the new model is depicted in Figure
\ref{fig:simp_mod}a-b, while the results from the C08 model with the
modified calculation of central dark matter fraction are shown in
panels c and d.  On average, the new model correctly predicts the
remnant sizes and velocity dispersions.  However, there is more spread
in the accuracy of predictions in the simple model.  This results from
variations in initial orbit in our simulation suite.  {\bf As stated above, we find that this variation is insignificant when mergers are calculated for the distribution
of orbits found in cosmological N-body simulations.}

{\bf We note that our merger model has primarily been calibrated for
  major mergers of disk-dominated galaxies.  An expansion of this
  model that includes a detailed treatment of minor mergers and
  mergers of elliptical galaxies is forthcoming (Porter et al., in
  prep), but here we do not attempt to model subsequent evolution of
  elliptical galaxies following the initial major merger because the
  current versions of the SAMs do not track bulge sizes using our
  dissipative model.  In limiting our model to major mergers, we are
  only examining one avenue of elliptical galaxy formation.  Recent
  works \citep{Naab06,Bournaud:2007a} have suggested that ellipticals
  can also form as a product of minor mergers, but we do not attempt
  to model this here.}

\subsection{Implementation of Merger Model with SAMs}

In order to make comparisons with observational data, and to explore
the evolution of scaling relations over cosmological time scales, we
apply the new merger model to progenitor properties from two SAMs: the
\citet{S08} SAM (hereafter S08) and a SAM based on the Millennium
Simulation \citep{Croton06}.  We implement the merger model externally
(i.e., by post-processing) rather than incorporating it within each of
the SAMs.  This provides an expedient means of exploring the
properties of new elliptical galaxies arriving into the population via
the merging of disk-dominated galaxies.  We restrict our analysis to
mergers of disk galaxies because the current generation SAMs do not
have a reliable method for calculating bulge sizes, and we need
stellar radii in order to calculate initial internal energies.
Furthermore, {\bf since the merger suite was designed to model mergers of disk galaxies, there were few bulge-dominated progenitors used to calibrate the merger model.  The performance of the model for mergers of bulge-dominated galaxies has not been tested.}  Consequently, we only analyze mergers in each SAM where
both progenitors have a stellar disk more massive than their bulge.
Additionally, since we are comparing the remnants to observations of
elliptical galaxy populations, we only include major mergers with a
mass ratio of 1:3 or greater, as these are the mergers expected to
create elliptical galaxies.

Each SAM directly provides the stellar mass, disk radius, and mass of
cold gas for each progenitor.  In addition, the model requires
information about the dark matter halo of each progenitor galaxy.
Specifically, we need to be able to calculate mass as a function of
radius in order to calculate both the halo half-mass radius and the
central dark matter mass.  S08 and Millennium provide the masses of
the dark matter halos, but specify the halo mass distributions using
different quantities.  S08 provides halo concentrations.  The
Millennium SAM provides $V_{\rm max}$ and $V_{\rm vir}$, from which
concentration can be calculated.  For both cases the concentration,
virial mass, and redshift are used to calculate the distribution of
mass within the halo.  For S08 the spherical top-hat collapse model is
used to calculate the virial overdensity, using the approximation from
\citet{Bryan98}, whereas for Millennium the virial overdensity is
assumed to be 200.  All of these calculations assume the $\Lambda$CDM
concordance cosmology, with $\Omega_{\rm m}=0.3$, $\Omega_{\rm
  \Lambda}=0.7$, and $h=0.7$.

\section{Systematic Exploration of the Merger Model}
\label{sec:toy}

To gain a better intuitive grasp of the behavior of the merger
model, we systematically explore the effects of variations in the
progenitor properties and model parameters.  For this study we
introduce a series of four idealized progenitor galaxy models.  In
order to isolate the effects of various progenitor properties we begin
with a fiducial progenitor and scale the progenitor properties in such
as way as to keep baryon fraction, average density, and halo
concentration constant.

The fiducial galaxy with which we begin our series is the G3 galaxy
from \citet{Cox08}, whose properties are designed to fit the average
properties of observed nearby disk galaxies in the Sloan Digital Sky
Survey.  The G3 has a dark matter halo mass of $1.1 \times 10^{12}
\msun$, a stellar mass of $5.0\times 10^{10} \msun$, an initial
half-mass radius of 3.8 kpc, and a halo concentration of 6.0.  Each
subsequent galaxy in the series is created by reducing the mass by
$1/3$, keeping the average densities, concentration, and baryon
fraction constant{\bf, and calculating $\sigma$ using the virial theorem.
}
\subsection{Effect of Varying Gas Fraction with Mass}

Previous work has shown that gas can have a significant effect on
merger remnants \citep{Barnes:1996a,RobertsonFP, Dekel06,Springel:2005c, Naabgas}.  \citet{Dekel06} suggest
that a systematic variation in gas fraction with mass could be
responsible for the tilt in the fundamental plane of elliptical
galaxies, and \citet{HopkinsFP} demonstrate observational evidence for
this hypothesis.  Thus, we examine the effect of progenitor gas
fraction, defined here as the ratio of gas to stellar mass, on the
properties of equal-mass merger remnants.

\begin{figure*}
\begin{center}
\subfigure[][]{\resizebox{8cm}{!}{\includegraphics{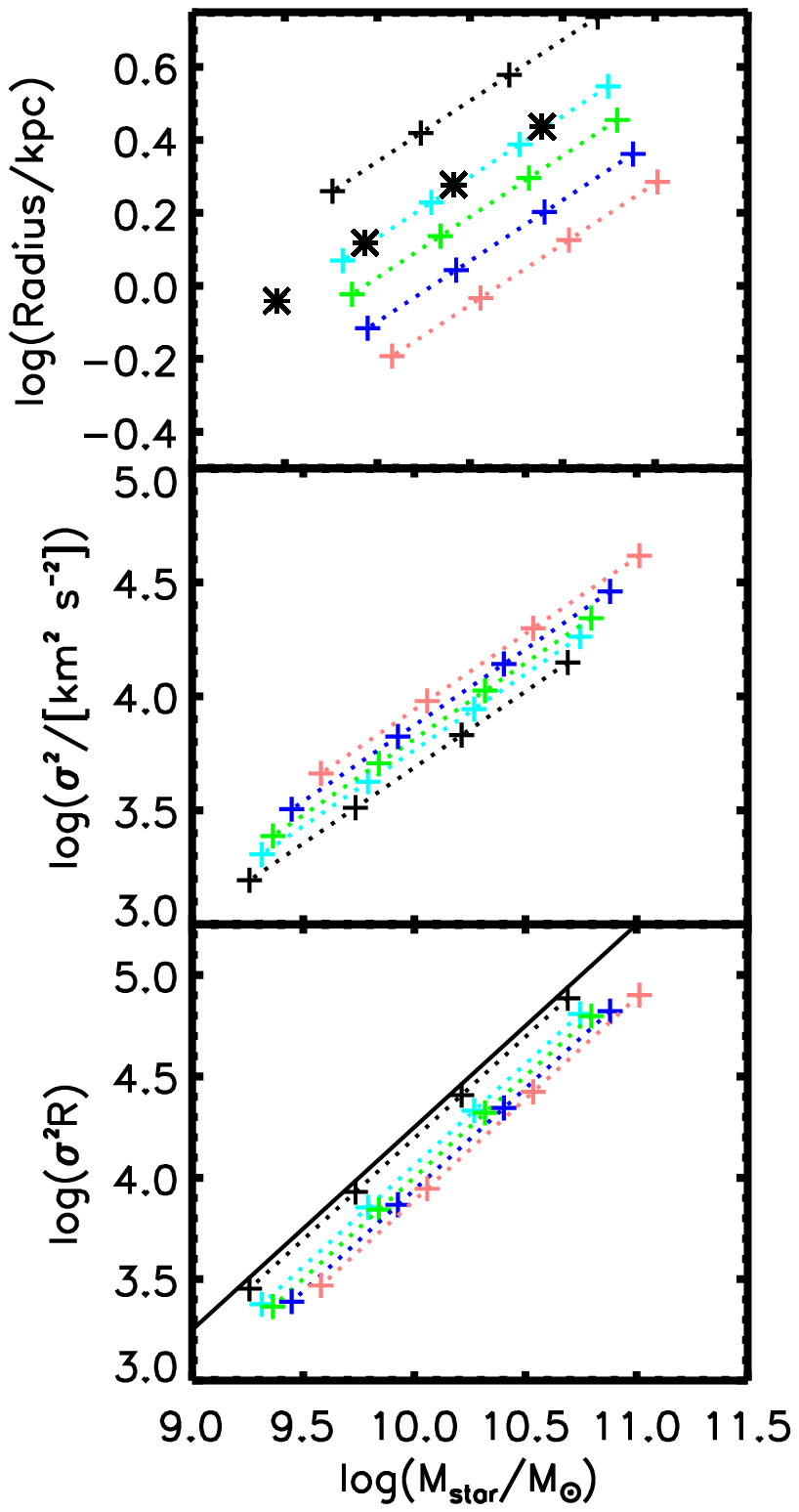}}\hspace{0.0cm}\label{fig:toyGasConst}}
\subfigure[][]{\resizebox{8cm}{!}{\includegraphics{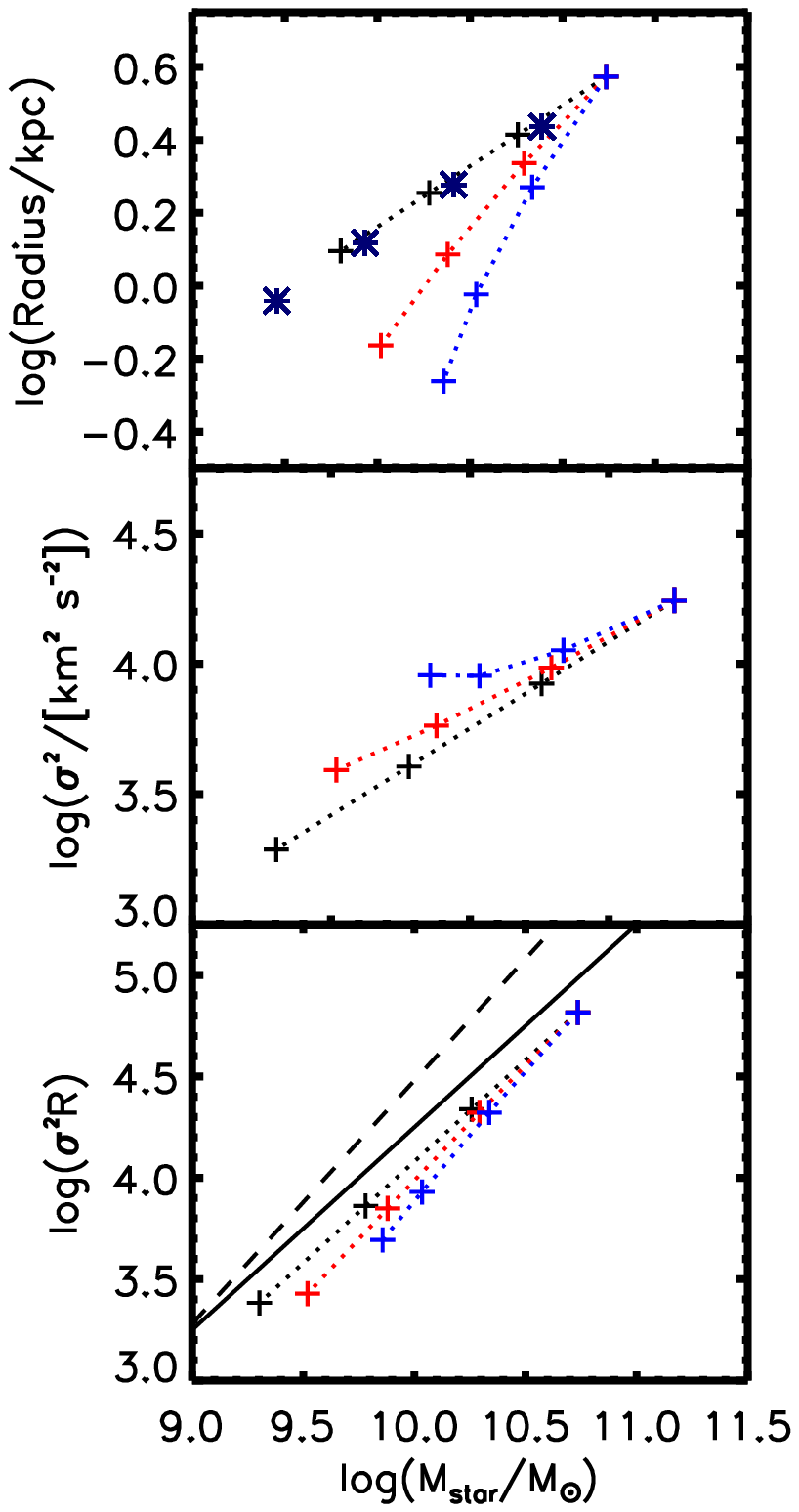}}\hspace{0.0cm}\label{fig:toyGasPower}}\\%
\caption{Scaling relations for {\bf  1:1} merger remnants produced by our model
  using {\bf  a series of idealized progenitors of different mass}.  The progenitors used in (a)
  have a range of initial gas fractions, which are held constant as a
  function of mass. For remnants, colors depict ratio of gas mass to
  stellar mass in the progenitors that created that remnant where
  black=0.0, light blue=0.25, green=0.5, blue=1.0, and red=2.0. Dotted
  lines connect remnants {\bf  resulting from} progenitors with {\bf  different mass but} the same gas
  fraction. The progenitors used in (b) have gas fractions that vary
  as a power law of baryonic mass ($M_{\rm gas}/M_{\rm baryons}\propto
  M_{\rm baryons}^{-\gamma}$). For remnants, colors depict the value
  of $\gamma$ used to set the progenitor gas fractions black=0.0,
  red=0.5, and blue=1.0. Dotted lines connect remnants with
  progenitors from the same gas fraction power law.  The top panels
  show the size-stellar mass relation of the merger remnants (crosses)
  and progenitors (stars), while the middle and lower panels show the
  stellar mass FJ relation and the fundamental plane for the merger
  remnants, respectively.  The solid line plotted in the lowest panels
  shows the slope for galaxies with a virial scaling, whereas the
  dashed line shows the observed tilt in the FP. Steepening in the
  size-mass relationship and tilt in the FP are produced by a decreasing
  gas fraction with increasing baryonic mass. }
\end{center}
\end{figure*}

In our first experiment we set the gas mass of each progenitor such
that gas fraction is constant as a function of mass.  We run the model
for several gas fractions, setting the ratio of gas mass to stellar
mass to 0.0, 0.25, 0.5, 1.0, and 2.0.  The size-stellar mass relation,
stellar-mass Faber-Jackson relation (FJ), and virial projection of the
fundamental plane (FP) of the merger remnants are shown in Figure
\ref{fig:toyGasConst}.

Several observations are worth noting about the size-mass relations plotted
in Figure \ref{fig:toyGasConst}.  First, for all gas fractions the remnant
relations (crosses) are just shifted horizontally and vertically from
the progenitor relation (stars) without any significant rotation
(i.e., change of slope).
That is, the vector between progenitor and remnant is constant for any
given gas fraction.  This means that $R_{\rm f}/R_{\rm i}$ and $M_{\rm
 star,f}/M_{\rm star,i}$ are also constant for any particular gas
fraction.  This is worth noting because the observed size-mass
relations for disks and ellipticals are significantly rotated from one
another with the relation for ellipticals being much steeper.  A
realistic mechanism for the production of ellipticals must account for
this steepening.

Remember that the progenitors are constructed so that they have a
constant density inside their half-mass radius.  Thus one can quickly
see from the plot that `dry' mergers, with no gas, will produce
remnants with densities less than their progenitors and sizes larger
than their progenitors, whereas gas-rich mergers will produce remnants
with densities higher than their progenitors.  The sizes of the
remnants of gas-rich mergers can be similar to or even smaller than
the sizes of their progenitors.  For this set of model parameters,
constant density evolution occurs at roughly a gas-to-stellar-mass
ratio of 0.25 (light-blue crosses).

The stellar-mass FJ plot, in the second panel of Figure
\ref{fig:toyGasConst}, demonstrates a similar effect.  Progenitors are
not shown because they have no values for $\sigma$, however one can
see that remnants of mergers with each gas fraction follow parallel
lines.  No rotation is introduced between one gas fraction series and
the next.

If elliptical galaxies followed an exact virial relation, then one
would expect that $M_{\rm star}\propto\sigma^2R$.  Thus plotting these
quantities against each other gives us a `virial' projection of the
fundamental plane.  Galaxies following the virial relation would fall
on a line with a slope of one.  Observed galaxies do not fall on the
expected virial relation.  This variance from virial scaling is the
so-called tilt of the fundamental plane.  In the third panel of Figure
\ref{fig:toyGasConst} we plot $M_{\rm star}$ versus $\sigma^2R$.  The
line plotted has a slope of one and therefore follows a virial
scaling.  The remnants from each gas fraction set fall on the same
virial scaling.  This demonstrates that our model produces no tilt in
the fundamental plane for remnants when the progenitors' properties
are scaled with the mass.  However, while each gas fraction series
follows the virial scaling, there is a shift between cases with
different gas fractions such that larger gas fraction progenitors
result in remnants with larger ratios of stellar mass to dynamical
mass.

In the next experiment we let gas fraction vary as a function of mass.
We fix the gas mass of the largest progenitor (G3) such that the ratio
of gas mass to stellar mass is 0.25.  Then we let the baryonic gas
fraction vary as a power law with the baryonic mass, $M_{\rm
  gas}/M_{\rm baryons}\propto M_{\rm baryons}^{-\gamma}$ (as suggested
in \citet{Dekel06}). We use values of $\gamma$ equal to 0, 0.5, and
1.0.  In Figure \ref{fig:toyGasPower} we show the scaling relations of
the remnants produced by this series of progenitors.  Each line of
remnants is fixed to the largest mass remnant, but the slope of the
size-mass and FJ relations clearly depends on the value chosen for
$\gamma$.  Non-zero values of $\gamma$ allow for significant rotations
in all of the plotted projections of the FP.  The rotation in the
size-mass relation is the direction of rotation required if one is to
create ellipticals from mergers of disks that follow the observed
relations.  Additionally, the fundamental plane relation rotates away
from the virial relation in the same direction as the observed
tilt. For reference, the slope of the observed tilt is shown with a
dashed line.  The tilt required is slightly overshot by the $\gamma=1$
case, suggesting that a slightly shallower power law slope would reproduce
the observed tilt.  This result is compatible with the gas fraction
power of 0.7 suggested by \citet{Dekel06}.  Thus within our model a
gas fraction gradient is capable of creating a tilt in the fundamental
plane.  

Our model relies on the virial relation to calculate sizes and
velocity dispersion.  However, in our model the central dark matter
fraction is calculated assuming that no dissipation occurs within the
dark matter halo, and this affects our calculation of $\sigma$.  The
break from virial scaling results from this changing central dark
matter fraction, with more gas rich progenitors producing a larger
difference between the dissipational baryons and dissipationless dark
matter resulting in a lower central dark matter fraction.

For large values of $\gamma$ some curvature is also introduced into
the scaling relations.  Since both the disk and elliptical scaling
relations are approximately power laws, this puts a constraint on the
strength of the gas fraction variation allowed by the model if we want
to reproduce the observed scaling relations.  However, the expected
value of $\gamma=0.7$ only produces a modest amount of curvature.

\subsection{Merger Model Parameters}

The merger model contains several parameters with uncertain values
because of the uncertainty in the feedback prescription used by the
merger simulations.  Because of this uncertainty, the values of the
model parameters might require adjustment.  In order to understand the
effect of these parameters we vary them systematically and examine the
results of these variations on the remnant scaling relations.

\begin{figure*}
\label{fig:cnew}
\begin{center}
\subfigure[][]{\resizebox{8.0cm}{!}{\includegraphics{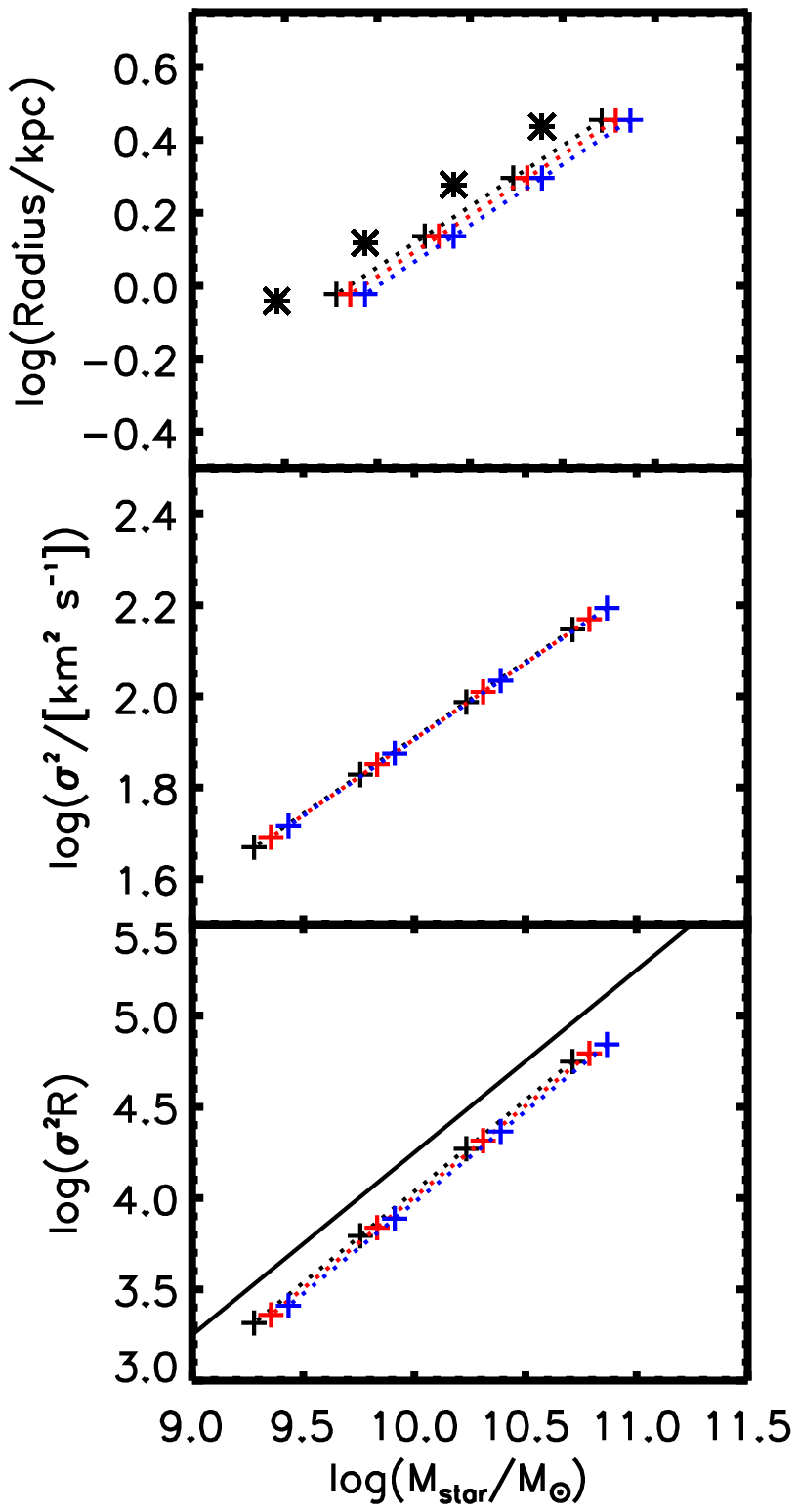}}\hspace{0.0cm}\label{fig:cnewConst}}
\subfigure[][]{\resizebox{8.0cm}{!}{\includegraphics{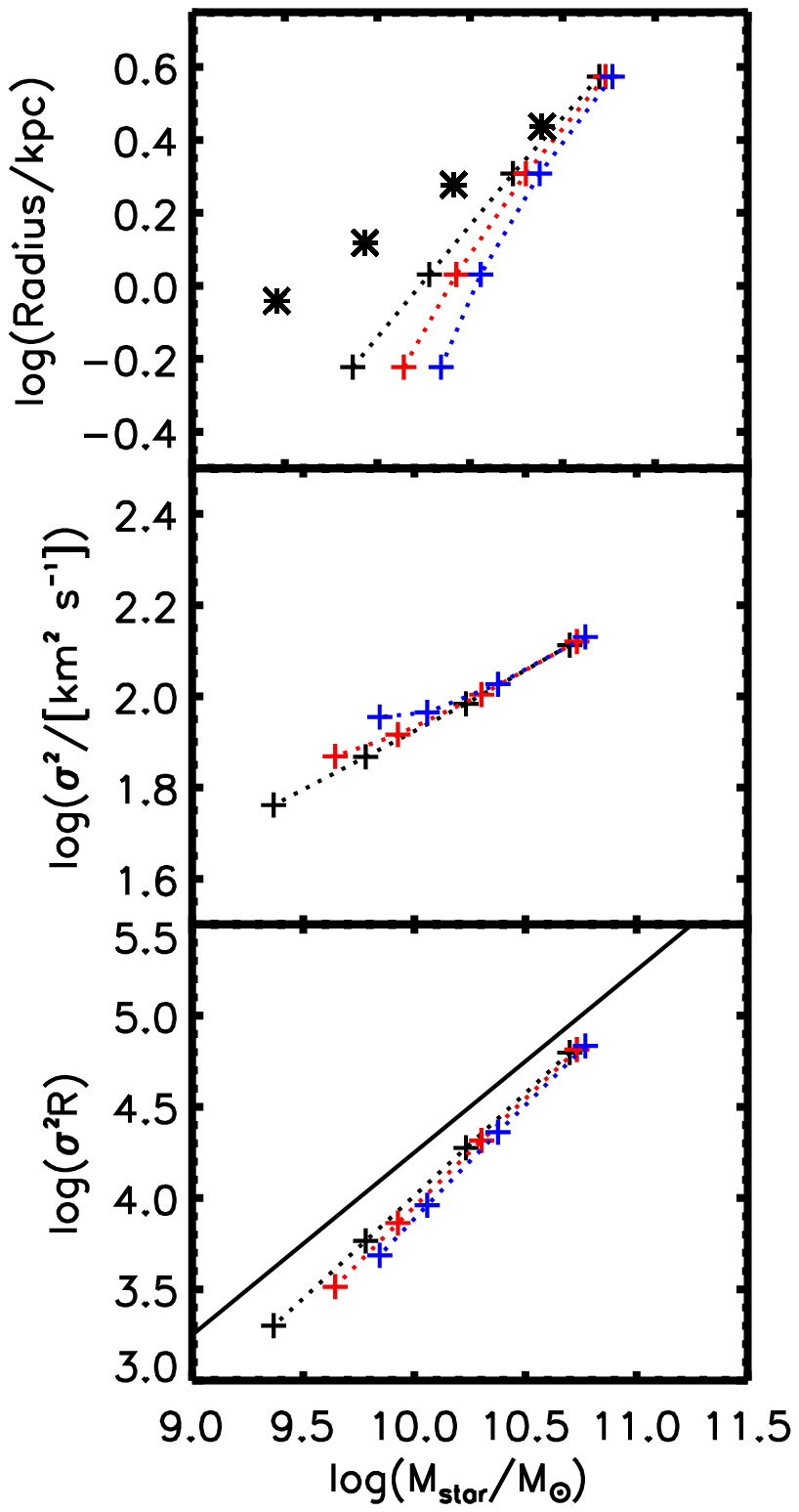}}\hspace{0.0cm}\label{fig:cnewPower}}\\%
\caption{Scaling relations for merger remnants produced by the model
  from our series of idealized progenitors with varying star formation
  parameter $e_{\rm 1:1}$, which takes values of 0.1 (black), 0.5
  (red), and 1.0 (blue).  0.55 is the calibrated value. Dotted lines
  connect cases with the same value of $e_{\rm 1:1}$. Progenitors in
  (a) have a constant gas to stellar mass ratio of 0.5, whereas
  progenitors in (b) have a changing gas fraction with mass
  ($\gamma=0.7$).  The top panels show the size-stellar mass relation
  of the merger remnants (crosses) and progenitors (stars), while the
  middle and lower panels show the stellar mass FJ relation and the
  fundamental plane for the merger remnants, respectively.  The solid
  line plotted in the lowest panels shows the FP slope for galaxies
  with a virial scaling. Changing $e_{\rm 1:1}$ affects the slopes of
  all three scaling relations, but only when gas fraction varies as a
  power law of stellar mass (b).}
\end{center}
\end{figure*}

We begin by varying the star formation efficiency parameter, $e_{\rm
  1:1}$.  The calibrated value of this parameter (from the
simulations) is 0.55, but for the experiments below we use values of
0.1, 0.5, and 1.0.  First, we assume a constant gas-to-stellar-mass
ratio of 0.5 across the idealized galaxy series.  Increasing $e_{\rm
  1:1}$ will increase the number of stars that form; however, it has
little effect on the remnant scaling relations (Figure
\ref{fig:cnewConst}).  Specifically, none of the relations experience
significant rotation as a result of the adjustment.  For all relations
an increase in $e_{\rm 1:1}$ simply results in a slight shift toward
higher stellar mass.

For the second experiment with $e_{\rm 1:1}$, we introduce a
mass-dependent gas fraction according to the power law relation used
above, using the expected value of $\gamma=0.7$.  For this series,
adjusting $e_{\rm 1:1}$ produces rotations of the scaling relations.
Specifically, changing $e_{\rm 1:1}$ has a greater effect on the
mergers with a larger gas fraction, resulting in rotations in the
size-mass and FJ relations and a slight tilting in the FP (Figure
\ref{fig:cnewPower}).

\begin{figure*}
\begin{center}
\subfigure[][]{\resizebox{8.0cm}{!}{\includegraphics{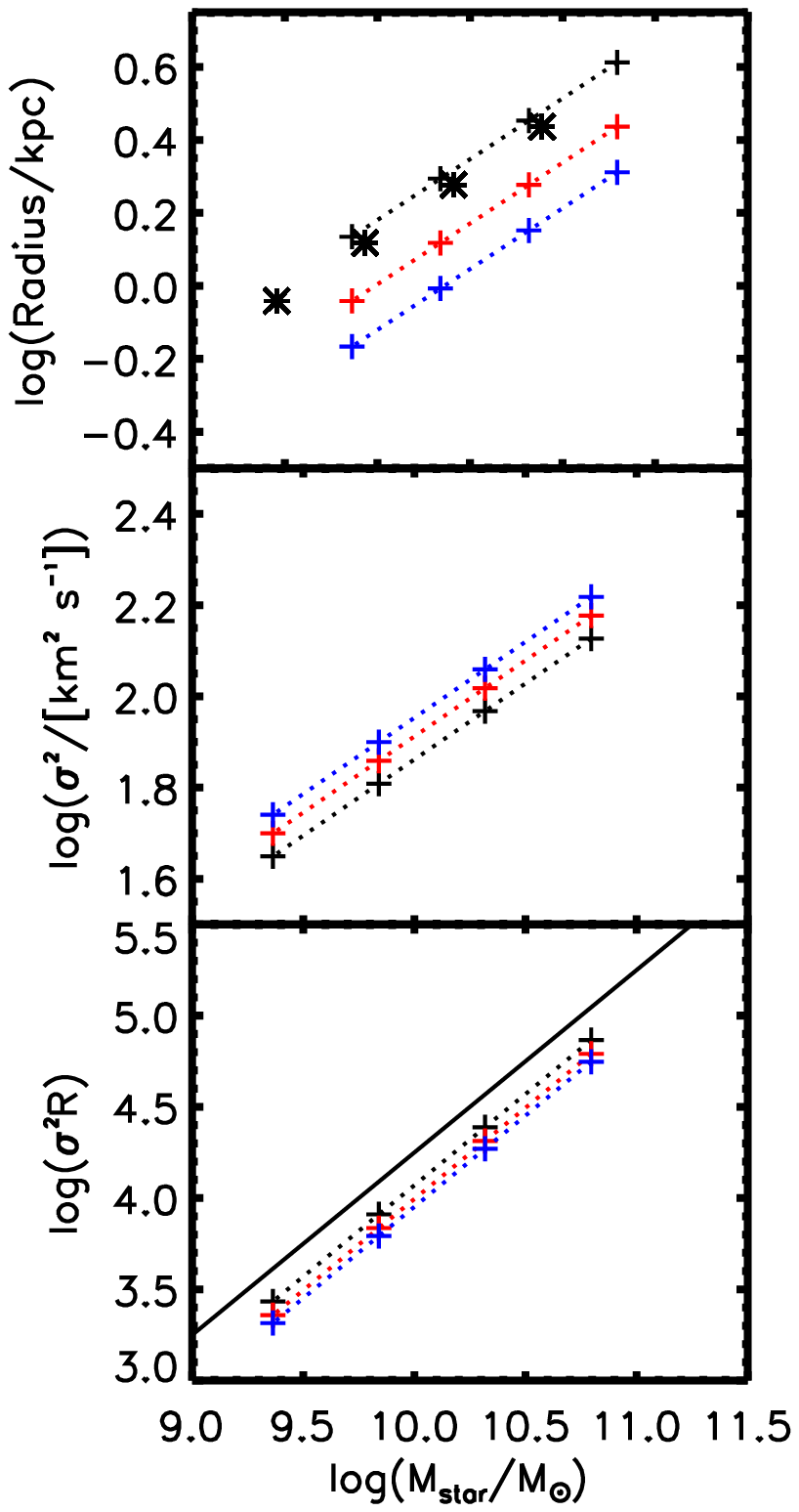}}\hspace{0.0cm}\label{fig:cradConst}}
\subfigure[][]{\resizebox{8.0cm}{!}{\includegraphics{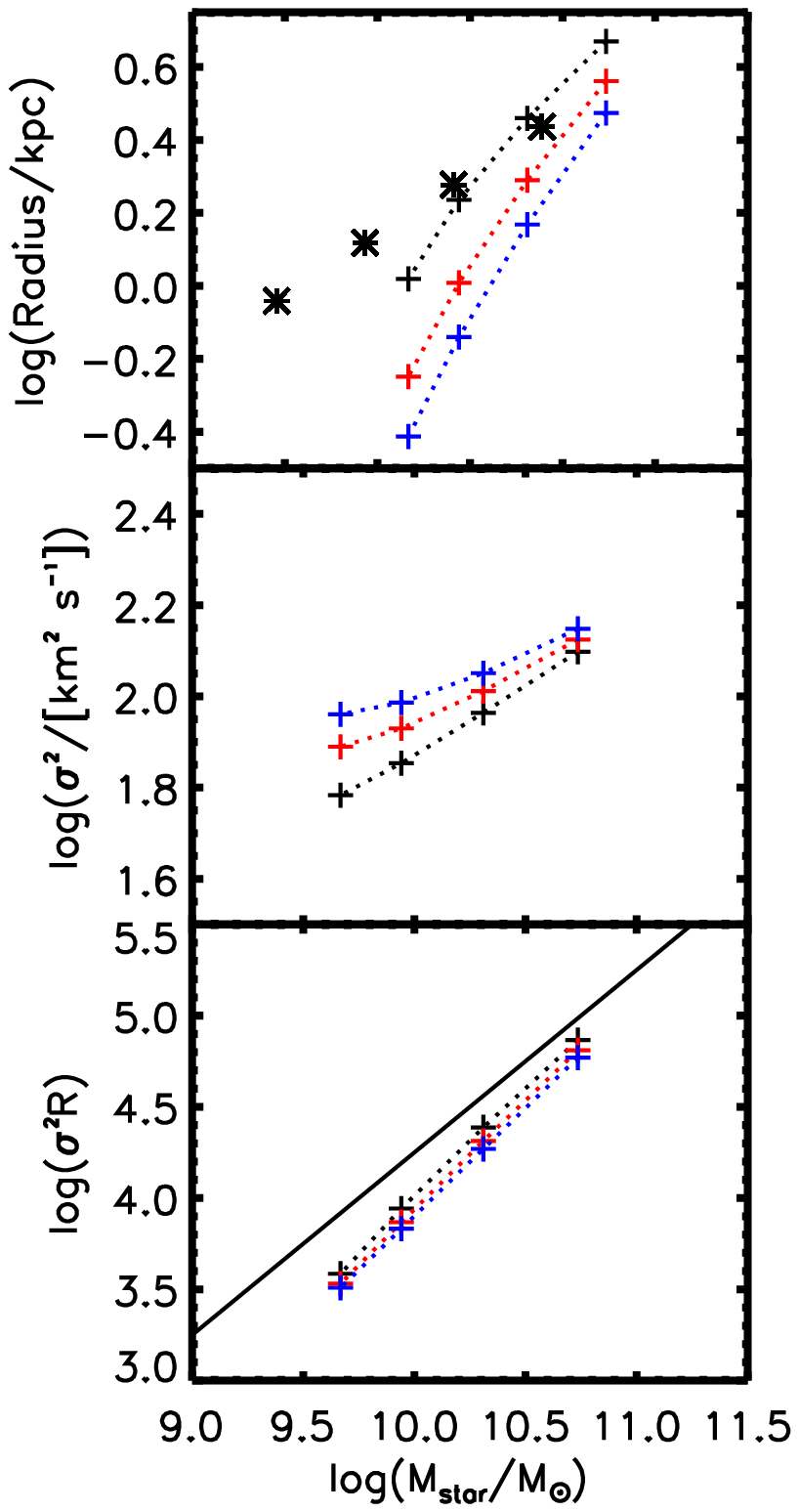}}\hspace{0.0cm}\label{fig:cradPower}}\\%
\caption{Scaling relations for merger remnants produced by the model
  from our series of idealized progenitors with a varying radiative
  energy loss parameter $C_{\rm rad}$, which takes values of 1.0
  (black), 3.0 (red), and 5.0 (blue).  $C_{\rm rad}=2.75$ is the
  calibrated value from the simulations.  Dotted lines connect cases
  with the same value of $C_{\rm rad}$. Progenitors in (a) have a
  constant gas to stellar mass ratio of 0.5, whereas progenitors in
  (b) have a changing gas fraction with mass ($\gamma=0.7$).  The top
  panels show the size-stellar mass relation of the merger remnants
  (crosses) and progenitors (stars), while the middle and lower panels
  show the stellar mass FJ relation and the fundamental plane for the
  merger remnants, respectively.  The solid line plotted in the lowest
  panels shows the FP slope for galaxies with a virial
  scaling. Changing $C_{\rm rad}$ affects the slopes of the size-mass
  relation and FJ relation for the cases where gas fraction varies as
  a function of stellar mass (b), but it does not effect the FP.  }
\end{center}
\end{figure*}

Another parameter for which feedback could produce some uncertainty is
$C_{\rm rad}$, which sets the importance of the radiative energy term.
Since the $C_{\rm rad}$ parameter is decoupled from the equation that
determines the number of new stars, adjusting $C_{\rm rad}$ results in
no difference in final mass.  However, an increase of the parameter
results in a significant reduction of size and increase in velocity
dispersion for the remnants.  The same constant gas fraction and
mass-dependent gas fraction merger series used for the star formation
parameter series are plotted in Figures \ref{fig:cradConst} and
\ref{fig:cradPower}, respectively, with $C_{\rm rad}$ taking values of
1.0 (black), 3.0 (red) and 5.0 (blue).  The value of $C_{\rm rad}$
determined by fitting to the merger simulations was 2.75.  As with
$e_{\rm 1:1}$, rotation in the scaling relations is only seen for the
series with a mass-dependent gas fraction.  Here, a significant
rotation is created in the size-mass and FJ relations, but adjusting
$C_{\rm rad}$ never introduces a tilt in the FP.  This is because the
model is built on the assumption of the virial theorem, and the only
portion of the model that violates this assumption is the formula for
calculating the change in central dark matter fraction, which is then
used to adjust the velocity dispersion.  Changing $C_{\rm rad}$ does
not affect this portion of the model.

\section{Summary of Observational Results}

The Sloan Digital Sky Survey (SDSS) \citep{York00} has provided
exquisite statistics on galaxy scaling relations in the local
universe.  \citet{Shen03} show that for local galaxies the size
distribution for each type of galaxy at a given stellar mass is
log-normal.  They provide fitting functions for the medians of the
distributions for both early- and late-type galaxies.  For late-type
galaxies the median (${\bar R}$) is described by
\begin{equation}
\label{eqn:latesize}
{\bar R}{\rm (kpc)}= \gamma \left(\frac{M}{\msun}\right)^{\alpha} \left(1+\frac{M}{M_0}\right)^{\beta - \alpha},
\end{equation}
where $\gamma=0.1$, $\alpha=0.14$, $\beta=0.39$, and $M_0=3.98 \times
10^{10}\msun$. A comparison of this distribution to that of
\citet{Barden05}, also from SDSS, demonstrates a discrepancy, despite
the fact that the samples contain significant overlap.  Specifically,
a comparison of the stellar-mass size ridge line in
\citet{Somerville08a} with the figure in Shen et al., correcting for
the conversion between disk scale length and half light radius, shows
that the Shen et al. distribution is a factor of $\sim 1.5$ smaller in
radius for a given mass. \citet{Dutton10} find a similar offset
between their re-analysis of SDSS and the results of Shen et al., and
argue that this was primarily due to their use of circular rather than
elliptical apertures. For this work, we scale the fitting function of
Shen et al. to match the normalization of \citet{Barden05} and
\citet{Dutton10}.

For early-type galaxies the median radius is described by
\begin{equation}
\label{eqn:earlysize}
{\bar R} {\rm (kpc)}= b \left(\frac{M}{\msun}\right)^a,
\end{equation}
where $a= 0.56$ and $b=3.47 \times 10^-5$. The scatter in the relation
is similar for both late- and early-types with the fit to late-type
galaxies being
\begin{equation}
\label{eqn:sizescatter}
\sigma_{\ln R}=\sigma_2 + \frac{(\sigma_1 - \sigma_2)}{1+(M/M_0)^2},
\end{equation}
where $\sigma_1=0.47$ and $\sigma_2=0.34$.

The fit for the stellar mass Faber-Jackson relation in SDSS can be
found in \citet{Gallazzi06}, and is given by
\begin{equation}
\label{eqn:fj}
\log (\sigma_v {\rm (km~s^{-1})}) = 0.286 \log (M_*/\msun) - 0.895,  
\end{equation}
where $\sigma_v$ is velocity dispersion and $M_*$ is the stellar
mass. The scatter in the relation is 0.071 dex.

The fundamental plane relation can be represented in a number of
projections.  However, recent work suggests that the tilt of the plane
results from a systematic change in the central dark matter fraction
\citep{Zaritsky07,HopkinsFP} as the result of a varying effect of dissipation
with mass.  A number of observational studies have examined the
relationship between central dynamical and stellar mass,
\begin{equation}
\label{eqn:dyn}
M_{\rm dyn}\propto M_{\rm star}^{1+\alpha},
\end{equation}
and have determined a value of $\alpha \approx 0.2$ \citep{Pahre98,
  Gerhard01, Padmanabhan04, Gallazzi06}.  

There is little published data on the evolution of the FJ and FP to
high redshift{\bf, though observations suggest that any evolution in the FP is minimal up to $z \sim 1$ \citep{Holden:2010a,Saglia:2010a,Herbert:2011a}}.  However, a number of studies have examined the
evolution of the size-mass relations between $z \sim 0$ and up to
$z\sim 3$ \citep{Barden05, McIntosh05, Trujillo06}.  The most
comprehensive of these studies, \citet{Trujillo06}, combines data from
SDSS, GEMS (Galaxy Evolution from Morphology and SEDs) and FIRES
(Faint Infrared Extragalactic Survey) in order to quantify evolution
between $z=0$ and $z=3$ for both early- and late-type galaxies.  They
differentiate galaxies by light concentration according to S\'{e}rsic
index $n$ and find that for both low-$n$ (late-type) and high-$n$
(early-type) galaxies the mean size at given mass at $z \sim 2.5$ is
$\sim 2 $ times smaller than today.  Specifically they find that at
given stellar mass the sizes of late-type galaxies evolve as
$(1+z)^{-0.40 \pm 0.06}$, whereas the sizes of early-types evolve as
$(1+z)^{-0.45 \pm 0.10}$.

\section{Comparison with Observations}

We now apply our merger model to {\bf  major mergers of disk-dominated galaxies} from the S08 and
Millennium SAMs.  These SAMs produce statistical samples of galaxies
with properties that are closely matched to those observed in the
universe.  Thus they provide an effective means for testing the merger
model in a cosmological framework. {\bf  At every time step in the SAMs, we use the data from all disk-disk major mergers, regardless of the previous or subsequent merger histories of the disks.  We emphasize that we are making no attempt to model subsequent evolution or mergers of the elliptical remnants.  Possible implications of this further activity are examined in Section 6.}  Since the modeled merger remnants
are a function of the {\bf  disk} progenitor properties, we begin by examining the
distributions of galaxy properties in the SAMs and making comparisons
to the observed distributions.

\subsection{Properties of the Progenitors}

\begin{figure*}
\begin{center}
\subfigure[][]{\resizebox{8.0cm}{!}{\includegraphics{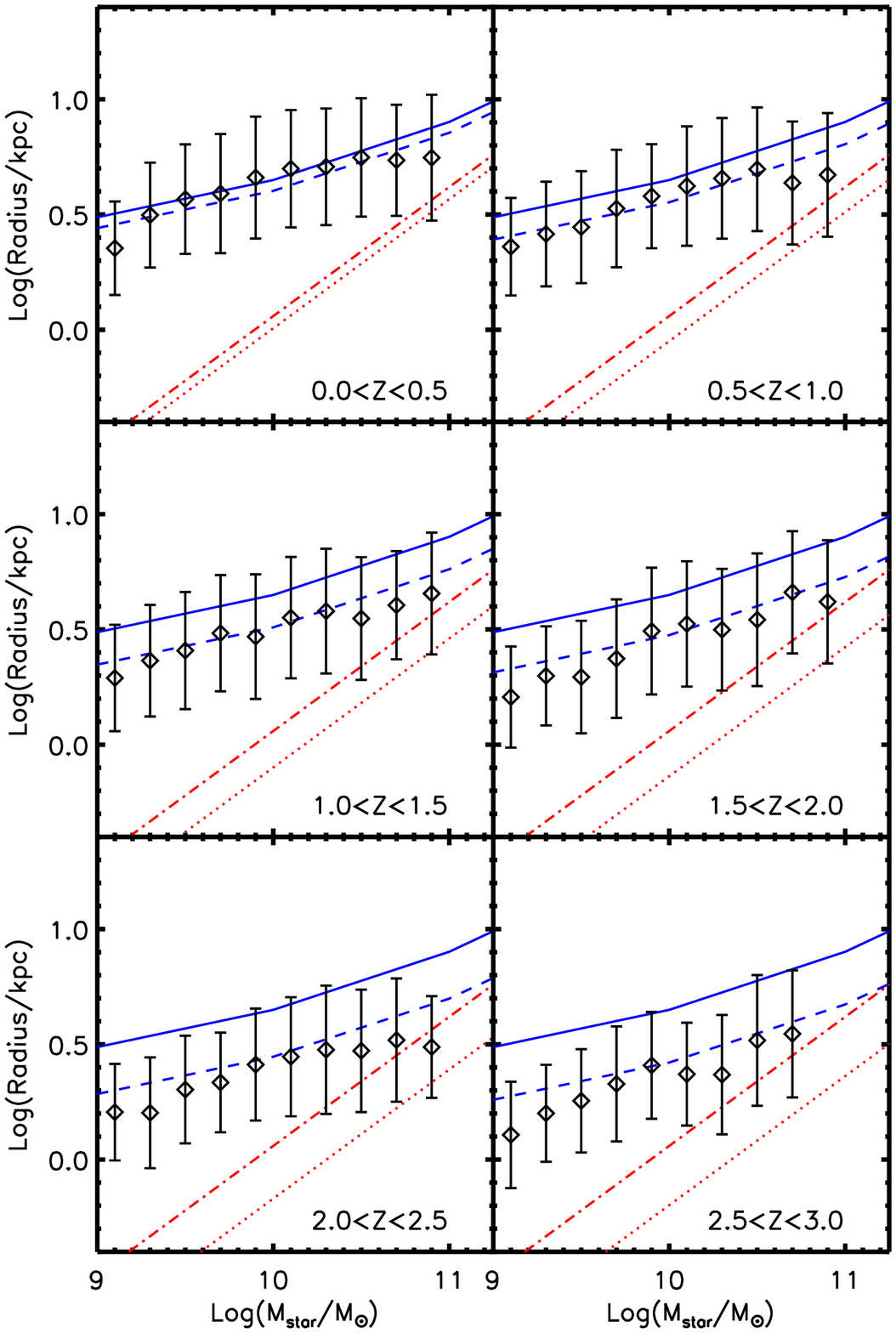}}\hspace{0.0cm}\label{fig:spfsizemass}}
\subfigure[][]{\resizebox{8.0cm}{!}{\includegraphics{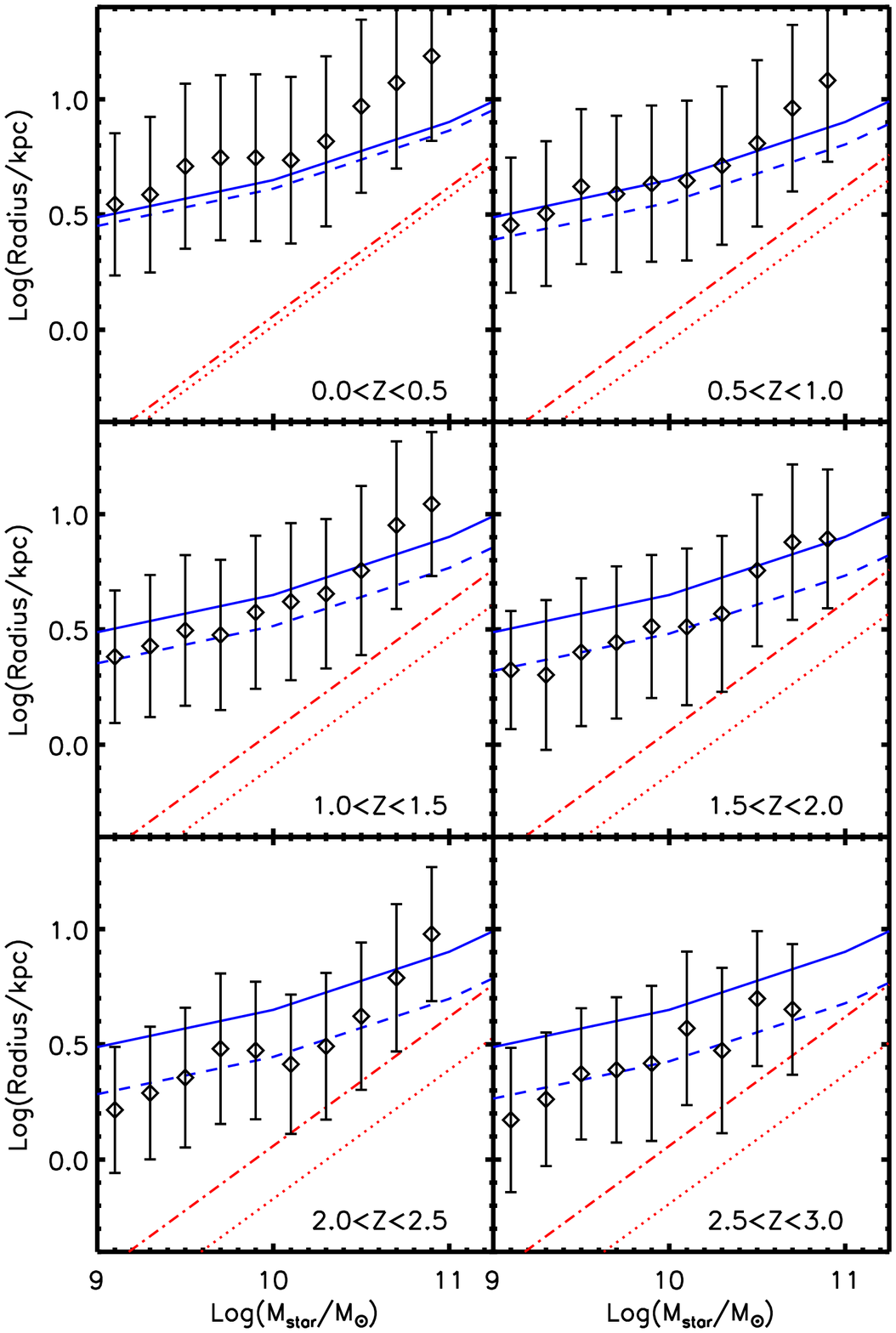}}\hspace{0.0cm}\label{fig:millsizemass}}
\caption[Size Mass relations for the progenitors in the S08 SAM (a) and the Millennium SAM (b).]{Size Mass relations for the progenitors in the S08 SAM (a) and the Millennium SAM,
  binned by redshift.  Symbols denote the median of the progenitor
  distribution in the SAM, and error bars depict the $1\sigma$ spread
  in the distribution.  The solid (blue) line and the dashed (red)
  line depict the redshift zero relations from \citet{Shen03} of the
  low-$n$ and high-$n$ populations respectively. The blue-dashed and
  red-dotted lines depict the evolution of the low-$n$ and high-$n$
  relations respectively with redshift as given by \citet{Trujillo06}.
  This is calculated using the median redshift for the progenitors in
  each bin. For this and subsequent scaling relations figures, blue and
  red lines denote observed scaling relations of disk and elliptical
  galaxies respectively.  }
\end{center}
\end{figure*}

In the model, the most important properties of the progenitors are
initial size, mass, and gas fraction.  Thus we begin by looking at the
size-mass distributions of progenitors in each SAM.  The relations are
plotted for S08 in Figure \ref{fig:spfsizemass} and for Millennium in
Figure \ref{fig:millsizemass}.  For each figure the progenitors are
separated into {\bf  one of} six redshift bins {\bf  according to the redshift of the merger}.  Within each bin, the progenitors
are divided into mass bins with a width of 0.2 in $\log (\msun)$.  The
local relations for low-$n$ (solid blue) and high-$n$ (dotted red)
galaxies are shown for comparison.  Additionally, the observed
redshift evolution of the size of the low-$n$ galaxies
\citep{Trujillo06} is depicted with the blue-dashed line.  This is
calculated using the median progenitor redshift in each redshift bin.

The size-mass relation produced by the progenitors in S08 reproduces
the observed relation quite nicely, including evolution with redshift.
The Millennium progenitors are also fairly close to the observed
relation, but in the lowest redshift bin they are $\sim50\% $ too
large on average.  This gap lessens with increasing redshift.  Also of
note is that the highest mass bin is typically systematically high.
It is also interesting to note the difference in slope between the
observed size-mass relations for early- and late-type galaxies. As
noted before, if a merger explanation of the elliptical size-mass
relation is to be successful, it must explain the rotation between the
two observed relations. We discuss this with respect to our model in
the next section.

Observing cold gas within galaxies is extremely challenging. However a
study by \citet{Kannappan04} used a photometric estimate of gas
fraction in order to calculate the changing ratio of gas to stellar
mass (G/S) with galaxies from SDSS.  \citet{Calura07} use star
formation rates from a large SDSS sample combined with theoretical
modeling to estimate (G/S) as a function of both stellar mass and
redshift.  Neither study provides fits to the observed relation, but
it can be seen from Figure 5 of \citet{Calura07} that, for $z<0.1$,
\begin{equation}
\label{eq:gas}
\log(G/S) \sim -0.5 \log(M/\msun) + C,
\end{equation}
where for blue galaxies $C\sim 4.6$ and for red galaxies $C\sim
4.1$.  The values of $\log(G/S)$ falls approximately in a range
between -4 and 1.5.

\begin{figure*}
\begin{center}
\subfigure[][]{\resizebox{8.0cm}{!}{\includegraphics{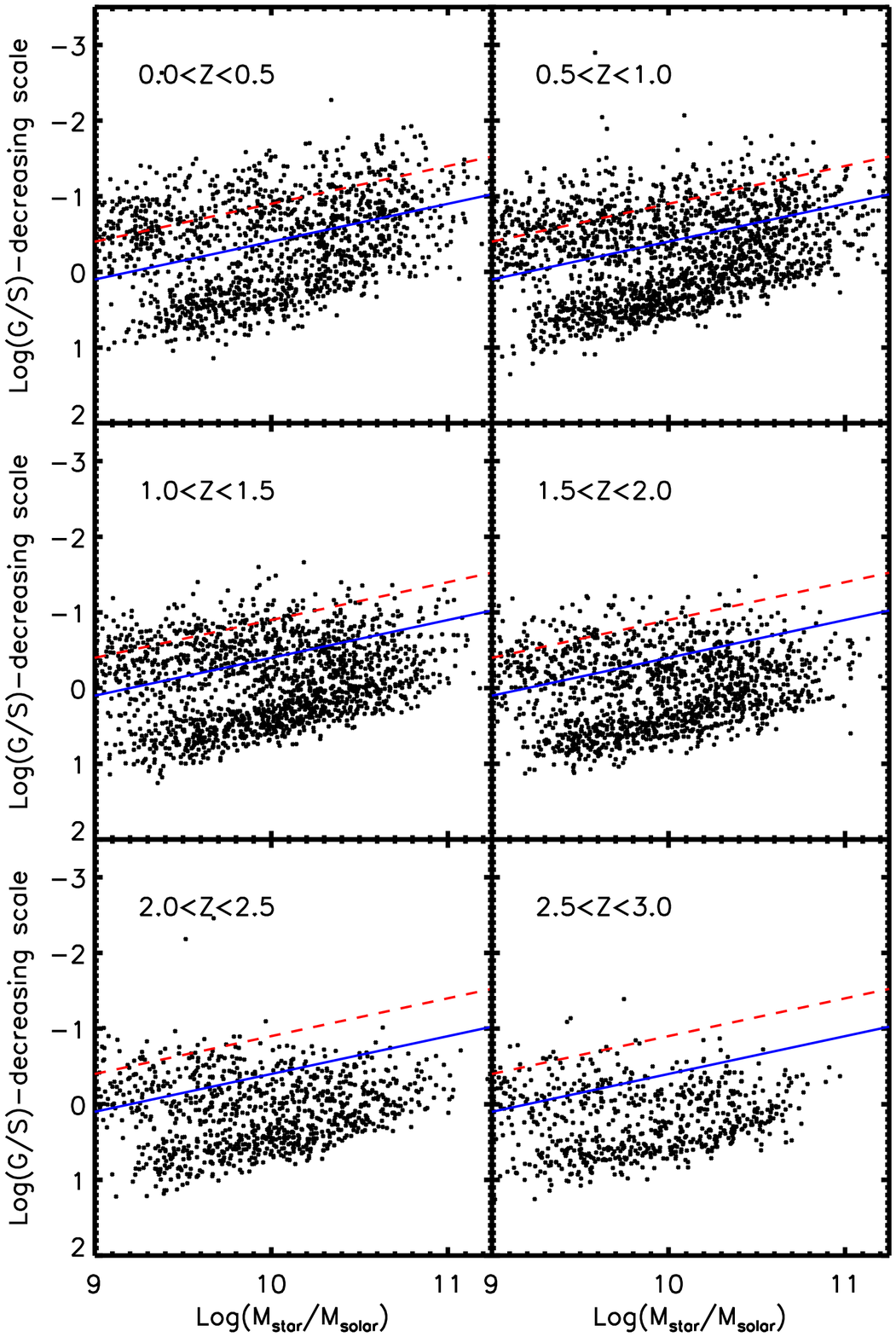}}\hspace{0.0cm}\label{fig:spfgas}}
\subfigure[][]{\resizebox{8.0cm}{!}{\includegraphics{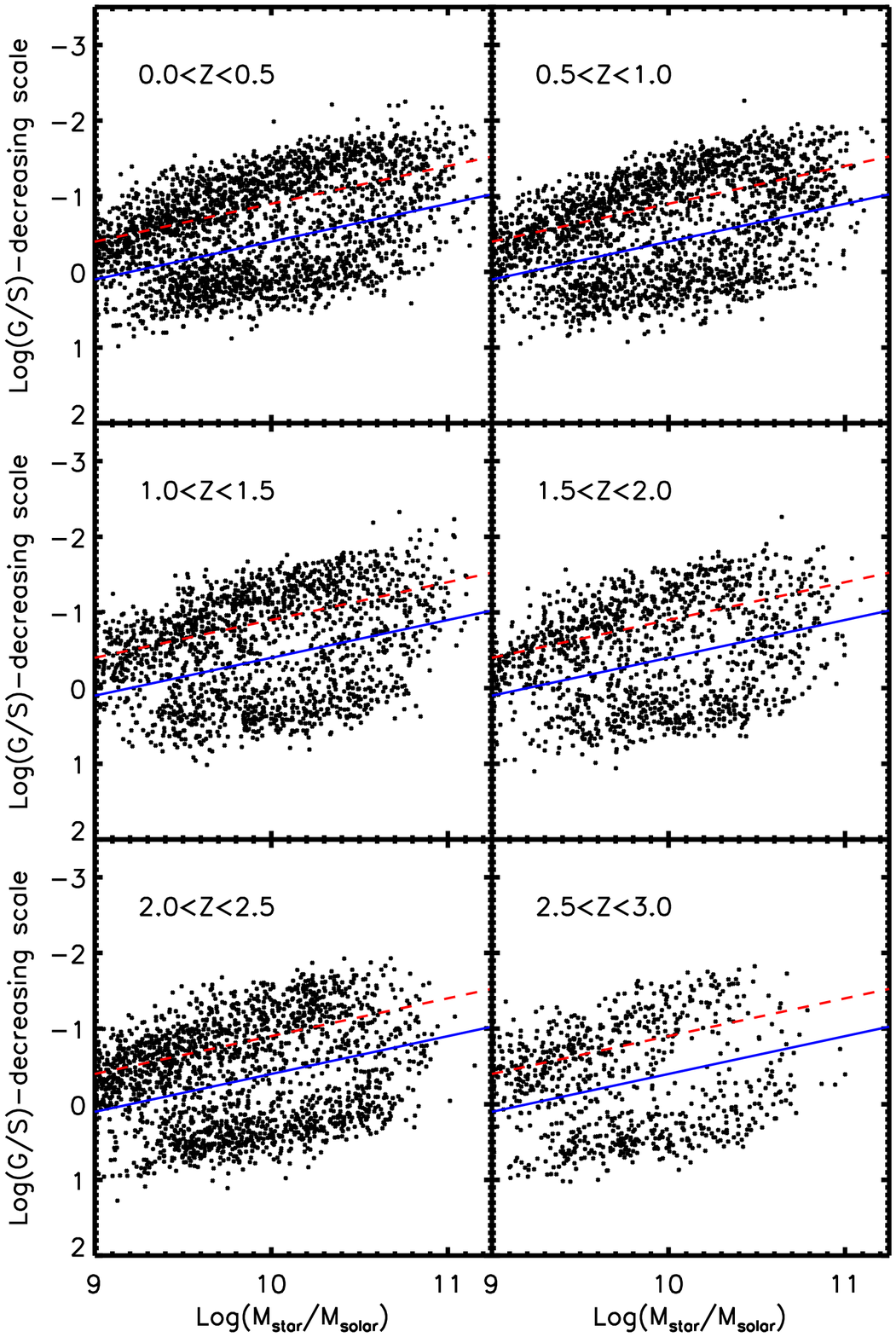}}\hspace{0.0cm}\label{fig:millgas}}
\caption[The relation between G/S and stellar mass for the progenitors
  in the S08 SAM.]{The relation between G/S and stellar mass for the
  progenitors in the S08 SAM (a) and the Millennium SAM (b).  Points
  are galaxies from the SAM.  The blue (solid) and red (dashed) lines
  are approximate fits to observations of nearby galaxies in the SDSS
  \citep{Kannappan04, Calura07}, with blue representing the relation
  for blue galaxies and red for red galaxies. $\log(G/S)$ is plotted
  using a decreasing scale as in the observational studies. The
  observed relations and SAM progenitors display similar slopes, but
  the relations are not expected to exactly coincide due to
  differences in redshift ranges and selection effects.}
\end{center}
\end{figure*}

The relation between G/S and stellar mass in the SAMs is shown in
Figures \ref{fig:spfgas} and \ref{fig:millgas}. The observational
relation (Equation \ref{eq:gas}) is also depicted to show that the
progenitors from the SAMs have gas fraction to stellar mass relations
with slopes similar to those observed.  However, the distribution of
progenitor gas fractions from the SAMs is not expected to exactly
align with the depicted observed relations for two reasons: 1) even
the lowest redshift bin ($0<z<0.3$) from the SAMs includes redshifts
significantly higher than those from the observations ($z<0.1$), and
2) the progenitors are a small subset of all galaxies.  Specifically,
the galaxies shown from the SAMs are those that undergo major mergers
during the specified redshift and are disk-dominated. The properties
of this subset may differ from both the red and blue populations in
the observations.  In fact, the models of S08 have been tuned to
reproduce the gas fraction vs. stellar mass observations of
\citet{Kannappan04} for spirals.

It can be seen that the distributions of G/S in both SAMs are bimodal
with a gas-rich (blue) and gas-poor (red) sequence.  While both SAMs
capture the slope of the observed relation, they have gas to baryonic
mass power law slopes which are significantly less than the
$\gamma=0.7$ value suggested in \citet{Dekel06} and the idealized
mergers presented above.  A majority of the mergers are gas-rich; 99\%
and 92\% of all mergers, for S08 and Millennium respectively, have
total progenitor gas fractions greater than 0.1.  The relation from
the SAMs only evolves very modestly over time, but for both SAMs the
fraction of mergers that are coming from high gas fraction progenitors
increases with increasing redshift. This is seen both as a
disappearance of the red sequence (in S08) and a decrease in the
typical mass of progenitors as redshift increases (both SAMs).  {\bf A
  recent study \citep{Tacconi:2010a} found that the gas fraction of
  star-forming galaxies increases by a factor of a few as the redshift
  increases from $z=0$ to $z \sim 2.3$.  While we have made no
  selection cuts to ensure that our spiral progenitors are
  star-forming, these results suggest that the gas fractions of major
  mergers might increase over time.}

\begin{figure*}
\begin{center}
\subfigure[][]{\resizebox{8.0cm}{!}{\includegraphics{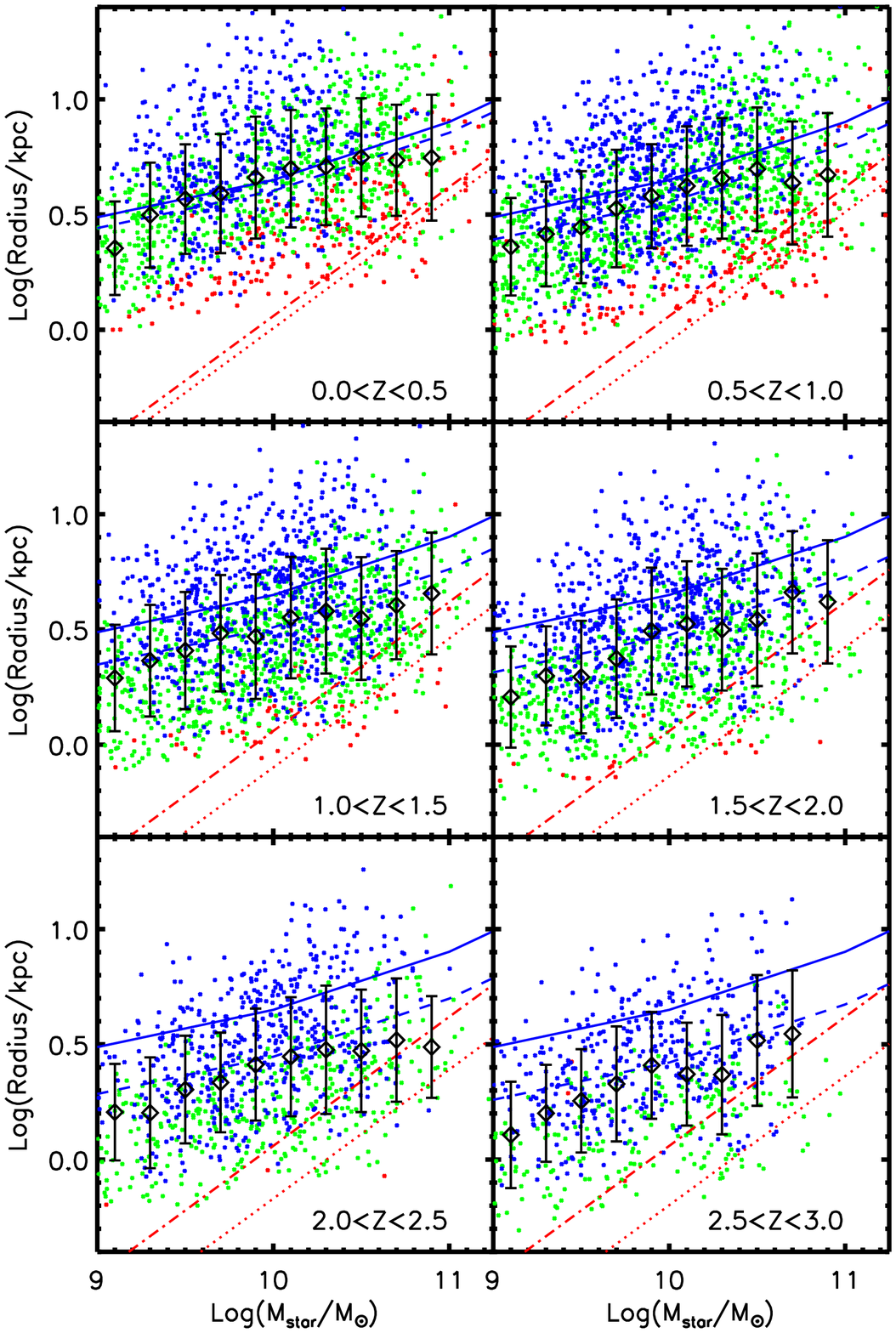}}\hspace{0.0cm}\label{fig:spfgfsizemass}}
\subfigure[][]{\resizebox{8.0cm}{!}{\includegraphics{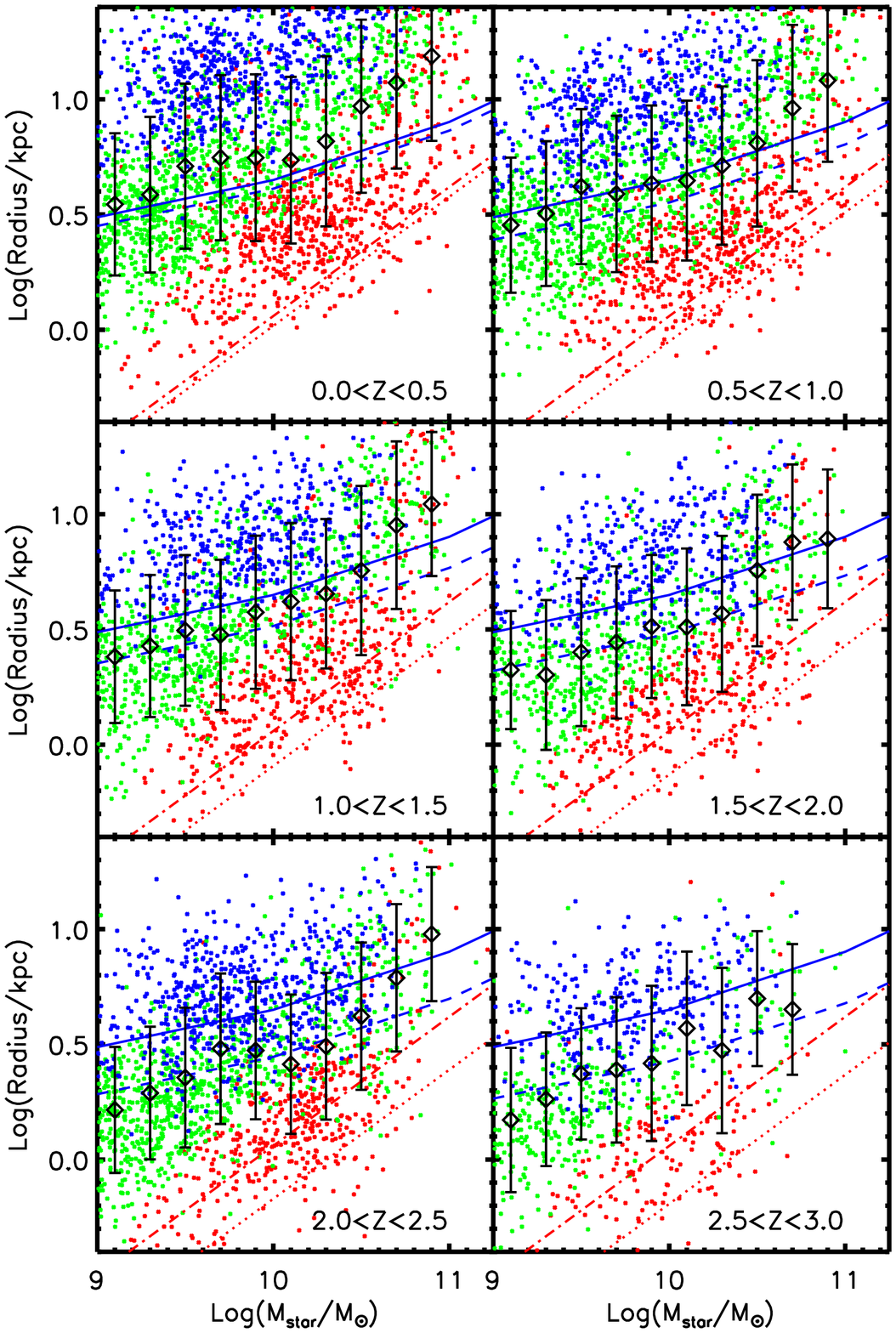}}\hspace{0.0cm}\label{fig:millgfsizemass}}
\caption[Size-Mass relations for the progenitors in the S08 SAM,
  including points for progenitors colored by gas fraction.]{Size-Mass
  relations for the progenitors in the S08 SAM (a) and Millennium SAM
  (b), binned by redshift and including points for progenitors colored
  by gas fraction.  Red denotes $-2.0<\log(G/S)<-1.0$. Green denotes
  $-1.0<\log(G/S)<0.0$. Blue denotes $0.0<\log(G/S)<1.0$. Other
  symbols and lines are as in Figure \ref{fig:spfsizemass}. Both SAMs
  produce systematic gradients in disk galaxy gas fractions as a
  function of size at given mass.  This results from higher star
  formation rates, and therefore faster gas consumption in denser
  galaxies.  When dissipative effects are included in a merger model,
  these systematic gradients result in smaller scatter in the merger
  remnant size-mass relation. }
\end{center}
\end{figure*}

There are also interesting correlations between gas fraction and size.
To demonstrate this, we replot the size-mass relation with points
drawn for progenitors colored by G/S (see Figures
\ref{fig:spfgfsizemass} and \ref{fig:millgfsizemass}).  Red points are
gas-poor with $-2.0<\log(G/S)<-1.0$.  Green points have intermediate
gas fractions with $-1.0<\log(G/S)<0.0$. Blue points are gas rich with
$0.0<\log(G/S)<1.0$.  For both SAMs, at given mass there is a
significant trend of gas fraction with size: {\it the larger radius
  progenitors have higher gas fractions}.  This is because these
galaxies have lower densities and therefore have lower star formation
rates and have consumed less of their gas, since both SAMs assume a
Kennicutt-Schmidt-type relationship for star formation (star formation
rate density is proportional to a power of the gas density).  {\bf A similar trend has been observed in a study of nearby galaxies  \citep{Catinella:2010a}. }This has
important implications for the evolution of scaling relations via
merging.  Remember that dry merging moves galaxies up and to the right
on the plot, that is, it increases both mass and size (Figure
\ref{fig:toyGasConst}).  Because of their location far above the
late-type scaling relation, dry merging would not take the larger
radius progenitors on to the desired relation for remnants.  However,
since these galaxies are gas-rich they can travel with a horizontal or
even somewhat downward vector when they merge.  Additionally, this
convergence of merging vectors within a single mass bin, such that the
gas-poor compact galaxies move to larger radii while the gas-rich low
density galaxies move to smaller radii, results in a reduction of
scatter for the remnant scaling relations.  {\it This effect may
  contribute to the small scatter in the observed early-type size-mass
  relation.}

\subsection{Properties of the Merger Remnants}

Given progenitor properties, the merger model allows us to predict the
sizes, stellar masses, and velocity dispersions of the merger
remnants.  We now exploit this ability to explore the buildup and
evolution of the remnant scaling relations as predicted by the models.
In this subsection we use the simplified merger model described in
Section 2.  For comparison we will show results on the size-mass
relation using the more elaborate C08 model and summing over merger
orbits in subsection 5.3 below.

\begin{figure*}
\begin{center}
\subfigure[][]{\resizebox{8.0cm}{!}{\includegraphics{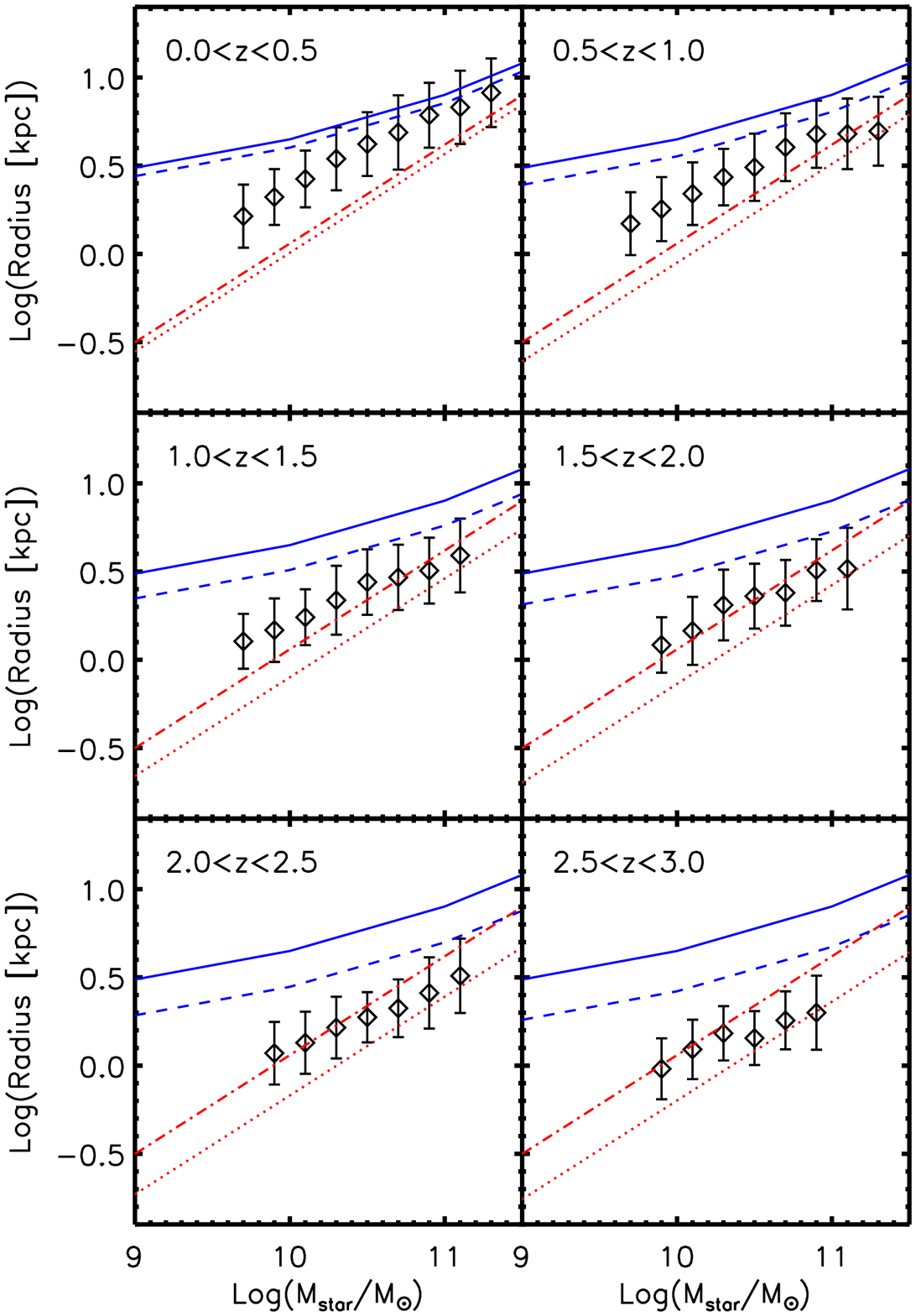}}\hspace{0.0cm}\label{fig:spfrem_sm}}
\subfigure[][]{\resizebox{8.0cm}{!}{\includegraphics{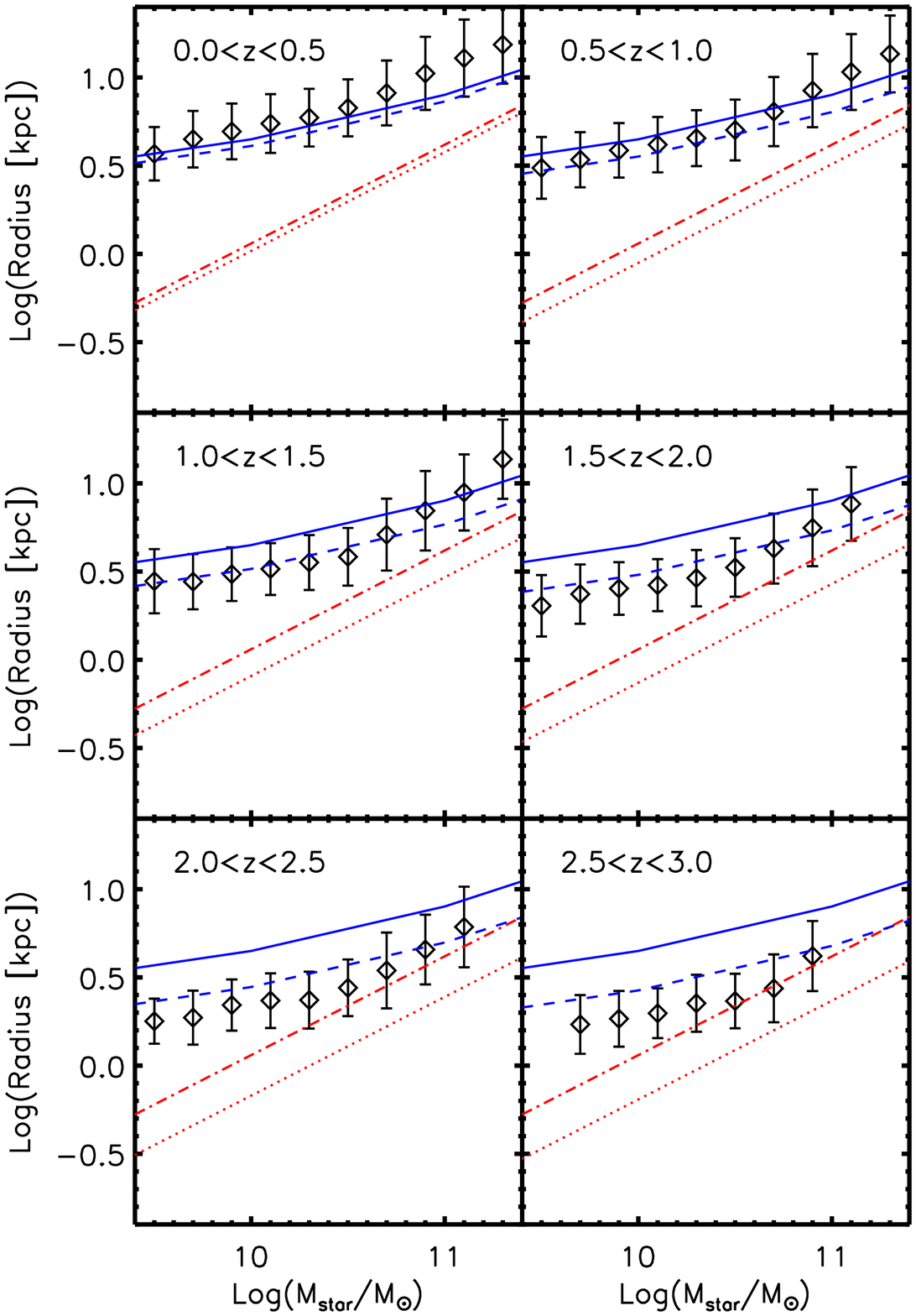}}\hspace{0.0cm}\label{fig:millrem_sm}}
\caption[Size-Mass relations for the remnants in the S08
  SAM]{Size-Mass relations for the remnants in the S08 SAM (a) and
  Millennium SAM (b), binned by redshift.  Lines are observed
  relations as in Figure \ref{fig:spfsizemass}. The S08 SAM captures
  most of the steepening that occurs between the observed late- and
  early-type scaling relations, whereas the slope produced by the
  Millennium SAM remains near that of the progenitors (disks).  Both
  SAMs show roughly the right amount of evolution in size between z=3
  and z=0, but both also overpredict the sizes for a given mass.}
\end{center}
\end{figure*}

The remnant size-mass relation for the SAMs is shown in Figures
\ref{fig:spfrem_sm} and \ref{fig:millrem_sm}.  Note that we are
comparing the size-mass relations for {\it all} observed early-type
galaxies at a given epoch with that for just the remnants of disk
major mergers that occurred in that redshift bin.  This neglects
early-types that formed at earlier epochs, and have not had a recent
merger, as well as early-types formed by mixed-morphology mergers.  In
addition, we only model binary mergers, and do not account for
multiple mergers {\bf or subsequent evolution following the formation event}.  Therefore, this may not be a fair comparison, but
it is the best we can do until we implement the merger model within
SAMs.

The model applied to the S08 SAM overpredicts the typical observed
sizes for a given mass by about 0.3 dex at low masses, and the slope
is steeper than observed.  However, the S08 results roughly capture
the magnitude of observed evolution in average sizes from smaller
sizes at high redshift to larger sizes at low redshift \citep[see also
][]{Trujillo06,Trujillo07,Franx:2008a}. The slope of the size-mass relation is slightly
shallower than the observed slope at all redshifts, but the merger
model has produced a significant steepening from the progenitor
size-mass relation as a result of the mass dependent gas fraction in
the progenitors.

The average predicted size for early-type galaxies in the Millennium
SAM is too large, sometimes by more than 0.5 dex.  Millennium produces
a different normalization for the size-mass relation than S08 for two
reasons: 1) the progenitors are larger, and 2) the progenitors have
less gas.  The slope of the relation is also much shallower than
observed.  This is likely the result of differences in progenitor gas
fractions between the two SAMs.  The evolution in the Millennium
size-mass relationship is such that in the highest redshift bin
($2.5<z<3.0$) the average size at a given mass is $\sim 3$ times
smaller than in the lowest redshift bin.  This is within the errors on
the observed evolution.

\begin{figure*}
\begin{center}
\subfigure[][]{\resizebox{8.0cm}{!}{\includegraphics{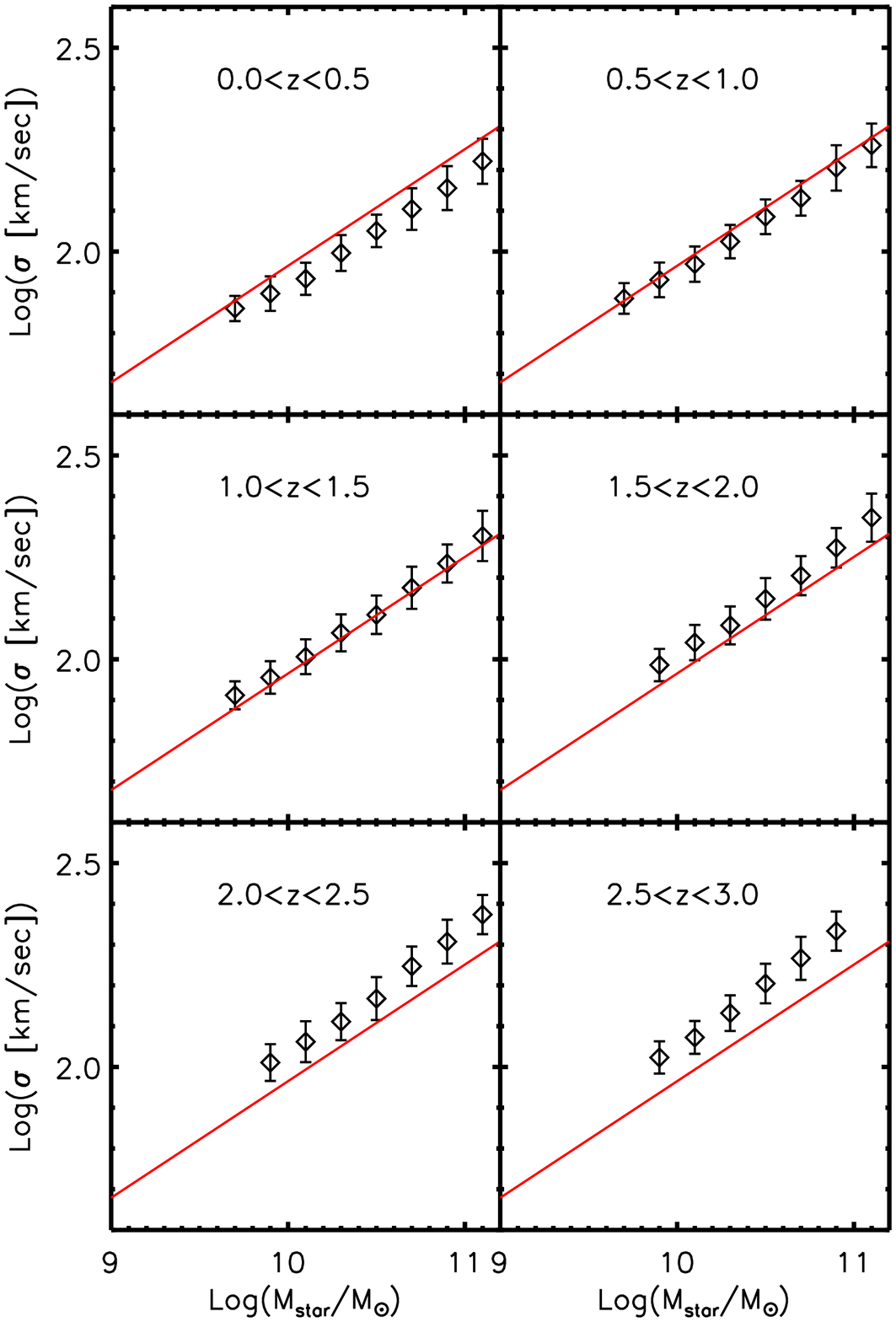}}\hspace{0.0cm}\label{fig:msig_rachel}}
\subfigure[][]{\resizebox{8.0cm}{!}{\includegraphics{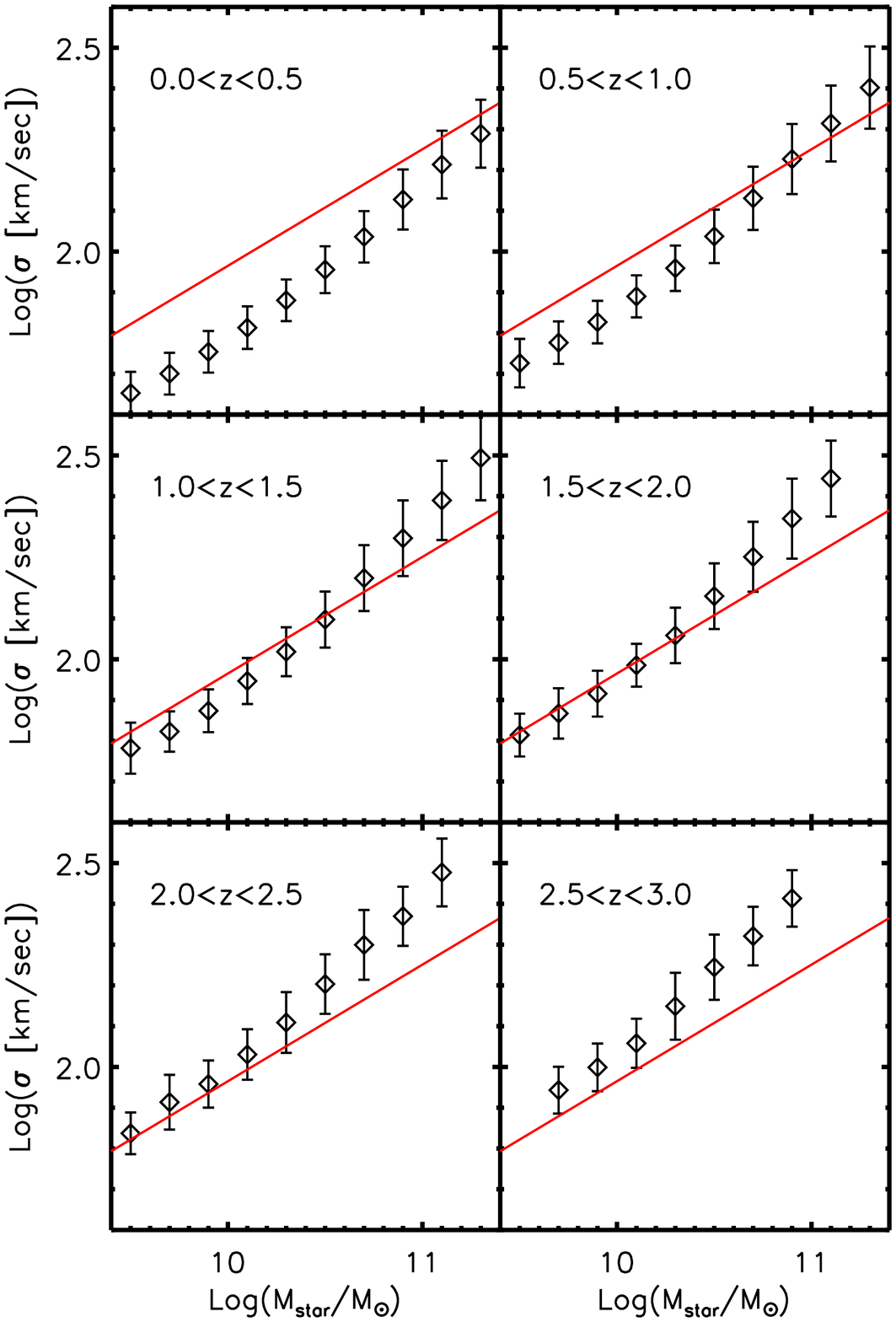}}\hspace{0.0cm}\label{fig:msig_mill}}
\caption[Faber-Jackson relations for the remnants in the S08
  SAM]{Faber-Jackson relations for the remnants in the S08 SAM (a) and
  Millennium SAM (b), binned by redshift. The red line is the observed
  relation at low redshift \citep{Gallazzi06}. The S08 SAM produces
  the correct slope for the FJ relation, whereas the Millennium SAM
  relation is somewhat steeper than observed.  Both SAMs also
  underpredict velocity dispersions for a given mass at z=0, though
  the problem is much more severe with the Millennium SAM.  Both SAMs
  show evolution in the normalization of the FJ relation since z=3.
  This is not yet constrained by observations.}
\end{center}
\end{figure*}

In addition to sizes, the merger model predicts values for velocity
dispersion ($\sigma$), so we also examine the stellar-mass
Faber-Jackson relation (FJ) (see Figures \ref{fig:msig_rachel} and
\ref{fig:msig_mill}). The S08 model roughly produces the correct slope
for the local FJ relation.  However, Millennium produces a slightly
steeper slope than observed in most redshift bins, particularly at
high mass.  S08 slightly underpredicts ($\sim 0.05$ dex) velocity
dispersions at $z=0$, whereas the offset for the Millennium relation
is larger ($\sim 0.2$ dex).  Since remnants from Millennium are
systematically too large, they also have systematically low velocity
dispersions.  Both SAMs show an evolution of the FJ relation with
redshift.  Within the model there are two possible mechanisms for
evolution in $\sigma$ at a given mass: evolving size and evolving dark
matter halo properties.  For S08 $\sigma$ at a given mass increases by
roughly a factor of 1.4 between $z=0$ and $z=3$; for Millennium
$\sigma$ increases by a factor of two.  These values are commensurate
with the decrease in size at a given mass for S08 and Millennium,
suggesting that changing dark matter properties have little effect on
the evolution.  It is not known yet whether this evolution is
consistent with the real universe.

\begin{figure*}
\begin{center}
\subfigure[][]{\resizebox{8.0cm}{!}{\includegraphics{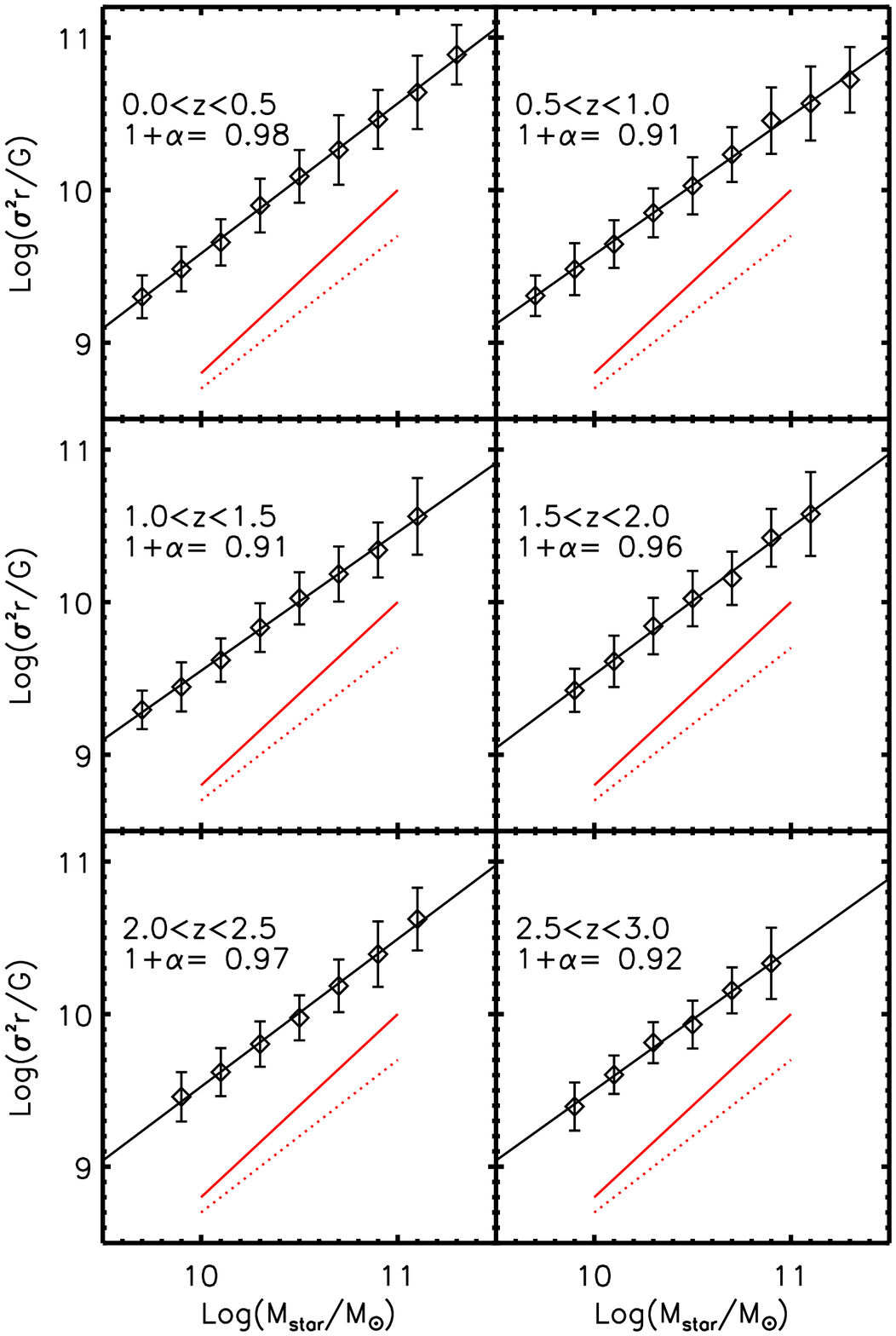}}\hspace{0.0cm}\label{fig:fp_rachel}}
\subfigure[][]{\resizebox{8.0cm}{!}{\includegraphics{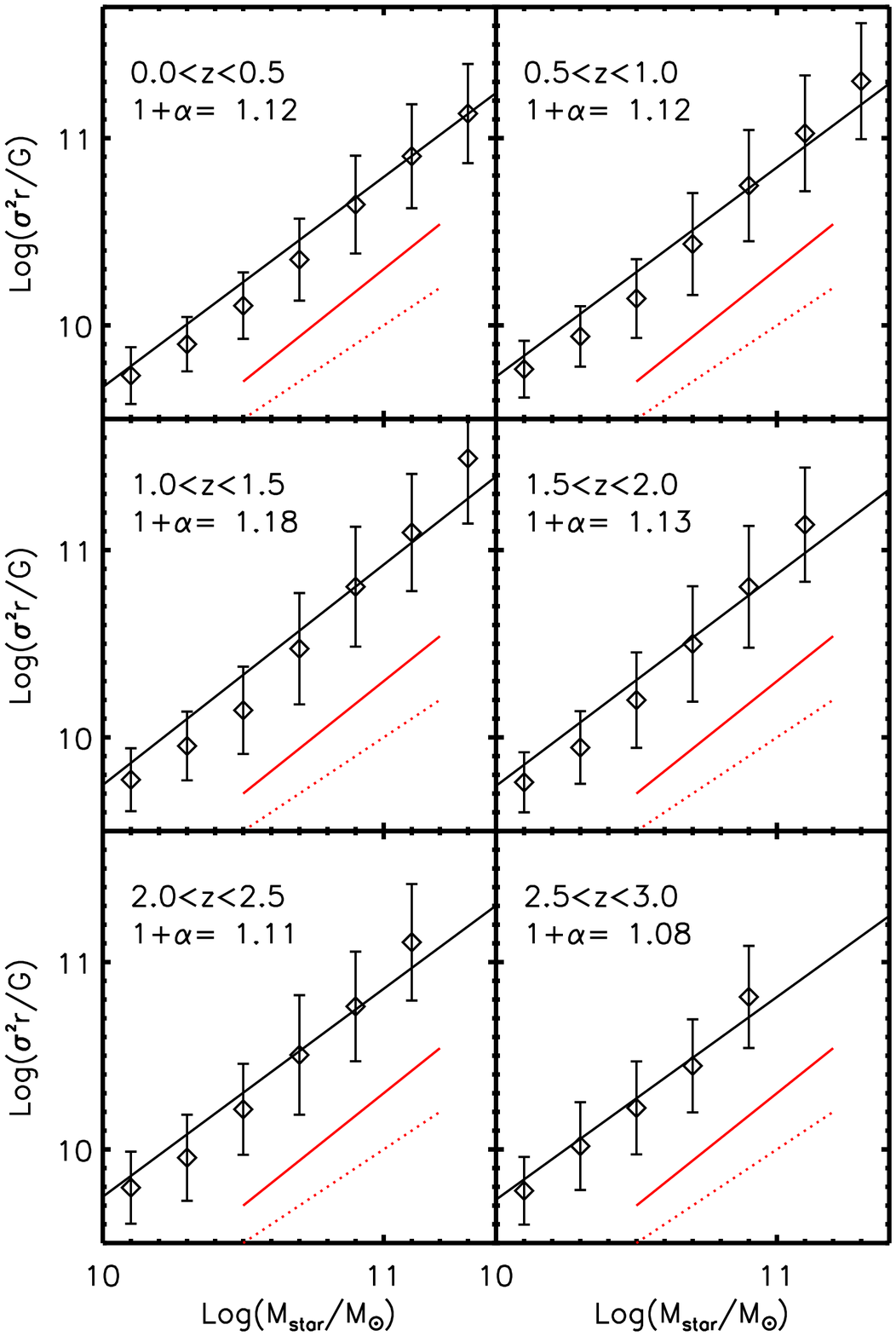}}\hspace{0.0cm}\label{fig:fp_mill}}
\caption[Fundamental plane plotted as $M_{\rm star}$ versus $M_{\rm
    dyn}$ for the remnants in the S08 SAM]{Fundamental plane plotted
  as $M_{\rm star}$ versus $M_{dyn}$ for the remnants in the S08 SAM
  (a) and Millennium SAM (b), binned by redshift. The solid red line
  shows the observed scaling of $M_{\rm dyn}\propto M_{\rm
    star}^{1.2}$, and the dotted red line shows the virial
  scaling. The black line is a fit to the SAM remnants with $M_{\rm
    dyn}\propto M_{\rm star}^{1+\alpha}$ and $1+\alpha$ is shown on
  the figure. Neither SAM completely captures the observed tilt of the
  FP.  S08 produces tilts that are somewhat less than virial, while
  Millennium is much closer to the observed tilt with $1+\alpha=1.12$
  at z=0.}
\end{center}
\end{figure*}

Finally, we examine the fundamental plane (FP) relation predicted by
the merger model. The model uses the virial relation to predict values
for $\sigma$, however the mass used in the model is a dynamical mass
that includes a contribution from dark matter.  Therefore, if a tilt
is produced by the model in the virial projection of the FP, then it
is purely a result of changing dark matter content in the galaxy
centers.  In Figures \ref{fig:fp_rachel} and \ref{fig:fp_mill} we plot
the relation between $\sigma^2r \propto M_{\rm dyn}$ and stellar mass.
The merger model applied to the S08 progenitors does not reproduce the
observed tilt in the FP, with slopes roughly the same as would be
expected from the virial relation.  For Millennium the FP is tilted in
the same sense as is suggested by observations, with dynamical mass
increasing faster than stellar mass, and the tilt produced is only
slightly less than observed.  This tilt results from the contrasting
dissipational and dissipationless evolution of the baryonic and dark
matter components combined with the gas gradient in the progenitors.
Lower mass progenitors have higher gas fractions and therefore produce
more compact stellar remnants with smaller central dark matter
fractions.  Higher mass progenitors have smaller gas fractions and
therefore end up with a sparser stellar center, resulting in larger
central dark matter fractions.

\begin{figure*}
\begin{center}
\subfigure[][]{\resizebox{8.0cm}{!}{\includegraphics{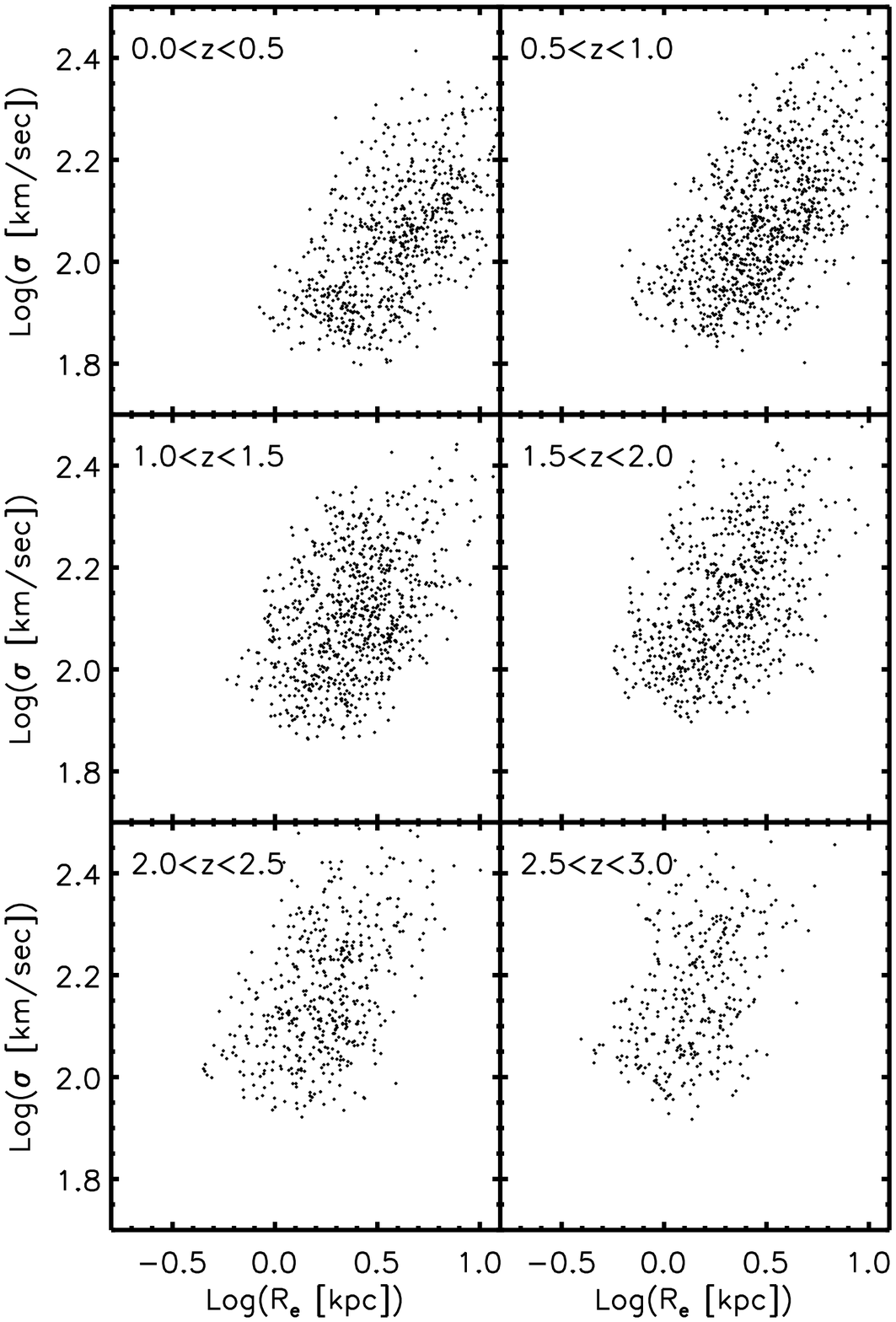}}\hspace{0.0cm}\label{fig:s08_face}}
\subfigure[][]{\resizebox{8.0cm}{!}{\includegraphics{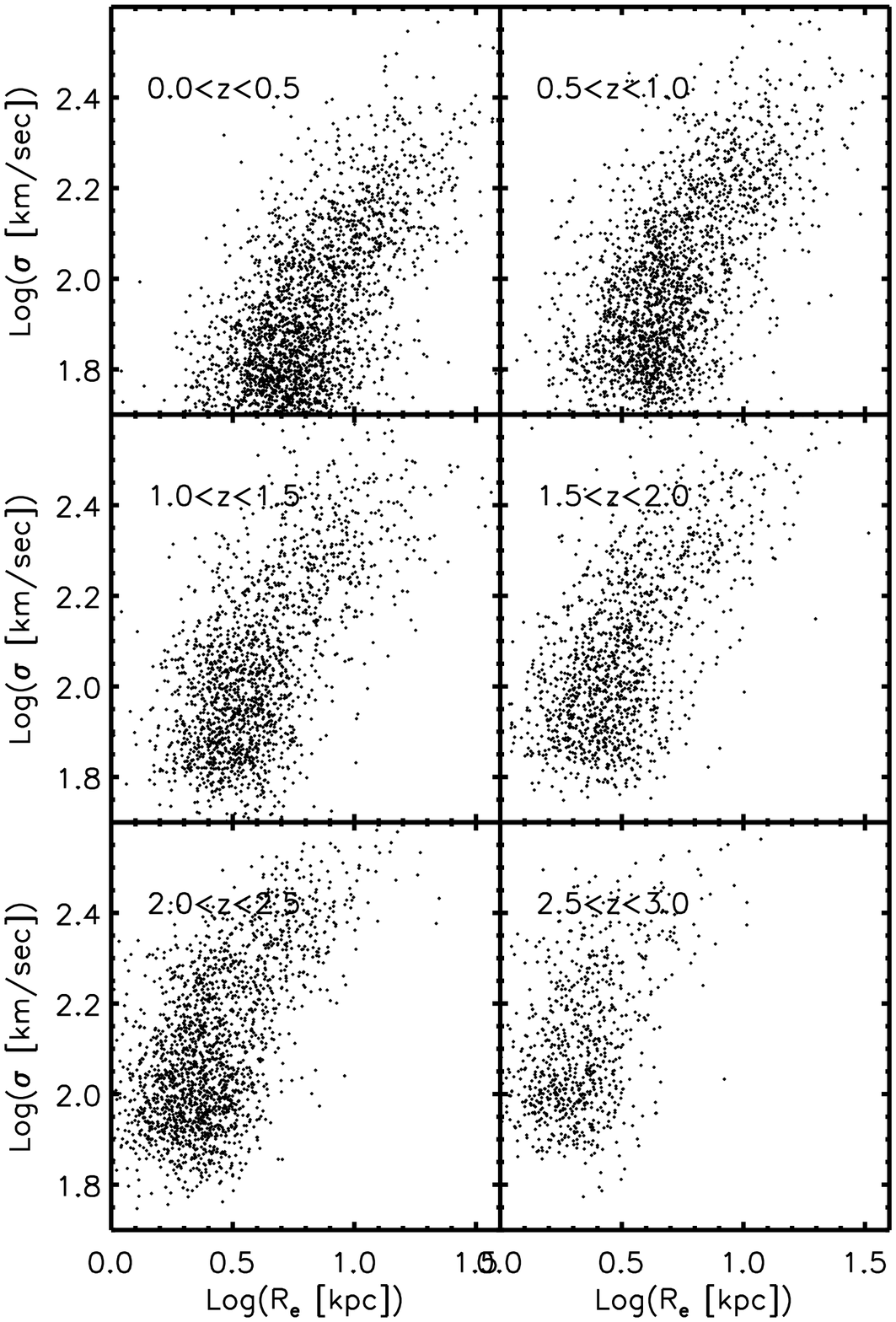}}\hspace{0.0cm}\label{fig:mill_face}}
\caption[Nearly face-on view of the fundamental plane plotted as $r$
  versus $\sigma$ for the remnants in the S08 SAM]{Nearly face-on view
  of the fundamental plane plotted as $r$ versus $\sigma$ for the
  remnants in the S08 SAM (a) and Millennium SAM (b), binned by
  redshift. Both SAMS show some positive correlation between velocity
  dispersion and size at all redshifts.  From z=3 to z=0 remnants also
  gradually migrate from higher dispersion and smaller size to lower
  dispersion and larger size.}
\end{center}
\end{figure*}

Plotting $\sigma$ versus $r$ gives a nearly face-on view of the FP,
and allows us to determine the portion of the FP being populated at a
given redshift by major mergers of disk galaxies.  We show this
relationship for both S08 (Figure \ref{fig:s08_face}) and Millennium
(Figure \ref{fig:mill_face}).  Both SAMs show a correlation between
$\sigma$ and $r$ such that galaxies with larger sizes also tend to
have larger velocity dispersions.  This is similar to a correlation
observed at low redshift in the SDSS \citep{Graves09}.  Most
interesting perhaps is the noticeable evolution with redshift across
the face of the FP.  At high redshift the mergers produce primarily
high $\sigma$ galaxies with relatively small sizes.  As time
progresses toward $z=0$, the typical $\sigma$ of a merger remnant
decreases and the typical size increases, as the correlation between
the size and velocity dispersion marches across the face of the plane.
\citet{Graves09} also see this correlation between $\sigma$ and age,
but not between size and age.  Also, at any given redshift there is a
stronger correlation between $\sigma$ and $r$ in our model than seen
in SDSS.  However, as discussed earlier, in reality each bin would
also contain remnants that were formed at earlier times, which would
weaken the correlation.  Furthermore, one would expect that the
remnants from higher redshifts would be modified by further merging,
which could also increase the radius and scatter across the plane.  An
interesting question that remains unanswered is whether there are
specific regions of the FP space that are not populated by major
merging.  We are working on further studies of the age and metallicity
predicted by our merging model implemented self-consistently within
the SAMs, and will report the results in a forthcoming paper (Porter et al.
in prep).

\subsection{Scaling relations produced with the C08 Model}

\begin{figure*}
\begin{center}
\subfigure[][]{\resizebox{8.0cm}{!}{\includegraphics{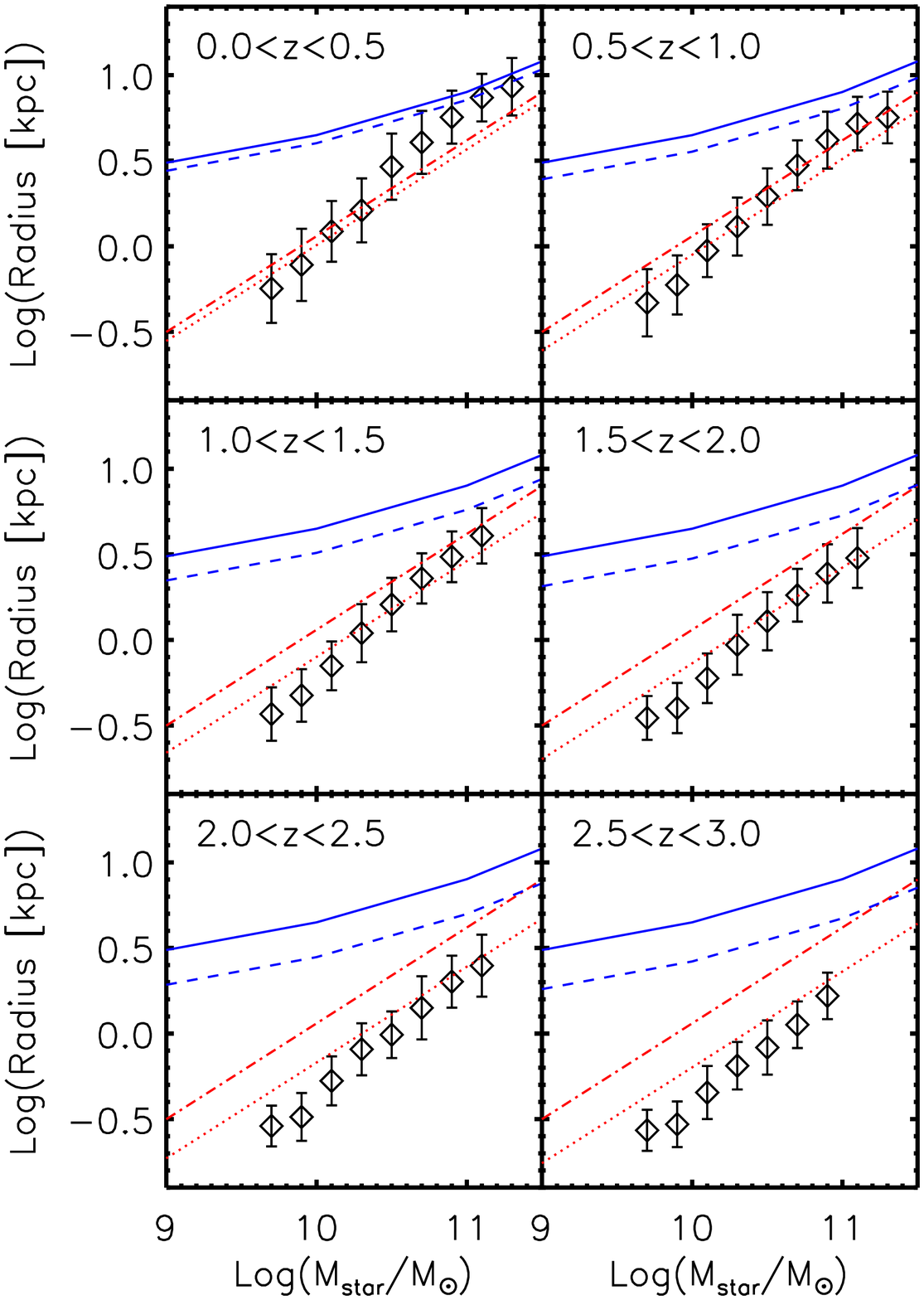}}\hspace{0.0cm}}
\subfigure[][]{\resizebox{8.0cm}{!}{\includegraphics{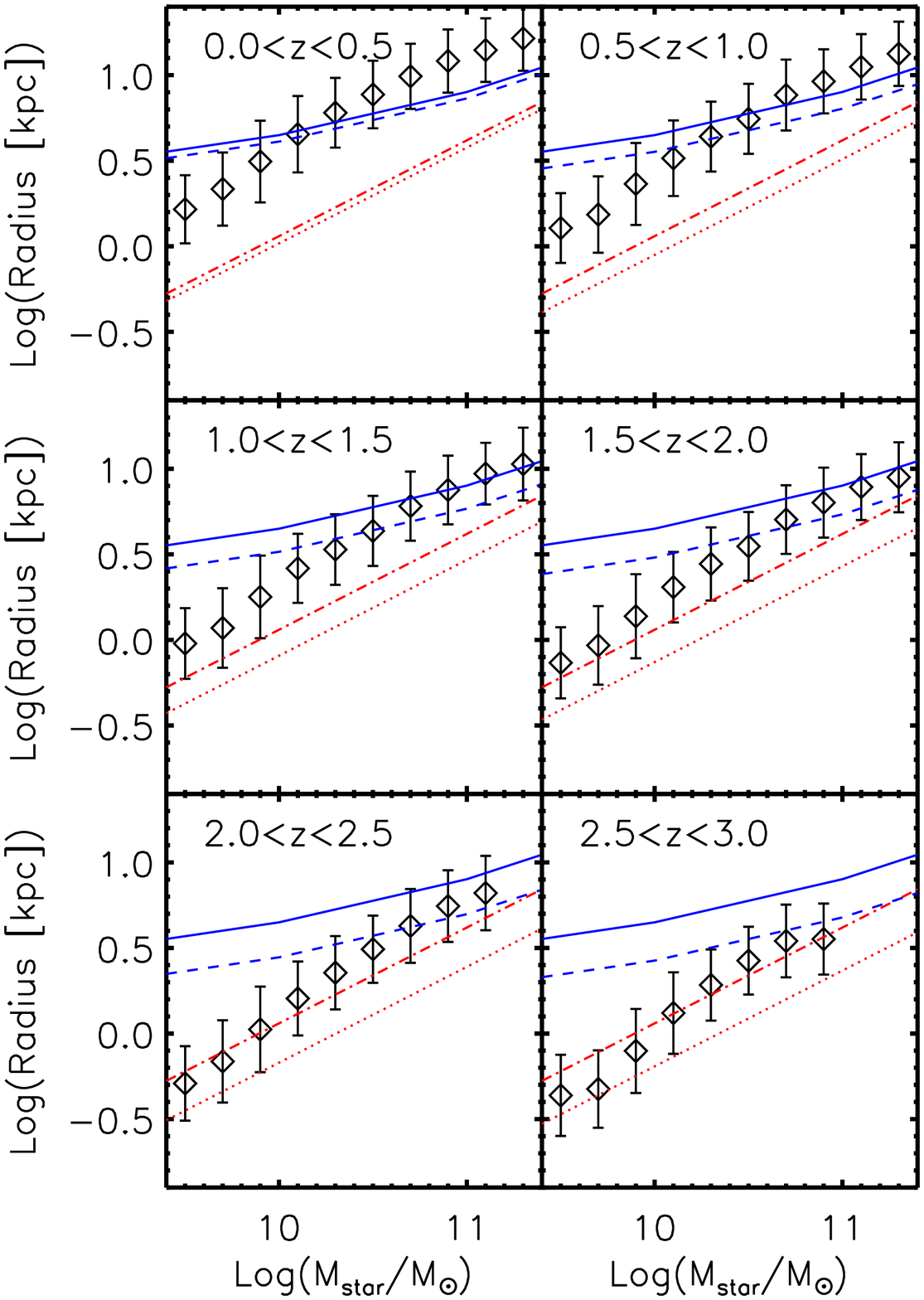}}\hspace{0.0cm}}
\caption{Size mass relations produced by the C08 model combined with
  the new model for central dark matter fractions, applied to S08 (a)
  and Millennium (b).  For both SAMs the relation is steeper than that
  produced by the simpler model.}
\label{fig:mr-old}
\end{center}
\end{figure*}

\begin{figure*}
\begin{center}
\subfigure[][]{\resizebox{8.0cm}{!}{\includegraphics{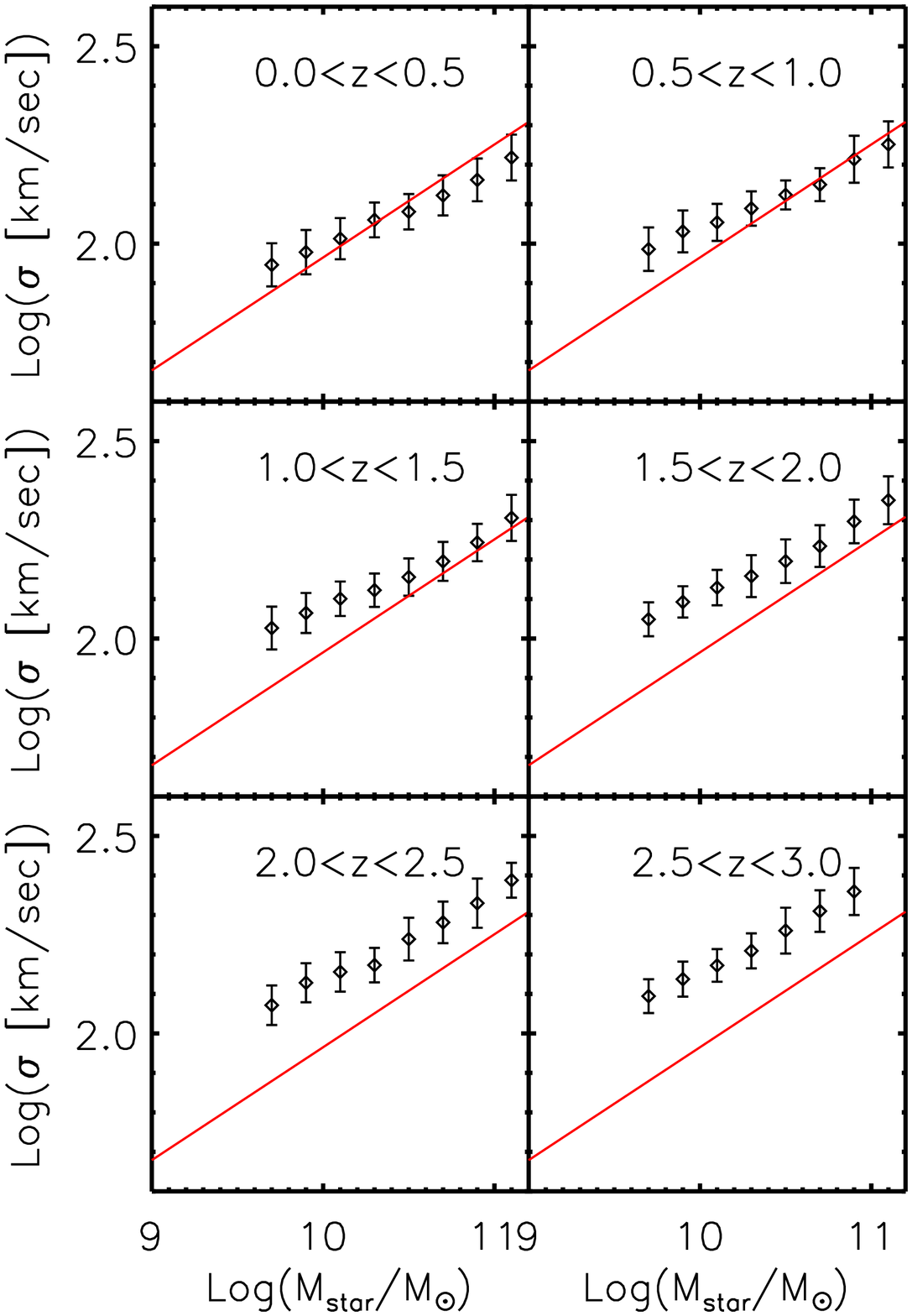}}\hspace{0.0cm}}
\subfigure[][]{\resizebox{8.0cm}{!}{\includegraphics{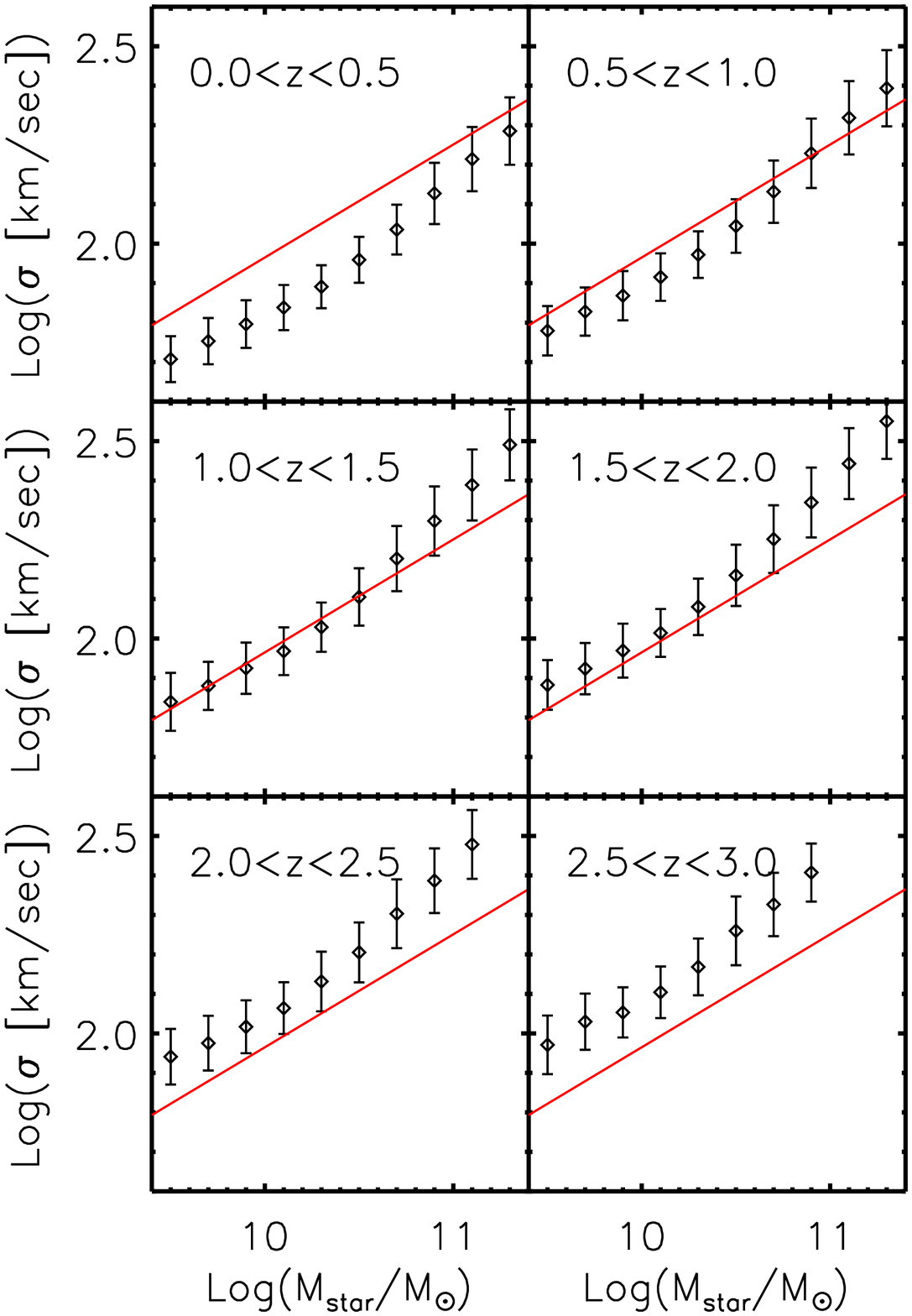}}\hspace{0.0cm}}
\caption{The Faber-Jackson relations produced by the C08 model combined with
  the new model for central dark matter fractions, applied to S08 (a)
  and Millennium (b).}
\label{fig:msig-old}
\end{center}
\end{figure*}

\begin{figure*}
\begin{center}
\subfigure[][]{\resizebox{8.0cm}{!}{\includegraphics{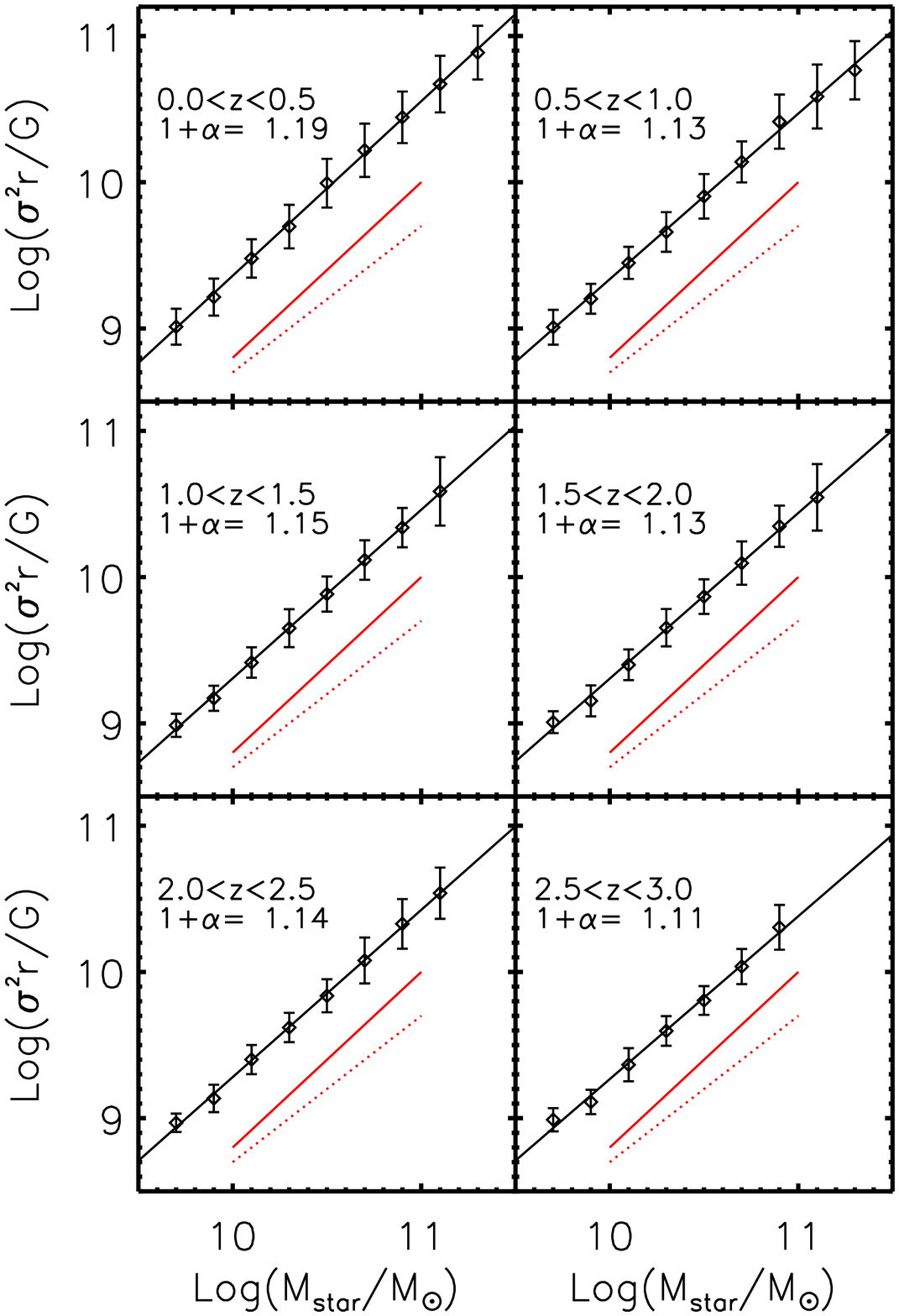}}\hspace{0.0cm}}
\subfigure[][]{\resizebox{8.0cm}{!}{\includegraphics{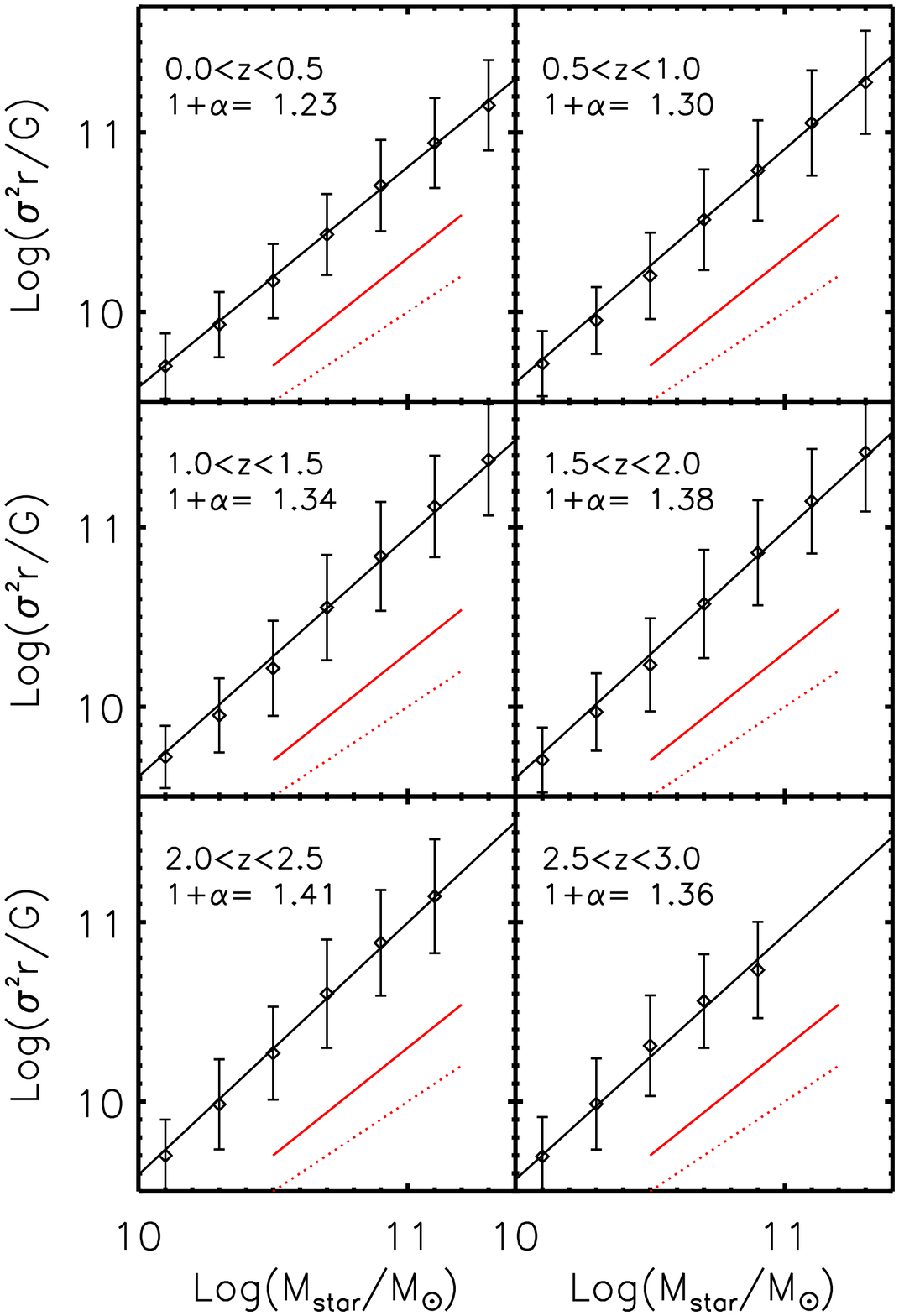}}\hspace{0.0cm}}
\caption{The fundamental plane produced by the C08 model combined with
  the new model for central dark matter fractions, applied to S08 (a)
  and Millennium (b).  The tilt in the fundamental plane is larger
  than created by the new simple model and for both SAMs it is comparable to the
  observed tilt.}
\label{fig:fp-old}
\end{center}
\end{figure*}

The new simplified merger model described above captures much of the
behavior of the C08 model.  However, when applied to the SAMs, the
results of the two models are not identical.  To illustrate the
differences, we show the scaling relations produced using the C08
model (Figures \ref{fig:mr-old}, \ref{fig:msig-old}, and
\ref{fig:fp-old}).  Aside from using the new, extended model for
central dark matter fraction (Appendix A), the model is implemented as
described in C08.  For both SAMs the C08 model results in a more
pronounced steepening of the size-mass relation.  In fact, at low
redshift, for S08 the relation produced has a slope that is too steep,
whereas for Millennium the slope is close to that observed.  Of the
three scaling relations, the size-mass relation is the one that shows
the largest difference between the two models.  However, C08 also
produces a stronger tilt in the FP relation for both SAMs, with both
SAMs having a tilt that is close to that observed.  Though we believe
these differences arise from variations in the gas fractions and
concentrations of the progenitors, we will refrain from a more
detailed comparison until the models are directly implemented within
the SAMs.

\section{Discussion}

A continuing question in the field of galaxy formation is the role
that mergers play in constructing elliptical galaxies.  \citet{Shen03}
have suggested that it is difficult to get the proper elliptical
size-mass relation from repeated major mergers of disk galaxies
because of the rotation between the early- and late-type relations.
Recent SAMs confirm this result, and also produce a scatter in the
remnant size-mass relation that is much too large \citep{Shankar10a,
  Guo10}.  However, these models do not take into account the effects
of dissipation and a possible dependence of gas fraction on progenitor
mass, {\bf as has been done in several high-resolution hydrodynamical simulations \citep{Dekel06, RobertsonFP,
  HopkinsFP}}.  Here we demonstrate that if the effect of dissipation is
included {\bf within SAMs} and if the progenitors have a decreasing gas fraction with
increasing mass, then this results in a steeper size-mass relation for
the merger remnants {\bf compared to the disk progenitors}.  Furthermore, since less dense disks have lower
star formation rates, the disk galaxy progenitors have increasing gas
fractions with increasing size at a given mass.  Applying a
dissipative merger model to progenitors with this trend in gas
fraction also reduces the scatter in the size-mass relation, as
explained at the end of Section 5.1.  In fact, our model, with a
typical dispersion of 0.2 dex in the size-mass relation, has less than
the 0.3-0.4 dex dispersion in the relation observed by \citet{Shen03}.
That our dispersion is too small might be expected, since we are only
considering major mergers of disk galaxies, a subset of the total
early-type population.  {\bf As noted above, we are also separating the size-mass relation according to the redshift of the merger.  In order to directly compare to observations, we would need to include all the mergers that have occurred at any given redshift.}  Comparison of the progenitor and remnant
properties suggests that the overall normalization of the early-type
size-mass relation is determined by a combination of the size-mass
relation of the progenitor disk galaxies and the normalization of the
gas fraction relation.

In order to capture the observed Faber-Jackson (FJ) relation, it is
necessary to track the changing central dark matter fraction.  The
results from our model suggest that the FJ relation may evolve
modestly over time such that early-type galaxies at a given mass have
larger velocity dispersions at higher redshift.  The merger model
applied to progenitors from the S08 SAM results in a slope close to
the virial slope, which is slightly shallower than the observed
slope. The model applied to progenitors from the Millennium SAM
reproduces the observed tilt in the fundamental plane (FP), but this
agreement appears to be a fortuitous cancellation between sizes that
are too large and velocity dispersions that are too small at a given
mass.  The tilt predicted by the model is the result of a changing
central dark matter fraction with mass that results from a mass
dependence of gas fractions in the progenitors.  Higher mass
progenitors have less gas and therefore less dissipation during their
mergers, resulting in remnants with a less concentrated stellar center
and consequently a higher dark matter fraction.  In the face-on view
of the FP, the remnants evolve from higher $\sigma$ and lower $r$ at
high redshift to lower $\sigma$ and higher $r$ at low redshift.

It is important to note that, since we calculate remnant properties
during post-processing, the scaling relations derived using the model
in this paper do not capture the entire elliptical population at each
redshift.  Rather they depict the scaling relations of elliptical
galaxies being added to the population via major mergers of disks at a
given redshift.  One would expect that the high-redshift galaxies
would undergo further accretion and merging, including minor merging
and dry merging in some cases {\bf \citep[see e.g.][]{Bezanson:2009a,Oser:2010a}.  In particular, minor mergers between a large, gas-poor elliptical and several small spirals have been shown to greatly increase the effective radius of the elliptical while leaving the velocity dispersion nearly unchanged \citep{Naab:2009a}. }  Moreover
some early-type galaxies may be formed via mergers involving one or
more early-type progenitors, which are also neglected here.  We find
that approximately one-third of the low-redshift spheroids at all
masses are accounted for by disk-disk major mergers, so it could be
misleading to compare our model outputs with all observed early-type
galaxies at various redshifts.

We also note that there are differences between the results of the
simple model used here and the previous model of C08.  Without fully
implementing the models within the SAMs and tuning the various model
parameters, it is difficult to know which model will ultimately
produce more favorable results.  The simple model requires fewer
unknown parameters, and therefore is preferred if it produces
reasonable results.  However, {\it the key conclusions of this work
  are that the steepening {\bf and reduced scatter of the size-mass relation from that of disk galaxies}, and the tilt in the fundamental plane can
  all be produced by accounting for dissipation during galaxy
  mergers.}  While the strength of these effects will vary by model,
the effects themselves are expected to be general features of any
realistic merger model that accounts for dissipation.

In future work, it will be useful to a incorporate a dissipational
merger model within semi-analytic models.  This will allow a more
comprehensive exploration of the evolution of galaxy scaling relations
via merging by allowing the inclusion of a wider variety of progenitor
types, and enable the study of correlations with other quantities such
as age and metallicity.  However, even the simple external processing
of SAM progenitors as carried out in this study shows that mergers are
a viable mechanism for the production of a large fraction of
elliptical galaxies.  Furthermore, we find that the elliptical scaling
relations are plausibly explained given the combination of progenitor
scaling relations and gas fractions.

\leftline{\bf Acknowledgments} MC, AD, and JRP acknowledge support
from NASA ATP grant NNX07AG94G and NSF grant 1010033 at UCSC.  DC
acknowledges receipt of a QEII Fellowship awarded by the Australian
government. The work of AD was partly supported by ISF grant 6/08, by
GIF grant G-1052-104.7/2009, and by a DIP grant. We also thank
T. J. Cox, Sandra Faber, Jenny Graves, Patrik Jonsson, and Thorsten
Naab for helpful conversations.  Finally, we thank the anonymous
referee for many suggestions and questions that helped us to clarify
the paper.

\bibliographystyle{mn2e}
\bibliography{matt,tj,patriks,new-refs,lauren}

\begin{thebibliography}{}

\bibitem[\protect\citeauthoryear{{Barden}, {Rix}, {Somerville}, {Bell},
  {H{\"a}u{\ss}ler}, {Peng}, {Borch}, {Beckwith}, {Caldwell}, {Heymans},
  {Jahnke}, {Jogee}, {McIntosh}, {Meisenheimer}, {S{\'a}nchez}, {Wisotzki} \&
  {Wolf}}{{Barden} et~al.}{2005}]{Barden05}
{Barden} M.,  {Rix} H.-W.,  {Somerville} R.~S.,  {Bell} E.~F.,
  {H{\"a}u{\ss}ler} B.,  {Peng} C.~Y.,  {Borch} A.,  {Beckwith} S.~V.~W.,
  {Caldwell} J.~A.~R.,  {Heymans} C.,  {Jahnke} K.,  {Jogee} S.,  {McIntosh}
  D.~H.,  {Meisenheimer} K.,  {S{\'a}nchez} S.~F.,  {Wisotzki} L.,    {Wolf}
  C.,  2005, \apj, 635, 959

\bibitem[\protect\citeauthoryear{{Barnes} \& {Hernquist}}{{Barnes} \&
  {Hernquist}}{1992}]{Barnes92}
{Barnes} J.~E.,  {Hernquist} L.,  1992, \araa, 30, 705

\bibitem[\protect\citeauthoryear{{Barnes} \& {Hernquist}}{{Barnes} \&
  {Hernquist}}{1996}]{Barnes:1996a}
{Barnes} J.~E.,  {Hernquist} L.,  1996, \apj, 471, 115

\bibitem[\protect\citeauthoryear{{Benson}}{{Benson}}{2005}]{Benson05}
{Benson} A.~J.,  2005, \mnras, 358, 551

\bibitem[\protect\citeauthoryear{{Benson}}{{Benson}}{2010}]{Benson10}
{Benson} A.~J.,  2010, ArXiv e-prints

\bibitem[\protect\citeauthoryear{{Benson} \& {Bower}}{{Benson} \&
  {Bower}}{2010}]{BensonBower10}
{Benson} A.~J.,  {Bower} R.,  2010, \mnras, 405, 1573

\bibitem[\protect\citeauthoryear{{Bernardi}, {Sheth}, {Annis}, {Burles},
  {Eisenstein}, {Finkbeiner}, {Hogg}, {Lupton}, {Schlegel}, {SubbaRao} \&
  {York}}{{Bernardi} et~al.}{2003a}]{Bernardi03a}
{Bernardi} M.,  {Sheth} R.~K.,  {Annis} J.,  {Burles} S.,  {Eisenstein} D.~J.,
  {Finkbeiner} D.~P.,  {Hogg} D.~W.,  {Lupton} R.~H.,  {Schlegel} D.~J.,
  {SubbaRao} M.,    {York} D.~G.,  2003a, \aj, 125, 1849

\bibitem[\protect\citeauthoryear{{Bernardi}, {Sheth}, {Annis}, {Burles},
  {Eisenstein}, {Finkbeiner}, {Hogg}, {Lupton}, {Schlegel}, {SubbaRao} \&
  {York}}{{Bernardi} et~al.}{2003b}]{Bernardi03b}
{Bernardi} M.,  {Sheth} R.~K.,  {Annis} J.,  {Burles} S.,  {Eisenstein} D.~J.,
  {Finkbeiner} D.~P.,  {Hogg} D.~W.,  {Lupton} R.~H.,  {Schlegel} D.~J.,
  {SubbaRao} M.,    {York} D.~G.,  2003b, \aj, 125, 1866

\bibitem[\protect\citeauthoryear{Bezanson, van Dokkum, Tal, Marchesini, Kriek,
  Franx \& Coppi}{Bezanson et~al.}{2009}]{Bezanson:2009a}
Bezanson R.,  van Dokkum P.~G.,  Tal T.,  Marchesini D.,  Kriek M.,  Franx M.,
    Coppi P.,  2009, \apj, 697, 1290

\bibitem[\protect\citeauthoryear{{Bolton}, {Burles}, {Treu}, {Koopmans} \&
  {Moustakas}}{{Bolton} et~al.}{2007}]{Bolton07}
{Bolton} A.~S.,  {Burles} S.,  {Treu} T.,  {Koopmans} L.~V.~E.,    {Moustakas}
  L.~A.,  2007, \apjl, 665, L105

\bibitem[\protect\citeauthoryear{{Bolton}, {Treu}, {Koopmans}, {Gavazzi},
  {Moustakas}, {Burles}, {Schlegel} \& {Wayth}}{{Bolton}
  et~al.}{2008}]{Bolton08}
{Bolton} A.~S.,  {Treu} T.,  {Koopmans} L.~V.~E.,  {Gavazzi} R.,  {Moustakas}
  L.~A.,  {Burles} S.,  {Schlegel} D.~J.,    {Wayth} R.,  2008, ArXiv e-prints,
  805

\bibitem[\protect\citeauthoryear{Bournaud, Chapon, Teyssier, Powell, Elmegreen,
  Elmegreen, Duc, Contini, Epinat \& Shapiro}{Bournaud
  et~al.}{2011}]{Bournaud:2011a}
Bournaud F.,  Chapon D.,  Teyssier R.,  Powell L.~C.,  Elmegreen B.~G.,
  Elmegreen D.~M.,  Duc P.-A.,  Contini T.,  Epinat B.,    Shapiro K.~L.,
  2011, \apj, 730, 4

\bibitem[\protect\citeauthoryear{Bournaud, Jog \& Combes}{Bournaud
  et~al.}{2007}]{Bournaud:2007a}
Bournaud F.,  Jog C.~J.,    Combes F.,  2007, A\&A, 476, 1179

\bibitem[\protect\citeauthoryear{Bournaud, Powell, Chapon \& Teyssier}{Bournaud
  et~al.}{2010}]{Bournaud:2010a}
Bournaud F.,  Powell L.~C.,  Chapon D.,    Teyssier R.,  2010, ArXiv e-prints,
  1012, 5227

\bibitem[\protect\citeauthoryear{{Bower}, {Benson}, {Malbon}, {Helly}, {Frenk},
  {Baugh}, {Cole} \& {Lacey}}{{Bower} et~al.}{2006}]{Bower06}
{Bower} R.~G.,  {Benson} A.~J.,  {Malbon} R.,  {Helly} J.~C.,  {Frenk} C.~S.,
  {Baugh} C.~M.,  {Cole} S.,    {Lacey} C.~G.,  2006, \mnras, 370, 645

\bibitem[\protect\citeauthoryear{{Bryan} \& {Norman}}{{Bryan} \&
  {Norman}}{1998}]{Bryan98}
{Bryan} G.~L.,  {Norman} M.~L.,  1998, \apj, 495, 80

\bibitem[\protect\citeauthoryear{{Buitrago}, {Trujillo}, {Conselice},
  {Bouwens}, {Dickinson} \& {Yan}}{{Buitrago} et~al.}{2008}]{Buitrago08}
{Buitrago} F.,  {Trujillo} I.,  {Conselice} C.~J.,  {Bouwens} R.~J.,
  {Dickinson} M.,    {Yan} H.,  2008, \apjl, 687, L61

\bibitem[\protect\citeauthoryear{{Calura}, {Jimenez}, {Panter}, {Matteucci} \&
  {Heavens}}{{Calura} et~al.}{2007}]{Calura07}
{Calura} F.,  {Jimenez} R.,  {Panter} B.,  {Matteucci} F.,    {Heavens} A.~F.,
  2007, ArXiv e-prints, 707

\bibitem[\protect\citeauthoryear{{Cappellari}, {Bacon}, {Bureau}, {Damen},
  {Davies}, {de Zeeuw}, {Emsellem}, {Falc{\'o}n-Barroso}, {Krajnovi{\'c}},
  {Kuntschner}, {McDermid}, {Peletier}, {Sarzi}, {van den Bosch} \& {van de
  Ven}}{{Cappellari} et~al.}{2006}]{Cappellari06}
{Cappellari} M.,  {Bacon} R.,  {Bureau} M.,  {Damen} M.~C.,  {Davies} R.~L.,
  {de Zeeuw} P.~T.,  {Emsellem} E.,  {Falc{\'o}n-Barroso} J.,  {Krajnovi{\'c}}
  D.,  {Kuntschner} H.,  {McDermid} R.~M.,  {Peletier} R.~F.,  {Sarzi} M.,
  {van den Bosch} R.~C.~E.,    {van de Ven} G.,  2006, \mnras, 366, 1126

\bibitem[\protect\citeauthoryear{{Catinella et al.}}{{Catinella et
  al.}}{2010}]{Catinella:2010a}
{Catinella et al.} 2010, \mnras, 403, 683

\bibitem[\protect\citeauthoryear{{Cimatti}, {Cassata}, {Pozzetti}, {Kurk},
  {Mignoli}, {Renzini}, {Daddi}, {Bolzonella}, {Brusa}, {Rodighiero},
  {Dickinson}, {Franceschini}, {Zamorani}, {Berta}, {Rosati} \&
  {Halliday}}{{Cimatti} et~al.}{2008}]{Cimatti08}
{Cimatti} A.,  {Cassata} P.,  {Pozzetti} L.,  {Kurk} J.,  {Mignoli} M.,
  {Renzini} A.,  {Daddi} E.,  {Bolzonella} M.,  {Brusa} M.,  {Rodighiero} G.,
  {Dickinson} M.,  {Franceschini} A.,  {Zamorani} G.,  {Berta} S.,  {Rosati}
  P.,    {Halliday} C.,  2008, \aap, 482, 21

\bibitem[\protect\citeauthoryear{{Ciotti}, {Lanzoni} \& {Volonteri}}{{Ciotti}
  et~al.}{2007}]{Ciotti07}
{Ciotti} L.,  {Lanzoni} B.,    {Volonteri} M.,  2007, \apj, 658, 65

\bibitem[\protect\citeauthoryear{{Cole}, {Aragon-Salamanca}, {Frenk}, {Navarro}
  \& {Zepf}}{{Cole} et~al.}{1994}]{Cole94}
{Cole} S.,  {Aragon-Salamanca} A.,  {Frenk} C.~S.,  {Navarro} J.~F.,    {Zepf}
  S.~E.,  1994, \mnras, 271, 781

\bibitem[\protect\citeauthoryear{{Cole}, {Lacey}, {Baugh} \& {Frenk}}{{Cole}
  et~al.}{2000}]{Cole00}
{Cole} S.,  {Lacey} C.~G.,  {Baugh} C.~M.,    {Frenk} C.~S.,  2000, \mnras,
  319, 168

\bibitem[\protect\citeauthoryear{{Cook}, {Barausse}, {Evoli}, {Lapi} \&
  {Granato}}{{Cook} et~al.}{2010}]{Cook10}
{Cook} M.,  {Barausse} E.,  {Evoli} C.,  {Lapi} A.,    {Granato} G.~L.,  2010,
  \mnras, 402, 2113

\bibitem[\protect\citeauthoryear{{Covington}, {Dekel}, {Cox}, {Jonsson} \&
  {Primack}}{{Covington} et~al.}{2008}]{remnants}
{Covington} M.,  {Dekel} A.,  {Cox} T.~J.,  {Jonsson} P.,    {Primack} J.~R.,
  2008, \mnras, 384, 94

\bibitem[\protect\citeauthoryear{{Covington}}{{Covington}}{2008}]{Covington08}
{Covington} M.~D.,  2008, PhD thesis, University of California, Santa Cruz

\bibitem[\protect\citeauthoryear{{Cox}}{{Cox}}{2004}]{thesis}
{Cox} T.~J.,  2004, PhD thesis, {UC Santa Cruz}

\bibitem[\protect\citeauthoryear{{Cox}, {Jonsson}, {Primack} \&
  {Somerville}}{{Cox} et~al.}{2006}]{Cox05}
{Cox} T.~J.,  {Jonsson} P.,  {Primack} J.~R.,    {Somerville} R.~S.,  2006,
  \mnras, 373, 1013

\bibitem[\protect\citeauthoryear{{Cox}, {Jonsson}, {Somerville}, {Primack} \&
  {Dekel}}{{Cox} et~al.}{2008}]{Cox08}
{Cox} T.~J.,  {Jonsson} P.,  {Somerville} R.~S.,  {Primack} J.~R.,    {Dekel}
  A.,  2008, \mnras, 384, 386

\bibitem[\protect\citeauthoryear{{Croton}, {Springel}, {White}, {De Lucia},
  {Frenk}, {Gao}, {Jenkins}, {Kauffmann}, {Navarro} \& {Yoshida}}{{Croton}
  et~al.}{2006}]{Croton06}
{Croton} D.~J.,  {Springel} V.,  {White} S.~D.~M.,  {De Lucia} G.,  {Frenk}
  C.~S.,  {Gao} L.,  {Jenkins} A.,  {Kauffmann} G.,  {Navarro} J.~F.,
  {Yoshida} N.,  2006, \mnras, 365, 11

\bibitem[\protect\citeauthoryear{{De Lucia}, {Springel}, {White}, {Croton} \&
  {Kauffmann}}{{De Lucia} et~al.}{2006}]{DeLucia06}
{De Lucia} G.,  {Springel} V.,  {White} S.~D.~M.,  {Croton} D.,    {Kauffmann}
  G.,  2006, \mnras, 366, 499

\bibitem[\protect\citeauthoryear{{Dekel} \& {Cox}}{{Dekel} \&
  {Cox}}{2006}]{Dekel06}
{Dekel} A.,  {Cox} T.~J.,  2006, \mnras, 370, 1445

\bibitem[\protect\citeauthoryear{{Djorgovski} \& {Davis}}{{Djorgovski} \&
  {Davis}}{1987}]{Djorgovski87}
{Djorgovski} S.,  {Davis} M.,  1987, \apj, 313, 59

\bibitem[\protect\citeauthoryear{{Dressler}, {Lynden-Bell}, {Burstein},
  {Davies}, {Faber}, {Terlevich} \& {Wegner}}{{Dressler}
  et~al.}{1987}]{Dressler87}
{Dressler} A.,  {Lynden-Bell} D.,  {Burstein} D.,  {Davies} R.~L.,  {Faber}
  S.~M.,  {Terlevich} R.,    {Wegner} G.,  1987, \apj, 313, 42

\bibitem[\protect\citeauthoryear{{Dutton}, {van den Bosch}, {Faber}, {Simard},
  {Kassin}, {Koo}, {Bundy}, {Huang}, {Weiner}, {Cooper}, {Newman}, {Mozena} \&
  {Koekemoer}}{{Dutton} et~al.}{2010}]{Dutton10}
{Dutton} A.~A.,  {van den Bosch} F.~C.,  {Faber} S.~M.,  {Simard} L.,  {Kassin}
  S.~A.,  {Koo} D.~C.,  {Bundy} K.,  {Huang} J.,  {Weiner} B.~J.,  {Cooper}
  M.~C.,  {Newman} J.~A.,  {Mozena} M.,    {Koekemoer} A.,  2010, ArXiv
  e-prints

\bibitem[\protect\citeauthoryear{{Faber} \& {Jackson}}{{Faber} \&
  {Jackson}}{1976}]{FJ}
{Faber} S.~M.,  {Jackson} R.~E.,  1976, \apj, 204, 668

\bibitem[\protect\citeauthoryear{{Fan}, {Lapi}, {Bressan}, {Bernardi}, {De
  Zotti} \& {Danese}}{{Fan} et~al.}{2010}]{Fan10}
{Fan} L.,  {Lapi} A.,  {Bressan} A.,  {Bernardi} M.,  {De Zotti} G.,
  {Danese} L.,  2010, \apj, 718, 1460

\bibitem[\protect\citeauthoryear{{Fontanot}, {De Lucia}, {Monaco}, {Somerville}
  \& {Santini}}{{Fontanot} et~al.}{2009}]{Fontanot09}
{Fontanot} F.,  {De Lucia} G.,  {Monaco} P.,  {Somerville} R.~S.,    {Santini}
  P.,  2009, \mnras, 397, 1776

\bibitem[\protect\citeauthoryear{{Franx}, {van Dokkum}, {Schreiber}, {Wuyts},
  {Labb{\'e}} \& {Toft}}{{Franx} et~al.}{2008}]{Franx:2008a}
{Franx} M.,  {van Dokkum} P.~G.,  {Schreiber} N.~M.~F.,  {Wuyts} S.,
  {Labb{\'e}} I.,    {Toft} S.,  2008, \apj, 688, 770

\bibitem[\protect\citeauthoryear{{Gallazzi}, {Charlot}, {Brinchmann} \&
  {White}}{{Gallazzi} et~al.}{2006}]{Gallazzi06}
{Gallazzi} A.,  {Charlot} S.,  {Brinchmann} J.,    {White} S.~D.~M.,  2006,
  \mnras, 370, 1106

\bibitem[\protect\citeauthoryear{{Gerhard}, {Kronawitter}, {Saglia} \&
  {Bender}}{{Gerhard} et~al.}{2001}]{Gerhard01}
{Gerhard} O.,  {Kronawitter} A.,  {Saglia} R.~P.,    {Bender} R.,  2001, \aj,
  121, 1936

\bibitem[\protect\citeauthoryear{{Graves}, {Faber} \& {Schiavon}}{{Graves}
  et~al.}{2009}]{Graves09}
{Graves} G.~J.,  {Faber} S.~M.,    {Schiavon} R.~P.,  2009, \apj, 698, 1590

\bibitem[\protect\citeauthoryear{{Guo}, {White}, {Boylan-Kolchin}, {De Lucia},
  {Kauffmann}, {Lemson}, {Li}, {Springel} \& {Weinmann}}{{Guo}
  et~al.}{2010}]{Guo10}
{Guo} Q.,  {White} S.,  {Boylan-Kolchin} M.,  {De Lucia} G.,  {Kauffmann} G.,
  {Lemson} G.,  {Li} C.,  {Springel} V.,    {Weinmann} S.,  2010, ArXiv
  e-prints

\bibitem[\protect\citeauthoryear{{Hatton}, {Devriendt}, {Ninin}, {Bouchet},
  {Guiderdoni} \& {Vibert}}{{Hatton} et~al.}{2003}]{Galics03}
{Hatton} S.,  {Devriendt} J.~E.~G.,  {Ninin} S.,  {Bouchet} F.~R.,
  {Guiderdoni} B.,    {Vibert} D.,  2003, \mnras, 343, 75

\bibitem[\protect\citeauthoryear{Herbert, Jarvis, Willott, McLure, Mitchell,
  Rawlings, Hill \& Dunlop}{Herbert et~al.}{2011}]{Herbert:2011a}
Herbert P.~D.,  Jarvis M.~J.,  Willott C.~J.,  McLure R.~J.,  Mitchell E.,
  Rawlings S.,  Hill G.~J.,    Dunlop J.~S.,  2011, \mnras, 410, 1360

\bibitem[\protect\citeauthoryear{Holden, van~der Wel, Kelson, Franx \&
  Illingworth}{Holden et~al.}{2010}]{Holden:2010a}
Holden B.~P.,  van~der Wel A.,  Kelson D.~D.,  Franx M.,    Illingworth G.~D.,
  2010, ArXiv e-prints, 1009, 4479

\bibitem[\protect\citeauthoryear{{Hopkins}, {Cox} \& {Hernquist}}{{Hopkins}
  et~al.}{2008}]{HopkinsFP}
{Hopkins} P.~F.,  {Cox} T.~J.,    {Hernquist} L.,  2008, ArXiv e-prints, 806

\bibitem[\protect\citeauthoryear{Hopkins, Cox, Younger \& Hernquist}{Hopkins
  et~al.}{2009}]{Hopkins:2009d}
Hopkins P.~F.,  Cox T.~J.,  Younger J.~D.,    Hernquist L.,  2009, \apj, 691,
  1168

\bibitem[\protect\citeauthoryear{{Kannappan}}{{Kannappan}}{2004}]{Kannappan04}
{Kannappan} S.~J.,  2004, \apjl, 611, L89

\bibitem[\protect\citeauthoryear{{Kauffmann}, {White} \&
  {Guiderdoni}}{{Kauffmann} et~al.}{1993}]{kwg1993}
{Kauffmann} G.,  {White} S.~D.~M.,    {Guiderdoni} B.,  1993, \mnras, 264, 201

\bibitem[\protect\citeauthoryear{{Khochfar} \& {Silk}}{{Khochfar} \&
  {Silk}}{2006}]{KhochfarSilk06}
{Khochfar} S.,  {Silk} J.,  2006, \apjl, 648, L21

\bibitem[\protect\citeauthoryear{{Kormendy}}{{Kormendy}}{1977}]{Kormendy77}
{Kormendy} J.,  1977, \apj, 218, 333

\bibitem[\protect\citeauthoryear{{Mancini}, {Daddi}, {Renzini}, {Salmi},
  {McCracken}, {Cimatti}, {Onodera}, {Salvato}, {Koekemoer}, {Aussel}, {Le
  Floc'h}, {Willott} \& {Capak}}{{Mancini} et~al.}{2010}]{Mancini10}
{Mancini} C.,  {Daddi} E.,  {Renzini} A.,  {Salmi} F.,  {McCracken} H.~J.,
  {Cimatti} A.,  {Onodera} M.,  {Salvato} M.,  {Koekemoer} A.~M.,  {Aussel} H.,
   {Le Floc'h} E.,  {Willott} C.,    {Capak} P.,  2010, \mnras, 401, 933

\bibitem[\protect\citeauthoryear{{McIntosh}, {Bell}, {Rix}, {Wolf}, {Heymans},
  {Peng}, {Somerville}, {Barden}, {Beckwith}, {Borch}, {Caldwell},
  {H{\"a}u{\ss}ler}, {Jahnke}, {Jogee}, {Meisenheimer}, {S{\'a}nchez} \&
  {Wisotzki}}{{McIntosh} et~al.}{2005}]{McIntosh05}
{McIntosh} D.~H.,  {Bell} E.~F.,  {Rix} H.-W.,  {Wolf} C.,  {Heymans} C.,
  {Peng} C.~Y.,  {Somerville} R.~S.,  {Barden} M.,  {Beckwith} S.~V.~W.,
  {Borch} A.,  {Caldwell} J.~A.~R.,  {H{\"a}u{\ss}ler} B.,  {Jahnke} K.,
  {Jogee} S.,  {Meisenheimer} K.,  {S{\'a}nchez} S.~F.,    {Wisotzki} L.,
  2005, \apj, 632, 191

\bibitem[\protect\citeauthoryear{{Mihos} \& {Hernquist}}{{Mihos} \&
  {Hernquist}}{1994}]{MH94dsc}
{Mihos} J.~C.,  {Hernquist} L.,  1994, ApJL, 437, L47

\bibitem[\protect\citeauthoryear{{Naab}, {Jesseit} \& {Burkert}}{{Naab}
  et~al.}{2006}]{Naabgas}
{Naab} T.,  {Jesseit} R.,    {Burkert} A.,  2006, \mnras, 372, 839

\bibitem[\protect\citeauthoryear{Naab, Johansson \& Ostriker}{Naab
  et~al.}{2009}]{Naab:2009a}
Naab T.,  Johansson P.~H.,    Ostriker J.~P.,  2009, ArXiv e-prints,
  astro-ph.CO

\bibitem[\protect\citeauthoryear{{Naab}, {Khochfar} \& {Burkert}}{{Naab}
  et~al.}{2006}]{Naab06}
{Naab} T.,  {Khochfar} S.,    {Burkert} A.,  2006, \apjl, 636, L81

\bibitem[\protect\citeauthoryear{{Nair}, {van den Bergh} \& {Abraham}}{{Nair}
  et~al.}{2010}]{Nair10}
{Nair} P.~B.,  {van den Bergh} S.,    {Abraham} R.~G.,  2010, \apj, 715, 606

\bibitem[\protect\citeauthoryear{{Nipoti}, {Treu} \& {Bolton}}{{Nipoti}
  et~al.}{2008}]{Nipoti08}
{Nipoti} C.,  {Treu} T.,    {Bolton} A.~S.,  2008, ArXiv e-prints, 806

\bibitem[\protect\citeauthoryear{Oser, Ostriker, Naab, Johansson \&
  Burkert}{Oser et~al.}{2010}]{Oser:2010a}
Oser L.,  Ostriker J.~P.,  Naab T.,  Johansson P.~H.,    Burkert A.,  2010,
  ArXiv e-prints, 1010, 1381

\bibitem[\protect\citeauthoryear{{Padmanabhan}, {Seljak}, {Strauss}, {Blanton},
  {Kauffmann}, {Schlegel}, {Tremonti}, {Bahcall}, {Bernardi}, {Brinkmann},
  {Fukugita} \& {Ivezi{\'c}}}{{Padmanabhan} et~al.}{2004}]{Padmanabhan04}
{Padmanabhan} N.,  {Seljak} U.,  {Strauss} M.~A.,  {Blanton} M.~R.,
  {Kauffmann} G.,  {Schlegel} D.~J.,  {Tremonti} C.,  {Bahcall} N.~A.,
  {Bernardi} M.,  {Brinkmann} J.,  {Fukugita} M.,    {Ivezi{\'c}} {\v Z}.,
  2004, New Astronomy, 9, 329

\bibitem[\protect\citeauthoryear{{Pahre}, {Djorgovski} \& {de
  Carvalho}}{{Pahre} et~al.}{1998}]{Pahre98}
{Pahre} M.~A.,  {Djorgovski} S.~G.,    {de Carvalho} R.~R.,  1998, \aj, 116,
  1591

\bibitem[\protect\citeauthoryear{Robertson, Bullock, Cox, Matteo, Hernquist,
  Springel \& Yoshida}{Robertson et~al.}{2006}]{Robertson:2006b}
Robertson B.,  Bullock J.~S.,  Cox T.~J.,  Matteo T.~D.,  Hernquist L.,
  Springel V.,    Yoshida N.,  2006, \apj, 645, 986

\bibitem[\protect\citeauthoryear{{Robertson}, {Cox}, {Hernquist}, {Franx},
  {Hopkins}, {Martini} \& {Springel}}{{Robertson} et~al.}{2006}]{RobertsonFP}
{Robertson} B.,  {Cox} T.~J.,  {Hernquist} L.,  {Franx} M.,  {Hopkins} P.~F.,
  {Martini} P.,    {Springel} V.,  2006, \apj, 641, 21

\bibitem[\protect\citeauthoryear{Saglia, Sanchez-Blazquez, Bender, Simard,
  Desai, Aragon-Salamanca, Milvang-Jensen, Halliday, Jablonka, Noll, Poggianti,
  Clowe, Lucia, Pello, Rudnick, Valentinuzzi, White \& Zaritsky}{Saglia
  et~al.}{2010}]{Saglia:2010a}
Saglia R.~P.,  Sanchez-Blazquez P.,  Bender R.,  Simard L.,  Desai V.,
  Aragon-Salamanca A.,  Milvang-Jensen B.,  Halliday C.,  Jablonka P.,  Noll
  S.,  Poggianti B.,  Clowe D.~I.,  Lucia G.~D.,  Pello R.,  Rudnick G.,
  Valentinuzzi T.,  White S. D.~M.,    Zaritsky D.,  2010, ArXiv e-prints,
  1009, 645

\bibitem[\protect\citeauthoryear{{Saracco}, {Longhetti} \& {Andreon}}{{Saracco}
  et~al.}{2009}]{Saracco09}
{Saracco} P.,  {Longhetti} M.,    {Andreon} S.,  2009, \mnras, 392, 718

\bibitem[\protect\citeauthoryear{{Shankar}, {Marulli}, {Bernardi},
  {Boylan-Kolchin}, {Dai} \& {Khochfar}}{{Shankar} et~al.}{2010}]{Shankar10a}
{Shankar} F.,  {Marulli} F.,  {Bernardi} M.,  {Boylan-Kolchin} M.,  {Dai} X.,
   {Khochfar} S.,  2010, \mnras, 405, 948

\bibitem[\protect\citeauthoryear{{Shankar}, {Marulli}, {Bernardi}, {Dai},
  {Hyde} \& {Sheth}}{{Shankar} et~al.}{2010}]{Shankar10b}
{Shankar} F.,  {Marulli} F.,  {Bernardi} M.,  {Dai} X.,  {Hyde} J.~B.,
  {Sheth} R.~K.,  2010, \mnras, 403, 117

\bibitem[\protect\citeauthoryear{Shen, Mo, White, Blanton, Kauffmann, Voges,
  Brinkmann \& Csabai}{Shen et~al.}{2003}]{Shen03}
Shen S.,  Mo H.~J.,  White S. D.~M.,  Blanton M.~R.,  Kauffmann G.,  Voges W.,
  Brinkmann J.,    Csabai I.,  2003, \mnras, 343, 978

\bibitem[\protect\citeauthoryear{{Somerville}, {Barden}, {Rix}, {Bell},
  {Beckwith}, {Borch}, {Caldwell}, {H{\"a}u{\ss}ler}, {Heymans}, {Jahnke},
  {Jogee}, {McIntosh}, {Meisenheimer}, {Peng}, {S{\'a}nchez}, {Wisotzki} \&
  {Wolf}}{{Somerville} et~al.}{2008}]{Somerville08a}
{Somerville} R.~S.,  {Barden} M.,  {Rix} H.-W.,  {Bell} E.~F.,  {Beckwith}
  S.~V.~W.,  {Borch} A.,  {Caldwell} J.~A.~R.,  {H{\"a}u{\ss}ler} B.,
  {Heymans} C.,  {Jahnke} K.,  {Jogee} S.,  {McIntosh} D.~H.,  {Meisenheimer}
  K.,  {Peng} C.~Y.,  {S{\'a}nchez} S.~F.,  {Wisotzki} L.,    {Wolf} C.,  2008,
  \apj, 672, 776

\bibitem[\protect\citeauthoryear{{Somerville}, {Hopkins}, {Cox}, {Robertson} \&
  {Hernquist}}{{Somerville} et~al.}{2008}]{S08}
{Somerville} R.~S.,  {Hopkins} P.~F.,  {Cox} T.~J.,  {Robertson} B.~E.,
  {Hernquist} L.,  2008, \mnras, 391, 481

\bibitem[\protect\citeauthoryear{{Somerville} \& {Primack}}{{Somerville} \&
  {Primack}}{1999}]{SP1999}
{Somerville} R.~S.,  {Primack} J.~R.,  1999, \mnras, 310, 1087

\bibitem[\protect\citeauthoryear{{Somerville}, {Primack} \&
  {Faber}}{{Somerville} et~al.}{2001}]{SPF}
{Somerville} R.~S.,  {Primack} J.~R.,    {Faber} S.~M.,  2001, \mnras, 320, 504

\bibitem[\protect\citeauthoryear{{Springel} \& {Hernquist}}{{Springel} \&
  {Hernquist}}{2003}]{SH03}
{Springel} V.,  {Hernquist} L.,  2003, \mnras, 339, 289

\bibitem[\protect\citeauthoryear{Springel \& Hernquist}{Springel \&
  Hernquist}{2005}]{Springel:2005c}
Springel V.,  Hernquist L.,  2005, \apj, 622, L9

\bibitem[\protect\citeauthoryear{{Springel}, {White}, {Jenkins}, {Frenk},
  {Yoshida}, {Gao}, {Navarro}, {Thacker}, {Croton}, {Helly}, {Peacock}, {Cole},
  {Thomas}, {Couchman}, {Evrard}, {Colberg} \& {Pearce}}{{Springel}
  et~al.}{2005}]{Springel05}
{Springel} V.,  {White} S.~D.~M.,  {Jenkins} A.,  {Frenk} C.~S.,  {Yoshida} N.,
   {Gao} L.,  {Navarro} J.,  {Thacker} R.,  {Croton} D.,  {Helly} J.,
  {Peacock} J.~A.,  {Cole} S.,  {Thomas} P.,  {Couchman} H.,  {Evrard} A.,
  {Colberg} J.,    {Pearce} F.,  2005, \nat, 435, 629

\bibitem[\protect\citeauthoryear{{Springel}, {Yoshida} \& {White}}{{Springel}
  et~al.}{2001}]{SpGad}
{Springel} V.,  {Yoshida} N.,    {White} S.~D.~M.,  2001, New Astronomy, 6, 79

\bibitem[\protect\citeauthoryear{Tacconi, Genzel, Neri, Cox, Cooper, Shapiro,
  Bolatto, Bouch{\'e}, Bournaud, Burkert, Combes, Comerford, Davis, Schreiber,
  Garcia-Burillo, Gracia-Carpio, Lutz, Naab, Omont, Shapley, Sternberg \&
  Weiner}{Tacconi et~al.}{2010}]{Tacconi:2010a}
Tacconi L.~J.,  Genzel R.,  Neri R.,  Cox P.,  Cooper M.~C.,  Shapiro K.,
  Bolatto A.,  Bouch{\'e} N.,  Bournaud F.,  Burkert A.,  Combes F.,  Comerford
  J.,  Davis M.,  Schreiber N. M.~F.,  Garcia-Burillo S.,  Gracia-Carpio J.,
  Lutz D.,  Naab T.,  Omont A.,  Shapley A.,  Sternberg A.,    Weiner B.,
  2010, \nat, 463, 781

\bibitem[\protect\citeauthoryear{{Toomre}}{{Toomre}}{1977}]{T77}
{Toomre} A.,  1977, in Evolution of Galaxies and Stellar Populations {Mergers
  and Some Consequences}.
p.~401

\bibitem[\protect\citeauthoryear{{Toomre} \& {Toomre}}{{Toomre} \&
  {Toomre}}{1972}]{TT72}
{Toomre} A.,  {Toomre} J.,  1972, \apj, 178, 623

\bibitem[\protect\citeauthoryear{{Trujillo}, {Conselice}, {Bundy}, {Cooper},
  {Eisenhardt} \& {Ellis}}{{Trujillo} et~al.}{2007}]{Trujillo07}
{Trujillo} I.,  {Conselice} C.~J.,  {Bundy} K.,  {Cooper} M.~C.,  {Eisenhardt}
  P.,    {Ellis} R.~S.,  2007, \mnras, 382, 109

\bibitem[\protect\citeauthoryear{{Trujillo et al.}}{{Trujillo et
  al.}}{2006}]{Trujillo06}
{Trujillo et al.} 2006, \apj, 650, 18

\bibitem[\protect\citeauthoryear{{van der Wel}, {Holden}, {Zirm}, {Franx},
  {Rettura}, {Illingworth} \& {Ford}}{{van der Wel} et~al.}{2008}]{vanderWel08}
{van der Wel} A.,  {Holden} B.~P.,  {Zirm} A.~W.,  {Franx} M.,  {Rettura} A.,
  {Illingworth} G.~D.,    {Ford} H.~C.,  2008, \apj, 688, 48

\bibitem[\protect\citeauthoryear{{van Dokkum}, {Whitaker}, {Brammer}, {Franx},
  {Kriek}, {Labb{\'e}}, {Marchesini}, {Quadri}, {Bezanson}, {Illingworth},
  {Muzzin}, {Rudnick}, {Tal} \& {Wake}}{{van Dokkum}
  et~al.}{2010}]{vanDokkum10}
{van Dokkum} P.~G.,  {Whitaker} K.~E.,  {Brammer} G.,  {Franx} M.,  {Kriek} M.,
   {Labb{\'e}} I.,  {Marchesini} D.,  {Quadri} R.,  {Bezanson} R.,
  {Illingworth} G.~D.,  {Muzzin} A.,  {Rudnick} G.,  {Tal} T.,    {Wake} D.,
  2010, \apj, 709, 1018

\bibitem[\protect\citeauthoryear{{Williams}, {Quadri}, {Franx}, {van Dokkum},
  {Toft}, {Kriek} \& {Labb{\'e}}}{{Williams} et~al.}{2010}]{Williams10}
{Williams} R.~J.,  {Quadri} R.~F.,  {Franx} M.,  {van Dokkum} P.,  {Toft} S.,
  {Kriek} M.,    {Labb{\'e}} I.,  2010, \apj, 713, 738

\bibitem[\protect\citeauthoryear{Wuyts, Cox, Hayward, Franx, Hernquist,
  Hopkins, Jonsson \& van Dokkum}{Wuyts et~al.}{2010}]{Wuyts:2010b}
Wuyts S.,  Cox T.~J.,  Hayward C.~C.,  Franx M.,  Hernquist L.,  Hopkins P.~F.,
   Jonsson P.,    van Dokkum P.~G.,  2010, \apj, 722, 1666

\bibitem[\protect\citeauthoryear{{York}, {Adelman}, {Anderson} Jr., {Anderson},
  {Annis}, {Bahcall}, {Bakken}, {Barkhouser}, {Bastian}, {Berman}, {Boroski},
  {Yanny} \& {Yasuda}}{{York} et~al.}{2000}]{York00}
{York} D.~G.,  {Adelman} J.,  {Anderson} Jr. J.~E.,  {Anderson} S.~F.,  {Annis}
  J.,  {Bahcall} N.~A.,  {Bakken} J.~A.,  {Barkhouser} R.,  {Bastian} S.,
  {Berman} E.,  {Boroski} W.~N.,  {Yanny} B.,    {Yasuda} N.,  2000, \aj, 120,
  1579

\bibitem[\protect\citeauthoryear{{Zaritsky}, {Zabludoff} \&
  {Gonzalez}}{{Zaritsky} et~al.}{2007}]{Zaritsky07}
{Zaritsky} D.,  {Zabludoff} A.~I.,    {Gonzalez} A.~H.,  2007, ArXiv e-prints,
  711

\end{thebibliography}

\section{Appendix A: New Model for Central Dark Matter Fraction}
\label{sec:appendix1}

In C08 we presented a formula for calculating the dark matter fraction
in the central part of the merger remnants.  We define the dark matter
fraction with a given radius as
\begin{equation}
f_{\rm dm}=\frac{M_{\rm dm}}{(M_{\rm dm} + M_{\rm stars})},
\end{equation}
where $M_{\rm dm}$ and $M_{\rm stars}$ are the dark matter and stellar
masses inside that radius, respectively.  For the merger simulations
we found that the dark matter fraction inside half of the stellar half
mass radius could be well-approximated by the following formula:
\begin{equation}
f_{\rm dm,f}= \frac{M_{\rm dm,1} + M_{\rm dm,2}}{M_{\rm dm,1} + M_{\rm dm,2} + C_{\rm stars}(M_{\rm 1} + M_{\rm 2} + M_{\rm new})}.
\end{equation}
$M_{\rm dm,1}$ and $M_{\rm dm,2}$ are the dark matter masses inside
half of the three-dimensional stellar half-mass radii of the
progenitors, $M_{\rm 1}$ and $M_{\rm 2}$ are the total stellar masses
of the progenitors, and $M_{\rm new}$ is the total mass of stars
formed during the merger. This expression simply assumes that the
inner region of the remnant contains the same amount of dark matter as
the sum of the inner regions of the progenitors, and that a fixed
fraction, $C_{\rm stars}$, of the final stellar mass is inside
one-half of the three-dimensional stellar half-mass radius.

While this formula works well for the simulated remnants, it fails
outside the regime where $R_{progenitor}\sim R_{remnant}$.  One of the
benefits of the formula is that since it puts all of the central dark
matter from the progenitors into the central portion of the remnant it
allows contraction of the halo as a result of the baryonic
dissipation.  However, this contraction is precisely what causes the
problem when the initial and final radii differ greatly.
Specifically, for high gas fraction progenitors with large radii, the
remnants can have much smaller radii and the previous formula predicts
extreme contraction of the dark matter.

In order to increase the range of applicability of the merger model,
we introduce a simpler, more physically intuitive model for predicting
the central dark matter fraction that does not suffer from the
deficiencies mentioned above.  We begin with the dark matter halo
masses and half-mass radii, and assume that the two halos merge
dissipationlessly so that
\begin{equation}
\frac{(M_{\rm 1,dm} + M_{\rm 2,dm})^2}{R_{\rm dm,f}} = \frac{M_{\rm 1,dm}^2}{R_{\rm 1,dm}} + \frac{M_{\rm 2,dm}^2}{R_{\rm 2,dm}}. 
\end{equation}
This equation can be solved for the final halo half-mass radius
($R_{\rm dm,f}$).  We use the final radius and mass to fit an
isothermal profile to the final halo. Then we calculate the mass
expected inside the stellar half-mass radius and this value is used to
calculate the final central dark matter fraction.  For this paper we
calculate the dark matter fraction inside the stellar half-mass radius
and use this to compute the velocity dispersion of the merger remnant.
In addition to extending the range of validity, this new method
removes one of the least certain parameters from the merger model.  A
check against the merger simulations demonstrates that the new method
produces a larger scatter between predicted and measured velocity
dispersion, with the fractional rms scatter increasing from 0.24 to
0.35.  However, most of the additional scatter results from variations
in orbit, and in this work we find that the distribution of orbits
found in N-body simulations is such that orbital variation plays a
minor role.

\end{document}